\documentclass[11pt]{article}
\usepackage{graphicx}
\newlength{\dgheight}
\usepackage[a4paper, width = 150mm, top = 30mm, bottom = 30mm]{geometry}
\usepackage{jheppub}
\usepackage{amsmath}
\usepackage{hyperref}
\usepackage{tikz-feynman}
\tikzfeynmanset{compat=1.1.0}
\usetikzlibrary{decorations.pathmorphing, calc}
\usepackage{float}
\usepackage{mathrsfs}
\usepackage{subcaption}
\usepackage[normalem]{ulem}

\usetikzlibrary{external}
\tikzsetfigurename{feyn}
\tikzset{external/mode=graphics if exists}

\tikzfeynmanset{
every plain={gray},
}

\tikzfeynmanset{ graviton/.style={thick,double, decorate, decoration={snake, amplitude=1.75pt, segment length=7pt} } }

\tikzfeynmanset{pertworld/.style={double, thick, gray} }

\newcommand{\scalarhalfangle}{50}
\newcommand{\gravskeletonp}{%
  \vertex (a) at (0,0);
  \vertex (d) at (1.5,0);
  \vertex (e) at (180-\scalarhalfangle:1.8);
  \vertex (f) at (180+\scalarhalfangle:1.8);
}

\newcommand{\legreach}{1.75}
\newcommand{\loopreach}{0.95}
\newcommand{\hardsep}{2.0}

\newcommand{\gravskeleton}{%
  \vertex (a) at (0,0);
  \vertex (d) at (\hardsep,0);
  \vertex (e) at ($(a)+(180-\scalarhalfangle:\legreach)$);
  \vertex (f) at ($(a)+(180+\scalarhalfangle:\legreach)$);
  \vertex (g) at ($(d)+(\scalarhalfangle:\legreach)$;
  \vertex (h) at ($(d)+(-\scalarhalfangle:\legreach)$);
  \vertex (x) at ($(a)+(180-\scalarhalfangle:\loopreach)$);
  \vertex (y) at ($(a)+(180+\scalarhalfangle:\loopreach)$);
  \vertex (z) at ($(d)+(\scalarhalfangle:\loopreach)$);
  \vertex (w) at ($(d)+(-\scalarhalfangle:\loopreach)$);
}

\newcommand{\gravskeletonpp}{%
  \gravskeleton
  \vertex (g) at ($(d)+(\scalarhalfangle:\legreach)$);
  \vertex (h) at ($(d)+(-\scalarhalfangle:\legreach)$);
}

\hypersetup{
    colorlinks=true,
    linkcolor=blue,
    filecolor=blue,      
    urlcolor=blue, 
    citecolor=blue
    }

\newcommand*\diff{\mathop{}\!\mathrm{d}}
\newcommand{\bas}{\begin{equation}\begin{split}}
\newcommand{\eas}{\end{split}}
\newcommand{\eeq}{\end{equation}}
\newcommand{\mc}{\mathcal}

\renewcommand{\[}{\begin{equation}\begin{aligned}}
\renewcommand{\]}{\end{aligned}\end{equation}}
\newcommand{\Ecal}{\mathcal E}
\newcommand{\Bcal}{\mathcal B}
\newcommand{\rh}{r_{\rm h}}
\def\dd{\mathrm{d}}

\title{Quantum Love Numbers are Non-Zero}

\author[a]{Asaad Elkhidir}
\affiliation[a]{Institut des Hautes Études Scientifiques, 91440 Bures-sur-Yvette, France}

\author[b]{Godwin Martin}
\affiliation[b]{International Centre for Theoretical Sciences, Tata Institute of Fundamental Research,\\
Bengaluru 560089, India}

\author[a]{Julio Parra-Martinez}

\author[b]{M. V. S. Saketh}

\emailAdd{elkhidir@ihes.fr, godwin.martin@icts.res.in, julio@ihes.fr, venkata.saketh@icts.res.in}

\abstract{
The static Love numbers of four-dimensional Schwarzschild black holes vanish classically. We show that this is not true in the quantum theory. Working in the worldline effective field theory, we compute the leading quantum correction to the Compton amplitude for massless scalars and photons scattering off a Schwarzschild black hole at first post-Minkowskian order, and find logarithmic divergences associated with the renormalization of the static Love numbers. We conclude that renormalization-group running necessarily generates Love numbers suppressed by the square of the Planck length in units of the horizon radius, $\ell_{\rm pl}^{2}/r_{\rm h}^{2}$. This is consistent with the classical vanishing being a consequence of an accidental symmetry of static general relativity, broken by quantum loops. Extrapolated to gravitational perturbations, our results suggest that the quadrupolar static Love numbers of Schwarzschild are of size $\lambda \sim M r_{\rm h}^{2} \ell_{\rm pl}^{2}$.
}

\begin{document}

\maketitle

\section{Introduction}
It is by now a classic result that the static tidal response coefficients, known as Love numbers, of a four-dimensional Schwarzschild black hole vanish identically for both scalar and tensor perturbations and for both linear and nonlinear tides \cite{Fang:2005qq, Damour:2009vw, Binnington:2009bb, Damour:2009va, Kol:2011vg, Gurlebeck:2015xpa, Hui:2020xxx, Poisson:2021yau, LeTiec:2020spy, LeTiec:2020bos, Chia:2020yla, Riva:2023Love, Iteanu:2024Love, Parra-Martinez:2025bcu}. This is a consequence of an emergent symmetry of time-independent solutions of General Relativity \cite{Penna:2018gfx, Charalambous:2021mea, Charalambous:2021kcz, Hui:2021vcv, BenAchour:2022uqo, Charalambous:2022rre, Katagiri:2022vyz, Berens:2022ebl, Charalambous:2023jgq, Sharma:2024hlz, Charalambous:2024tdj, Rai:2024lho, Charalambous:2024gpf, Gray:2024qys, Combaluzier-Szteinsznaider:2024sgb, Gounis:2024hcm, Charalambous:2025ekl, Berens:2025okm, Sharma:2025xii, Guevara:2025psg, DeLuca:2025zqr, Cvetic:2026wht,Ghosh:2026vig}. It is not known, however, whether this vanishing is an accident of the classical limit or a robust property of the theory~\cite{Porto:2016zng,Parra-Martinez:2025bcu}. After all, static perturbations can source finite-frequency quantum fluctuations of the gravitational field which violate the symmetry and could lead to a non-zero Love number. Indeed, higher-derivative corrections to General Relativity are known to induce non-vanishing Love numbers~\cite{Barbosa:2025uau, Barbosa:2026qcv}. Here we instead consider quantum effects within General Relativity itself, which are universal. We address this question in a simple setting by studying the scalar and electromagnetic Love numbers of a Schwarzschild black hole at quantum one-loop and first post-Minkowskian (1PM) order. This corresponds to linear order in the black-hole mass, $GM$, and one order in $\hbar=G\ell_{\rm pl}^2$ beyond the classical result.

Worldline effective field theory~\cite{Goldberger:2004jt,Goldberger:2005cd,Goldberger:2022ebt} provides a systematic framework for defining and computing the tidal response of compact objects. In this approach, the compact object is modeled as a worldline and its internal ultraviolet (UV) physics is encoded in Wilson coefficients that affect the worldline equations of motion. These coefficients describe finite-size effects such as induced multipole moments and tidal response. Once they have been constrained by the UV physics, the resulting dynamics is immediately apparent from the effective worldline action. Well-established EFT methods can then connect this action to the dynamics of a two-body system and subsequently to gravitational waveforms~\cite{Porto:2016pyg, Levi:2018nxp, Barack:2023oqp}.

The EFT framework also permits the use of efficient quantum-field-theory (QFT) techniques developed in particle physics, including recent advances in multiloop scattering-amplitude calculations. Amplitude-based methods have proved particularly powerful for constraining the Wilson coefficients governing tidal response~\cite{Bern:2020uwk,Cheung:2020sdj,Haddad:2020que,Kalin:2020lmz,Bini:2020flp,Saketh:2023bul,Saketh:2022wap,Saketh:2024juq,Ivanov:2022qqt,Ivanov:2024sds,Jakobsen:2023pvx,Mandal:2023hqa,Ivanov:2026icp,Bautista:2026qse,Brunello:2026lzf}. In this framework, one studies the scattering of waves off an isolated compact object and computes the associated gravitational Raman or Compton amplitude~\cite{Creci:2021rkz,Bautista:2021wfy,Bautista:2022wjf,Bautista:2023sdf,Caron-Huot:2025tlq,Bjerrum-Bohr:2025bqg,Bjerrum-Bohr:2026fhs,Correia:2026utp}. Since this amplitude is an observable, it can be matched between different descriptions within their common regime of validity. In particular, the same amplitude can be computed in black-hole perturbation theory and matched to the EFT result to constrain the Wilson coefficients~\cite{Ivanov:2022hlo,Ivanov:2024sds,Wang:2026qst,Rodriguez:2026iot, Chakraborty:2026qru}. In the classical static limit, this comparison forces the Wilson coefficients associated with the Love numbers to vanish. Beyond the static limit, a black hole has a dynamical tidal response~\cite{Chakrabarti:2013lua,Steinhoff:2016rfi,Saketh:2023bul,Perry:2023wmm,Chakraborty:2023zed,Kosmopoulos:2025rfj,Solon:2026ubm,Chakraborty:2026dox}, including dissipation through its horizon~\cite{Goldberger:2005cd,Chia:2020yla,Charalambous:2021mea,Goldberger:2020fot,Saketh:2022xjb}.

The worldline EFT also lets us determine the running of the scalar and electromagnetic Love numbers from the ultraviolet divergences of the corresponding scattering amplitudes. A divergence with the structure of a Love-number operator requires a counterterm for its Wilson coefficient and therefore signals renormalization-group running. The EFT predicts all logarithms associated with this running at the order considered, irrespective of the microscopics of the compact object. Indeed, in this work we find nonzero beta functions even at vanishing tidal couplings, so the classical vanishing of the Love numbers cannot persist at all scales. Determining the finite values of the Love numbers at a reference scale requires matching to a one-loop computation in a black-hole background, which we leave for future work.

This paper is organized as follows: Section~\ref{sec:WorldlineAction} introduces the worldline action and Feynman rules. Section~\ref{sec:LoveRunning} computes the scalar and electromagnetic running and discusses the gravitational extension. Section~\ref{sec:WorldlineVsQFT} compares the scalar result with the soft limit of a massive-scalar field theory.

We work in mostly minus metric signature $\eta_{\mu\nu} = \mathrm{diag}(+,-,-,-)$ and in natural units $\hbar = c =1$.

\section{Worldline Effective Theory}\label{sec:WorldlineAction}
We use the worldline effective theory to describe the scattering of long-wavelength fields off compact objects, in our case Schwarzschild black holes. When the wavelength of the probe is much larger than the size of  the compact object, its short-distance structure cannot be resolved and can be encoded in local operators supported on its worldline~\cite{Goldberger:2004jt}. 

A Schwarzschild black hole is characterized by its Schwarzschild radius $r_{\mathrm{h}} =2 GM$, which, for a macroscopic black hole, is much larger than its Compton wavelength $\lambda_{\rm C} =1/M$. In terms of the Planck mass, $M_{\rm pl} = 1/\sqrt{G}$, this implies $M_{\rm pl} \ll M$. Furthermore, the scattered wave introduces an additional length scale characterized by its wavelength $L=1/\omega$ where $\omega$ is the wave's frequency. An effective point-particle description requires that the scattered waves cannot resolve the black-hole length scale, enforcing $r_{\mathrm{h}} \ll L$ or equivalently $G M\omega \ll 1$. Our effective description therefore rests on the following hierarchy of scales
\[\label{eq:Goldilocks}
\frac{\omega}{M} \ll GM\omega \ll 1.
\]
This regime contains both classical and quantum contributions to wave-particle scattering. Restoring $\hbar$ in the Compton wavelength, $\lambda_{\rm C}=\hbar/M$, identifies the expansion parameters $GM\omega$ and $\hbar\omega/M$. The latter controls quantum corrections.

\subsection{Worldline Action}
We describe the Schwarzschild black hole of mass $M$ by a worldline $z^\mu(\tau)$ embedded in a dynamical spacetime with metric $g_{\mu\nu}$. We couple the worldline and metric to a real massless scalar $\phi$. The effective action  for the theory is
\begin{equation}
\label{eq:FullAction}
\begin{split}
    S &= -M\int \diff \tau+ \int \diff^4 x \sqrt{-g} \left[\frac{1}{2}\, g^{\mu\nu}
    \partial_\mu \phi\, \partial_\nu \phi - \frac{2}{\kappa^2} R \right]
     \\
    &\quad +  \sum_{\ell=0}^\infty \sum_{n=0}^\infty\lambda_{\ell}^{(n)} \kappa^2 \frac{M}{2}\int \diff\tau \ 
    \big[\partial_{\parallel}^n \partial_{\langle \nu\rangle _\ell \perp} \phi(z(\tau))\big]
    \big[\partial_{\parallel}^n \partial_\perp^{\langle \nu\rangle _\ell} \phi(z(\tau))\big]  \ ,
\end{split}
\end{equation}
where $\kappa$ is the graviton coupling defined as
\begin{equation}
    \kappa^2 \equiv 32\pi G = \frac{32\pi}{M_{\rm pl}^2} = 32\pi \ell_{\rm pl}^2 \ .
\end{equation}
 
The first line of Eq.~\eqref{eq:FullAction} is the sum of the point-particle action, the bulk scalar kinetic term, and the Einstein-Hilbert action. In the point-particle action, the proper time is defined by
\begin{equation}
    \diff\tau = \sqrt{g_{\mu\nu}\diff z^\mu \diff z^\nu} \ .
\end{equation}
The second line contains the operators describing the body's scalar tidal response. The symbol $\partial_{\parallel}$ denotes the derivative parallel to the four velocity $v^\mu = \diff z^\mu/\diff\tau$ of the worldline, while $\partial_{\langle \nu \rangle_{\ell}\perp}$ denotes the symmetric trace-free product of $\ell$ derivatives orthogonal to the four velocity. More precisely,
\begin{equation}\label{eq:STFnotationv}
    \begin{split}
        \partial_{\parallel} &\equiv v^{\mu} \partial_\mu \ ,\\
        \partial_{\langle \mu \rangle_{\ell}\perp} & \equiv \text{STF} \left[\prod_{i=1}^\ell \mathbb{P}_{\mu_i}^{\ \nu_i} \partial_{\nu_i}\right]=\text{STF}\left[\prod_{i=1}^\ell(\delta_{\mu_i}^{\nu_i} - v_{\mu_i} v^{\nu_i}) \partial_{\nu_i}\right] \ .
    \end{split}
\end{equation}
The projector $\mathbb{P}_{\mu}^{\ \nu}$ projects orthogonally to the worldline, i.e., onto the spatial slice in the instantaneous rest frame of the body. STF denotes the symmetric trace-free projection. For $\ell=2$,
\begin{equation}
    \partial_{\langle \mu \rangle_2 \perp}=  \left(\mathbb{P}^{\nu_1}_{\mu_1}\mathbb{P}^{\nu_2}_{\mu_2} -\frac{1}{D-1} \mathbb{P}_{\mu_1\mu_2}\mathbb{P}^{\nu_1\nu_2}\right)\partial_{\nu_1}\partial_{\nu_2} \ .
\end{equation}

The two sums in Eq.~\eqref{eq:FullAction} organize the response of the body by the two independent ways in which a derivative can act on the external scalar field at the location of the worldline. The label $\ell$ counts \emph{transverse} derivatives $\partial_\perp^\mu$, where the subscript $\perp$ denotes projection with the projector $\mathbb{P}_{\mu}^{\ \nu}$.  Because of the STF projection, $\partial_{\langle \mu \rangle_{\ell}\perp}$ computes the rank-$\ell$ symmetric-trace-free spatial tidal moment of $\phi$ measured in the rest frame of the body, so the operator (in the action) carrying label $\ell$ is the square of the $\ell$-pole tidal field, and $\ell$ is the same multipole label that appears on the black-hole perturbation theory side. The label $n$, on the other hand, counts \emph{longitudinal} derivatives, $\partial_\parallel = v\cdot\partial$, which reduce to $\diff/\diff\tau$ acting on the field profile seen by the body.

Accordingly, $\lambda^{(0)}_\ell$ is precisely the static, or adiabatic, scalar Love number of multipole order $\ell$. Operators with $n > 0$ measure the response to a \emph{time-dependent} tidal environment and their coefficients $\lambda^{(n)}_\ell$ are the dynamical, frequency-dependent Love numbers. The superscript
denotes the exponent of the frequency $\omega$. In frequency space each $\partial_\parallel$ supplies a power of $\omega$, so resumming $n$ reconstructs the full $\omega$-dependence of the $\ell$-pole response function, with $\lambda^{(n)}_\ell$
its Taylor coefficients about zero frequency. Trace operators need not be included separately: they reduce either to $\Box\phi$, which vanishes by the free scalar equation of motion, or to lower-rank structures already present in the sum.

We assume that the scalar has a shift
symmetry $\phi(x) \to \phi(x) + \phi_0$, where $\phi_0$ is a constant. Since the $n = \ell = 0$ operator is simply $\phi^2$ evaluated on the worldline, shift symmetry enforces
\begin{equation}
    \lambda_{0}^{(0)} = 0 \ .
\end{equation}
All other operators have $\ell+n \ge 1$ and carry at least one derivative on each scalar and are thus automatically invariant. We determine which of the remaining couplings $\lambda_{\ell}^{(n)}$ require counterterms at the leading quantum order. We define the dimensionless Love numbers $\hat{\lambda}_{\ell}^{(n)}$ through the rescaling:
\[
\lambda_{\ell}^{(n)} = \hat{\lambda}_{\ell}^{(n)}\, r_{\rm h}^{2(\ell+n)}.
\]
We focus on the sector with $\ell+n=1$.

\subsection{Expansion about flat space and a static source}
\label{sec:Expansion}

We use the Polyakov (einbein) form of the point-particle action to make the flat-space expansion polynomial:
\begin{equation}
\label{eq:PolyakovAction}
\begin{split}
    S &= -\frac{M}{2}\int \diff\tau \left(1 + g_{\mu\nu} v^\mu v^\nu\right)
    + \int \diff^4 x \sqrt{-g}\left[\frac{1}{2}\, g^{\mu\nu}\partial_\mu \phi\,
    \partial_\nu \phi - \frac{2}{\kappa^2}R\right] \\
    &\quad + \frac{M \kappa^2}{2}\sum_{\ell,n=0}^\infty \hat{\lambda}_{\ell}^{(n)}\, r_{\rm h}^{2(\ell+n)} \int \diff\tau\,
    \big[\partial_{\parallel}^n \partial_{\langle \nu\rangle _\ell \perp} \phi(z(\tau))\big]
    \big[\partial_{\parallel}^n \partial_\perp^{\langle \nu\rangle _\ell} \phi(z(\tau))\big]  \ .
\end{split}
\end{equation}
We fix the einbein to unity and impose its equation of motion, $g_{\mu\nu}v^\mu v^\nu=1$. The parameter $\tau$ then equals proper time, and the point-particle action agrees with Eq.~\eqref{eq:FullAction} on shell. The tidal couplings are written in terms of the dimensionless coefficients $\hat\lambda_\ell^{(n)}$.

We now expand around flat spacetime with vanishing scalar background,
\begin{equation}
    g_{\mu\nu} = \eta_{\mu\nu} + \kappa\, h_{\mu\nu} \ ,
    \qquad
    z^\mu(\tau) = u^\mu \tau + \delta z^\mu(\tau) \ ,
\end{equation}
where $h_{\mu\nu}$ is the canonically normalized graviton fluctuation, $u^\mu$ is the constant four-velocity of the unperturbed worldline, normalized as $\eta_{\mu\nu}u^\mu u^\nu = 1$, and $\delta z^\mu(\tau)$ is the displacement, or recoil, of the worldline away from the straight trajectory. Since the worldline is displaced only in response to the field at its location, $\delta z^\mu$ is first order in the metric perturbation, $\delta z^\mu = \mc{O}(\kappa h)$. At the order in $\kappa$ at which we work it is therefore sufficient to keep the worldline action to quadratic order in $\delta z^\mu$.

The tidal operators must in principle also be expanded in $h_{\mu\nu}$ and $\delta z^\mu$, since both $v^\mu$ and the projectors depend on the metric and on the trajectory. For the process of interest, i.e., the scattering of a light scalar probe or photon off the heavy worldline, these perturbations may be dropped at $\mc{O}(M M_{\rm pl}^{-2})$. For scalar and photon scattering at this order, we evaluate the tidal operators on the unperturbed worldline, with $v^\mu\to u^\mu$ and $z^\mu\to u^\mu\tau$.

Keeping the terms relevant to massless-scalar scattering from the worldline at one quantum loop gives the following action:
\begin{equation}
\label{eq:ActionExpanded}
\begin{split}
    S \supset& -M\!\int\! \diff\tau
    - \frac{M\kappa}{2}\!\int\! \diff\tau\, h_{\mu\nu}(u\tau)\, u^\mu u^\nu
    - M \!\int\! \diff\tau \left(\delta z^\alpha F_\alpha
    + \frac{1}{2}\, \eta_{\mu\nu}\, \dot{\delta z}^\mu \dot{\delta z}^\nu \right) \\
    &+ \frac{M \kappa^2}{2}\sum_{\ell = 0, n}^\infty \hat{\lambda}_{\ell}^{(n)}\, r_{\rm h}^{2(\ell+n)} \int \diff\tau\,
    \big[\bar{\partial}_{\parallel}^n \bar{\partial}_{\langle \nu\rangle _\ell \perp} \phi(z(\tau))\big]
    \big[\bar{\partial}_{\parallel}^n \bar{\partial}_\perp^{\langle \nu\rangle _\ell} \phi(z(\tau))\big] \\
    &+ \int \diff^4 x \sqrt{-g} \left[\frac{1}{2} g^{\mu\nu}\partial_\mu \phi\,
    \partial_\nu \phi - \frac{2}{\kappa^2} R \right]\ ,
\end{split}
\end{equation}
where the overdot denotes differentiation with respect to proper time,
\begin{equation}
    \dot{z} \equiv \frac{\diff z}{\diff \tau} \ ,
\end{equation}
and the bar on the derivative operators denotes that they are defined with the constant four velocity $u^{\mu}$, i.e.,
\begin{equation}\label{eq:STFnotationu}
    \begin{split}
        \bar{\partial}_{\parallel} &\equiv u^{\mu} \partial_\mu \ ,\\
        \bar{\partial}_{\langle \mu \rangle_{\ell}\perp} & \equiv {\rm STF} \left[\prod_{i=1}^\ell \bar{\mathbb{P}}_{\mu_i}^{\ \nu_i} \partial_{\nu_i}\right]=\text{STF}\left[\prod_{i=1}^\ell(\delta_{\mu_i}^{\nu_i} - u_{\mu_i} u^{\nu_i}) \partial_{\nu_i}\right] \ .
    \end{split}
\end{equation}
The symbol $F^{\alpha}$ denotes the geodesic force per unit mass acting on the unperturbed trajectory,
\begin{equation}
\label{eq:ForceDef}
    F_\alpha  = -\frac{\kappa}{2} u^{\mu} u^{\nu} (\partial_\mu h_{\nu \alpha}+\partial_\nu h_{\mu \alpha}- \partial_\alpha h_{\mu \nu}) \ .
\end{equation}
Note that the object in parentheses is proportional to the Christoffel symbol at linear order in $\kappa$.

The free-worldline term contributes only to the normalization. The second term is the coupling $-\tfrac{\kappa}{2}h_{\mu\nu}T^{\mu\nu}$ to the stress tensor
\begin{equation}
T^{\mu\nu} = M\!\int\!\diff\tau\, u^\mu u^\nu \delta^{(4)}(x-u\tau) 
\end{equation}
of a static point mass. The third term governs the displacement $\delta z^\mu$ away from the unperturbed geodesic.
The second line is the tidal operator quadratic in $\phi$, so a scalar is absorbed and re-emitted at a single point of the worldline.

The static source preserves translations along $u^\mu$ but breaks spatial translation invariance. Thus each worldline vertex conserves the frequency $\omega=u\cdot k$, while the heavy body can absorb spatial momentum. Bulk vertices conserve four-momentum.

\subsection{Feynman Rules}

To invert the graviton kinetic operator, we impose linearized de Donder gauge and include Faddeev--Popov ghosts. Following the conventions of \cite{Latosh:2025vax}, the gauge-fixing action is
\begin{equation}
    S_{\rm gf} = -\frac{2}{\xi}\int \diff^4 x \ \eta_{\mu \nu}
    \left(\partial_{\rho} h^{\rho \mu} - \frac{1}{2} \partial^\mu h\right)
    \left(\partial_{\sigma} h^{\sigma \nu} - \frac{1}{2} \partial^\nu h\right) \ ,
\end{equation}
with $\xi$ the gauge-fixing parameter and $h \equiv \eta^{\mu\nu}h_{\mu\nu}$. The accompanying ghost action is obtained by varying this gauge condition along a diffeomorphism $\delta_\Lambda h_{\mu\nu} = \partial_\mu \Lambda_\nu +
\partial_\nu \Lambda_\mu + \kappa\, \mathcal{L}_\Lambda h_{\mu\nu}$ and
replacing $\Lambda^\mu$ by the anticommuting vector ghost $c^\mu$,
\begin{equation}
\label{eq:GhostAction}
    S_{\rm gh} = \int \diff^4 x\, \bar{c}_\mu \left[ \Box\, c^\mu
    + \kappa \left( \partial_\rho\, \mathcal{L}_c h^{\rho\mu}
    - \frac{1}{2}\, \partial^\mu\, \eta^{\alpha\beta} \mathcal{L}_c h_{\alpha\beta}
    \right) \right] \ ,
\end{equation}
where
\begin{equation}
    \mathcal{L}_c h_{\mu\nu} = c^\rho \partial_\rho h_{\mu\nu}
    + h_{\rho\nu}\partial_\mu c^\rho + h_{\mu\rho}\partial_\nu c^\rho \ .
\end{equation}
The first term in Eq.~\eqref{eq:GhostAction} produces the ghost propagator. The second term is the single $\bar{c}ch$ vertex, and no higher ghost couplings are generated because the de Donder condition is linear in $h_{\mu\nu}$.

The solid line denotes the unperturbed worldline, and the double line denotes its fluctuation propagator. Dashed, dotted, and double wavy lines denote scalars, ghosts, and gravitons, respectively. We list the bulk and worldline rules separately.

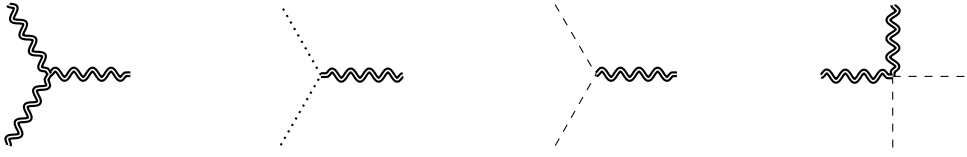
\begin{figure}[H]
    \centering
    \begin{subfigure}{0.24\textwidth}
    \centering
\feynmandiagram [scale=0.75,baseline=(current bounding box.center),horizontal=a to b] {
a -- [graviton] b,
e -- [graviton] a,
f -- [graviton] a,
};
\end{subfigure}%
    \begin{subfigure}{0.24\textwidth}
    \centering
\feynmandiagram [baseline=(current bounding box.center),scale=0.75,horizontal=a to c] {
a -- [ghost] b,
e -- [ghost] a,
c -- [graviton] a,
}; 
\end{subfigure}%
\begin{subfigure}{0.24\textwidth}
    \centering
\feynmandiagram [scale=0.75,baseline=(current bounding box.center),horizontal=a to b] {
a -- [graviton] b,
e -- [scalar] a,
f -- [scalar] a,
};  
\end{subfigure}%
    \begin{subfigure}{0.24\textwidth}
    \centering
     \feynmandiagram [baseline=(current bounding box.center),scale=0.6,horizontal=a to b] {
a -- [scalar] b,
e -- [scalar] a,
f -- [graviton] a,
c -- [graviton] a,
}; 
\end{subfigure}%
    \caption{Feynman vertices for a shift-symmetric scalar interacting with gravity. Conventionally, we take all the momenta at a vertex to be ingoing. The double wavy line denotes the graviton, the dotted line the ghost, and the dashed line the massless scalar.}
    \label{fig:VerticesEFT}
\end{figure}

The bulk vertices are shown in Fig.~\ref{fig:VerticesEFT}; we use the expressions in \texttt{FeynGrav 4.0}~\cite{Latosh:2025vax}. We retain an arbitrary de Donder gauge-fixing parameter $\xi$ to check gauge-parameter independence. The graviton propagator is
\begin{equation}
\begin{split}
\frac{i}{q^2}P_{\mu\nu\rho\sigma}(q)
=&
\frac{i}{q^2}\frac{1}{2}
\left(
\eta^{\mu\sigma}\eta^{\nu\rho}
+\eta^{\mu\rho}\eta^{\nu\sigma}
-\frac{2}{D-2}\eta^{\mu\nu}\eta^{\rho\sigma}
\right)
\\
&+\frac{(\xi-2)}
{4 q^4}
\Big[
q^\nu
\left(
\eta^{\mu\sigma}q^\rho
+\eta^{\mu\rho}q^\sigma
\right)
+q^\mu
\left(
\eta^{\nu\sigma}q^\rho
+\eta^{\nu\rho}q^\sigma
\right)
\Big].
\end{split}
\label{eq:graviton-propagator}
\end{equation}
Here, we have kept the spacetime dimension $D$ explicit, since we will soon be doing our loop calculations in dimensional regularization.

The worldline fluctuation propagator follows from Eq.~\eqref{eq:ActionExpanded} and depends only on the frequency $\omega=u\cdot k$. Note that we have suppressed the $i0^{+}$ prescription in our propagators. This causes no ambiguity since our calculation focuses on UV divergences which are not sensitive to this choice. Reading off the vertices, we obtain the rules in Table~\ref{tab:VerticesWorldline} with
\begin{equation}
    \begin{split}
        V_{\alpha \mu \nu}^{h\delta}(k,q) &= -\frac{\kappa}{2}M[q_\alpha u_\mu u_\nu - 2 u_\mu \eta_{\alpha \nu}(q \cdot u)]\\
        V^{h\delta \delta}_{\mu \nu \alpha \beta}(q,k_1,k_2) & = \frac{i\kappa}{2}M
    \left[q_{\mu}u_{\alpha} - (k_1\cdot u)\eta_{\mu\alpha}\right]
    \left[q_{\nu}u_{\beta} - (k_2\cdot u)\eta_{\nu\beta}\right]\ .
    \end{split}
\end{equation}

\begin{table}[h]
    \centering
    \begin{tabular}{|cc|cc|}
    \hline
           \feynmandiagram [baseline=(current bounding box.center),horizontal=a to b] {
a -- [pertworld] b,
};  &  $-\frac{i}{M \omega^2}\eta^{\mu \nu}$  &  \feynmandiagram [baseline=(current bounding box.center),horizontal=a to b] {
a -- [graviton] b [desired at={(0, 0)}],
b -- c [desired at={(0, 1)}],
b -- d [desired at={(0, -1)}],
};  &   $-i \frac{\kappa}{2} M u^\mu u^\nu$\\
\feynmandiagram [baseline=(current bounding box.center),horizontal=a to b] {
a -- [graviton] b [desired at={(0, 0)}],
b -- [pertworld] c [desired at={(0, 1)}],
b -- d [desired at={(0, -1)}],
};  & $V_{\alpha \mu \nu}^{h\delta}(k,q)$ & \feynmandiagram [baseline=(current bounding box.center),horizontal=a to b] {
a -- [graviton] b [desired at={(0, 0)}],
b -- [pertworld] c [desired at={(0, 1)}],
b -- [pertworld] d [desired at={(0, -1)}],
};  &  $V^{h\delta \delta}_{\mu \nu \alpha \beta}(q,k_1,k_2)$\\
\hline
    \end{tabular}
    \caption{Feynman rules for a worldline of mass $M$ interacting with gravity. The solid line denotes the unperturbed worldline and the double line its fluctuation propagator. Worldline momenta are denoted $k$, with $\omega=u\cdot k$; graviton momenta are denoted $q$. All momenta are incoming.}
    \label{tab:VerticesWorldline}
\end{table}

\section{Running of Love Numbers}\label{sec:LoveRunning}

\subsection{Scalar Love numbers}\label{sec:ScalarLoveRunning}

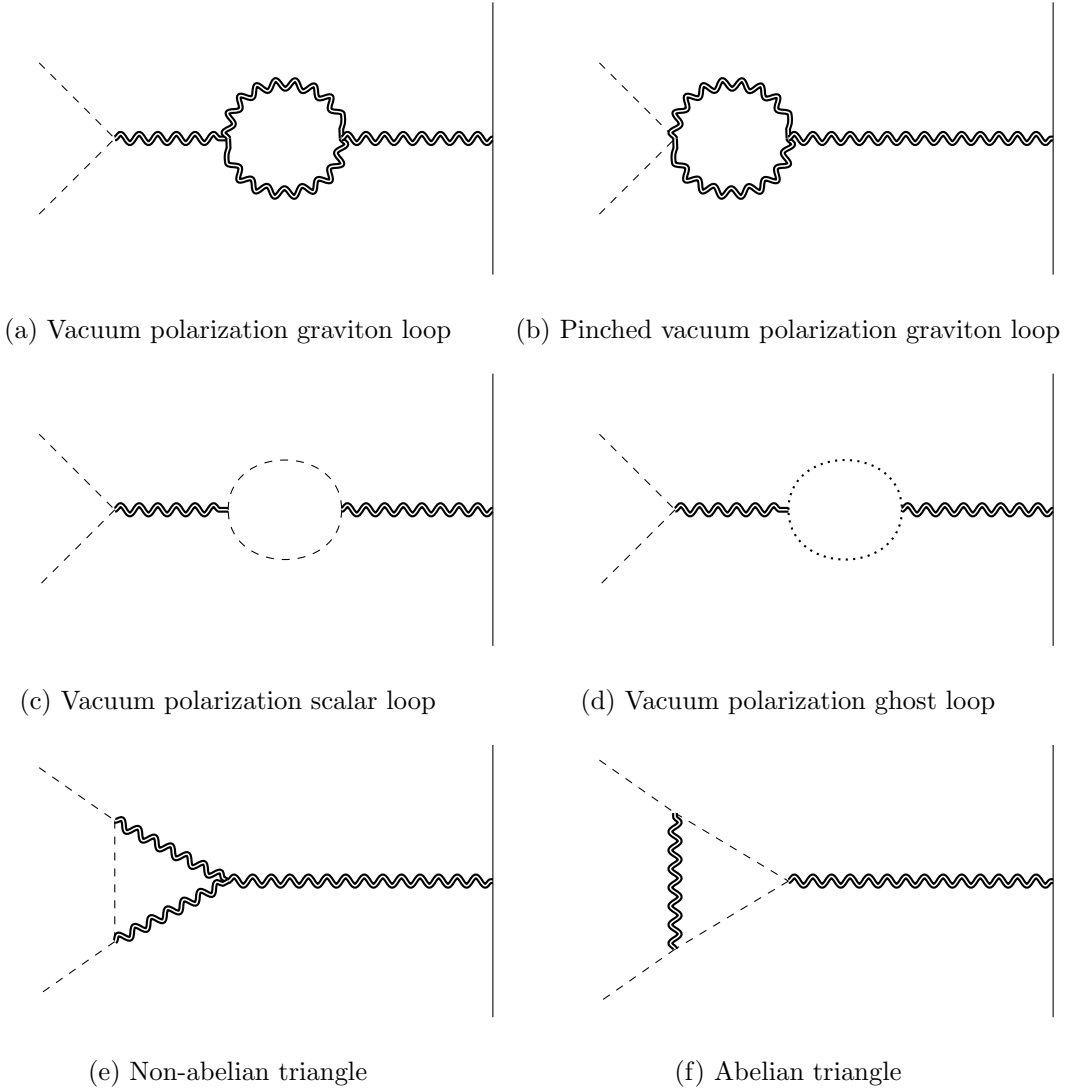
\begin{figure}[h]
    \centering
        \begin{subfigure}{0.49\textwidth}
        \raggedleft
        \begin{tikzpicture}
  \useasboundingbox (-3.2,-2.1) rectangle (3.2,2.1);
  \begin{feynman}
    \vertex (f) at (-3,  1);
    \vertex (e) at (-3, -1);
    \vertex (a) at (-2,  0);
    \vertex (b) at (-0.5, 0);
    \vertex (c) at ( 1,  0);
    \vertex (d) at ( 3,  0);
    \vertex (y) at ( 3,  1.8);
    \vertex (x) at ( 3, -1.8);
    \diagram* {
      (e) -- [scalar] (a),
      (f) -- [scalar] (a),
      (a) -- [graviton] (b),
      (b) -- [graviton, half left] (c),
      (c) -- [graviton, half left] (b),
      (c) -- [graviton] (d),
      (x) -- (d) -- (y),
    };
  \end{feynman}
\end{tikzpicture}
    \caption{Vacuum polarization graviton loop}
    \label{subfig:VacPolGrav}
    \end{subfigure}%
    \begin{subfigure}{0.49\textwidth}
    \raggedleft
       \begin{tikzpicture}
  \useasboundingbox (-3.2,-2.1) rectangle (3.2,2.1);
  \begin{feynman}
    \vertex (f) at (-3,   1);
    \vertex (e) at (-3,  -1);
    \vertex (a) at (-2,   0);
    \vertex (c) at (-0.5, 0);
    \vertex (d) at ( 3,   0);
    \vertex (y) at ( 3,   1.8);
    \vertex (x) at ( 3,  -1.8);
    \diagram* {
      (e) -- [scalar] (a),
      (f) -- [scalar] (a),
      (a) -- [graviton, half left] (c),
      (c) -- [graviton, half left] (a),
      (c) -- [graviton] (d),
      (x) -- (d) -- (y),
    };
  \end{feynman}
\end{tikzpicture}
    \caption{Pinched vacuum polarization graviton loop}
    \label{subfig:PinVacPolGrav}
    \end{subfigure}\\
    \begin{subfigure}{0.49\textwidth}
    \raggedleft
        \begin{tikzpicture}
  \useasboundingbox (-3.2,-2.1) rectangle (3.2,2.1);
  \begin{feynman}
    \vertex (f) at (-3,   1);
    \vertex (e) at (-3,  -1);
    \vertex (a) at (-2,   0);
    \vertex (b) at (-0.5, 0);
    \vertex (c) at ( 1,   0);
    \vertex (d) at ( 3,   0);
    \vertex (y) at ( 3,   1.8);
    \vertex (x) at ( 3,  -1.8);
    \diagram* {
      (f) -- [scalar] (a) -- [scalar] (e),
      (a) -- [graviton] (b),
      (b) -- [scalar, half left] (c),
      (c) -- [scalar, half left] (b),
      (c) -- [graviton] (d),
      (x) -- (d) -- (y),
    };
  \end{feynman}
\end{tikzpicture}
    \caption{Vacuum polarization scalar loop}
    \label{subfig:VacPolScalar}
    \end{subfigure}%
    \begin{subfigure}{0.49\textwidth}
    \raggedleft
        \begin{tikzpicture}
  \useasboundingbox (-3.2,-2.1) rectangle (3.2,2.1);
  \begin{feynman}
    \vertex (f) at (-3,   1);
    \vertex (e) at (-3,  -1);
    \vertex (a) at (-2,   0);
    \vertex (b) at (-0.5, 0);
    \vertex (c) at ( 1,   0);
    \vertex (d) at ( 3,   0);
    \vertex (y) at ( 3,   1.8);
    \vertex (x) at ( 3,  -1.8);
    \diagram* {
      (f) -- [scalar] (a) -- [scalar] (e),
      (a) -- [graviton] (b),
      (b) -- [ghost, half left] (c),
      (c) -- [ghost, half left] (b),
      (c) -- [graviton] (d),
      (x) -- (d) -- (y),
    };
  \end{feynman}
\end{tikzpicture}
        \caption{Vacuum polarization ghost loop}
        \label{subfig:VacPolGhost}
        \end{subfigure}\\
        \begin{subfigure}{0.49\textwidth}
        \raggedleft
        \begin{tikzpicture}
  \useasboundingbox (-3.2,-2.1) rectangle (3.2,2.1);
  \begin{feynman}
    \vertex (p1) at (-3,    1.5);
    \vertex (p2) at (-3,   -1.5);
    \vertex (t2) at (-2,    0.8);
    \vertex (t3) at (-2,   -0.8);
    \vertex (t1) at (-0.5,  0);
    \vertex (d)  at ( 3,    0);
    \vertex (y)  at ( 3,    1.8);
    \vertex (x)  at ( 3,   -1.8);
    \diagram* {
      (p1) -- [scalar] (t2) -- [scalar] (t3) -- [scalar] (p2),
      (t2) -- [graviton] (t1),
      (t3) -- [graviton] (t1),
      (t1) -- [graviton] (d),
      (x) -- (d) -- (y),
    };
    \end{feynman}
  \end{tikzpicture}
    \caption{Non-abelian triangle}
    \label{subfig:NonAbTri}
    \end{subfigure}%
            \begin{subfigure}{0.49\textwidth}
        \raggedleft
            \begin{tikzpicture}
  \useasboundingbox (-3.2,-2.1) rectangle (3.2,2.1);
  \begin{feynman}
    \vertex (a) at (-3,    1.6);
    \vertex (c) at (-2,    0.9);
    \vertex (b) at (-2,   -0.9);
    \vertex (d) at (-3,   -1.6);
    \vertex (e) at (-0.5,  0);
    \vertex (g) at ( 3,    0);
    \vertex (y) at ( 3,    1.8);
    \vertex (x) at ( 3,   -1.8);
    \diagram* {
      (a) -- [scalar] (c) -- [scalar] (e) -- [scalar] (b) -- [scalar] (d),
      (b) -- [graviton] (c),
      (e) -- [graviton] (g),
      (x) -- (g) -- (y),
    };
  \end{feynman}
\end{tikzpicture}
        \caption{Abelian triangle}
        \label{subfig:AbTri}
        \end{subfigure}\\
    \caption{Bulk-loop contributions to scalar scattering from the worldline. Dashed and dotted lines denote scalar and ghost propagators, respectively.}
    \label{fig:ScalWorldBulkLoops}
\end{figure}

With the Feynman rules in hand, we can now assemble the amplitude for a massless scalar scattering off the worldline. We organize the calculation by first enumerating the contributing diagrams. Since we work to one quantum loop and to first post-Minkowskian (PM) order, each diagram carries exactly one insertion of the worldline and one loop of bulk fields. It is convenient to separate the topologies into those in which the loop is confined to the bulk and those in which it attaches to the worldline. We begin with the former, shown in Fig.~\ref{fig:ScalWorldBulkLoops}. These are the loop corrections one would encounter in any gravity--scalar computation, the worldline entering only as a static source. Diagrams that are scaleless, and hence vanish in dimensional regularization, are not displayed.

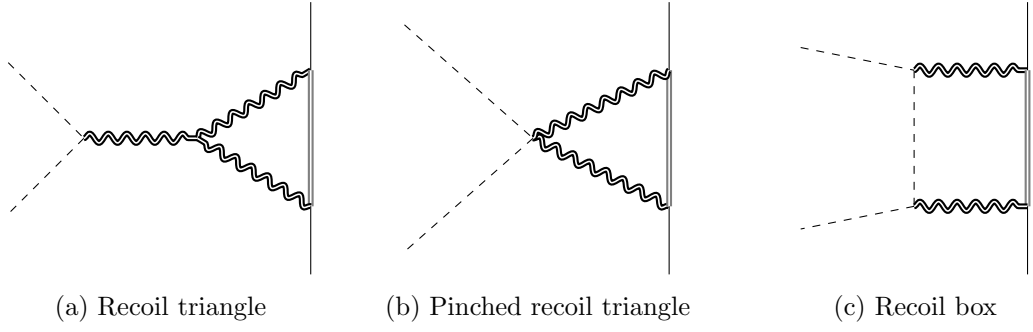
\begin{figure}[h]
    \centering
    \begin{subfigure}{0.33\textwidth}
        \centering
           \begin{tikzpicture}
  \begin{feynman}
    \vertex (e) at (-2,  1);
    \vertex (f) at (-2, -1);
    \vertex (a) at (-1,   0);
    \vertex (b) at (0.5, 0);
    \vertex (c) at ( 2,   0.9);
    \vertex (d) at ( 2,  -0.9);
    \vertex (y) at ( 2,   1.8);
    \vertex (x) at ( 2,  -1.8);

    \diagram* {
      (e) -- [scalar] (a) -- [scalar] (f),
      (a) -- [graviton] (b),
      (b) -- [graviton] (c),
      (c) -- [pertworld] (d),
      (d) -- [graviton] (b),
      (y) -- (c),
      (d) -- (x),
    };
  \end{feynman}
\end{tikzpicture}
        \caption{Recoil triangle}
        \label{subfig:Recoil}
        \end{subfigure}%
        \begin{subfigure}{0.33\textwidth}
    \centering
       \begin{tikzpicture}
  \begin{feynman}
    \vertex (e) at (-1.5,  1.5);
    \vertex (f) at (-1.5, -1.5);
    \vertex (b) at (0.2, 0);
    \vertex (c) at ( 2,   0.9);
    \vertex (d) at ( 2,  -0.9);
    \vertex (y) at ( 2,   1.8);
    \vertex (x) at ( 2,  -1.8);

    \diagram* {
      (e) -- [scalar] (b) -- [scalar] (f),
      (b) -- [graviton] (d),
      (c) -- [pertworld] (d),
      (c) -- [graviton] (b),
      (y) -- (c),
      (d) -- (x),
    };
  \end{feynman}
\end{tikzpicture}
    \caption{Pinched recoil triangle}
    \label{subfig:PinRecoil}
        \end{subfigure}%
        \begin{subfigure}{0.33\textwidth}
        \centering
        \begin{tikzpicture}
  \begin{feynman}
    \vertex (p1) at (-0.5,  1.2);
    \vertex (p2) at (-0.5, -1.2);
    \vertex (t2) at (1,  0.9);
    \vertex (t3) at (1, -0.9);
    \vertex (t1) at ( 2.5,   0.9);
    \vertex (t0) at ( 2.5,  -0.9);
    \vertex (y)  at ( 2.5,   1.8);
    \vertex (x)  at ( 2.5,  -1.8);

    \diagram* {
      (p1) -- [scalar] (t2) -- [scalar] (t3) -- [scalar] (p2),
      (t2) -- [graviton] (t1),
      (t0) -- [pertworld] (t1),
      (t3) -- [graviton] (t0),
      (y) -- (t1),
      (t0) -- (x),
    };
  \end{feynman}
\end{tikzpicture}
    \caption{Recoil box}
    \label{subfig:RecoilTri}
    \end{subfigure}
    \caption{Contributions to scalar scattering from the worldline involving the recoil propagator. Dashed lines denote scalar propagators; double lines denote worldline fluctuation propagators.}
    \label{fig:WorldFlucLoops}
\end{figure}

Figure~\ref{fig:WorldFlucLoops} shows the contributing diagrams with worldline fluctuations. The omitted scaleless topologies follow from two properties of the worldline rules. The fluctuation propagator depends only on the energy $\omega$; its factor of $M$ is an overall normalization and introduces no scale into the denominator. Attachments to the static trajectory impose $\int\diff\tau\,e^{i(k\cdot u)\tau}=2\pi\delta(k\cdot u)$, so the source transfers no energy.

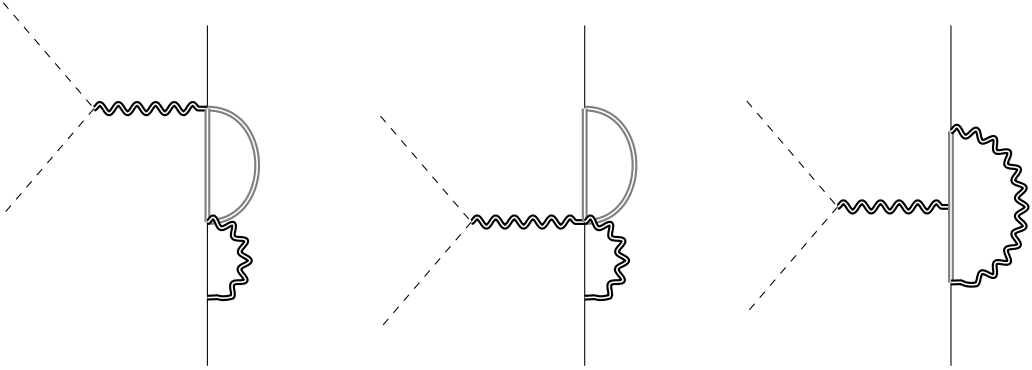
\begin{figure}[h]
    \centering
    \begin{subfigure}{0.33\textwidth}
  \centering
  \begin{tikzpicture}
    \begin{feynman}
      \vertex (e)  at (-0.7,  1.9);
      \vertex (f)  at (-0.7, -0.9);
      \vertex (a)  at ( 0.5,  0.5);
      \vertex (y)  at ( 2,    1.6);
      \vertex (v1) at ( 2,    0.5);
      \vertex (v2) at ( 2,   -1);
      \vertex (x)  at ( 2,   -2.9);
      \vertex (d)  at ( 2,   -2);
      \diagram* {
        (e)  -- [scalar] (a) -- [scalar] (f),
        (a)  -- [graviton] (v1),
        (y)  -- (v1),
        (v1) -- [pertworld, half left]  (v2),
        (v1) -- [pertworld] (v2),
        (v2) -- (x),
        (v2) -- [graviton, half left] (d),
      };
    \end{feynman}
  \end{tikzpicture}
\end{subfigure}%
\begin{subfigure}{0.33\textwidth}
  \centering
  \begin{tikzpicture}
    \begin{feynman}
      \vertex (e)  at (-0.7,  0.4);
      \vertex (f)  at (-0.7, -2.4);
      \vertex (a)  at ( 0.5, -1);
      \vertex (y)  at ( 2,    1.6);
      \vertex (v1) at ( 2,    0.5);
      \vertex (v2) at ( 2,   -1);
      \vertex (x)  at ( 2,   -2.9);
      \vertex (d)  at ( 2,   -2);
      \diagram* {
        (e)  -- [scalar] (a) -- [scalar] (f),
        (a)  -- [graviton] (v2),
        (y)  -- (v1),
        (v1) -- [pertworld, half left]  (v2),
        (v1) -- [pertworld] (v2),
        (v2) -- (x),
        (v2) -- [graviton, half left] (d),
      };
    \end{feynman}
  \end{tikzpicture}
\end{subfigure}%
\begin{subfigure}{0.33\textwidth}
  \centering
  \begin{tikzpicture}
    \begin{feynman}
      \vertex (e) at (-0.7,  1.4);
      \vertex (f) at (-0.7, -1.4);
      \vertex (a) at ( 0.5,  0);
      \vertex (y) at ( 2,    2.4);
      \vertex (c) at ( 2,    1.0);
      \vertex (b) at ( 2,    0);
      \vertex (d) at ( 2,   -1.0);
      \vertex (x) at ( 2,   -2.1);
      \diagram* {
        (e) -- [scalar] (a) -- [scalar] (f),
        (a) -- [graviton] (b),
        (y) -- (c),
        (c) -- [pertworld] (b),
        (b) -- [pertworld] (d),
        (d) -- (x),
        (c) -- [graviton, half left] (d),
      };
    \end{feynman}
  \end{tikzpicture}
\end{subfigure}
    \caption{One-loop diagrams involving worldline fluctuations that vanish in dimensional regularization due to scalelessness.}
    \label{fig:WorldScaleless}
\end{figure}

The third diagram in Fig.~\ref{fig:WorldScaleless} illustrates why these integrals have no scale. The two external scalars meet at a single bulk vertex, and the resulting graviton carries $q = p_i - p_f$ to the worldline, where it attaches at a single point. Here $p_i$ and $p_f$ are the momenta of the incoming and the outgoing scalars respectively. The loop is closed by a graviton exchanged between two further worldline vertices, joined back to the attachment point by a pair of fluctuation propagators.

Route the loop momentum $\ell$ through the internal graviton. The static source enforces $\delta(q\cdot u)$ at the external attachment, so no energy is injected into the worldline, and the two fluctuation lines are forced to carry the same worldline energy. Energy conservation at the remaining two vertices fixes that energy to be $\ell\cdot u$. The loop therefore contributes
\begin{equation}
\label{eq:ScalelessIntegral}
    \int\frac{\diff^D \ell}{(2\pi)^D}
    \frac{N^{\mu\nu\cdots}(\ell, u, q)}{\ell^{2}\,(\ell\cdot u)^{4}} \ ,
\end{equation}
the quartic pole arising from the two factors of $(\ell\cdot u)^{-2}$ in the fluctuation propagators.

The denominator contains no external scale. The source absorbs the spatial momentum transfer, while $q\cdot u=0$. Polynomial $q$-dependence from the vertices factors out of the loop integration, leaving scaleless integrals that vanish in dimensional regularization.

Tensor reduction~\cite{Passarino:1978jh} expresses the contributing bulk and worldline loop integrals in the following scalar family:
\begin{equation}
    J_{n_1n_2n_3n_4} \equiv \int\!\frac{\diff^D l}{(2\pi)^D}\,
    \frac{1}{[l^2]^{n_1}\,[(l+p_i-p_f)^2]^{n_2}\,[l\cdot u]^{n_3}\,[(l-p_f)^2]^{n_4}}
    \ ,
    \label{eq:DefMaster}
\end{equation}
where the incoming and outgoing momenta are $p_i$ and $p_f$, respectively, and
\[ \label{eq:kinematics}
 p_i^2=p_f^2=0,
\qquad
 p_i\cdot u=p_f\cdot u\equiv\omega.
\]
Since $u^\mu$ is constant, the only invariants available are $p_i\cdot p_f$ and $\omega$, so the members of Eq.~\eqref{eq:DefMaster} depend on at most two scales. We work throughout in dimensional regularization with $D = 4-2\epsilon$. Explicit results for the integrals appearing in the calculation, together with their derivations and their divergent parts as $\epsilon \to 0$, are collected in Appendix~\ref{sec:AppMasters}.

Summing the diagrams in Figs.~\ref{fig:ScalWorldBulkLoops} and \ref{fig:WorldFlucLoops} and reducing to the master basis in Eq.~\eqref{eq:DefMaster} gives the one-loop amplitude:
\begin{equation}
\label{eq:AmplitudeOneLoop}
\begin{split}
    \mathcal{M}_{\phi\phi} = -\frac{\kappa^{4} M}{4} &\bigg[\frac{D-1}{2}q^2\omega^{3} J_{1111}
    +(D-2)\omega^{3} J_{1110}\\
    &+\frac{(D-3)(D-2)}{2(D-4)}\omega J_{0011}
    +[c_{q}q^2 + c_{\omega}\omega^{2} ] J_{1100}
    \,\bigg] \ ,
\end{split}
\end{equation}
where $q\equiv p_i-p_f$, and the two rational functions of the spacetime dimension $D$ multiplying $J_{1100}$ are
\begin{equation}
\label{eq:AmplitudeCoefficients}
\begin{split}
    c_q &= \frac{3D\bigl[D(D-4)(D+3)-4\bigr]}{128(D-2)(D^{2}-1)}\ , \\
    c_\omega &= \frac{3D\bigl[D\bigl(D(D-1)(4D-15)+28\bigr)+4\bigr]
                     -48}{16(D-4)(D-2)(D^{2}-1)}\ .
\end{split}
\end{equation}

The double poles from $J_{0011}$, $J_{1100}$, and $J_{1111}$ cancel in the sum. Although $J_{1111}$ is ultraviolet finite, its infrared double pole is needed for this cancellation. The remaining master, $J_{1110}$, is finite in both regimes. The master integrals and their $\epsilon$ expansions are given in Appendix~\ref{sec:AppMasters}.

Expanding the master integrals and using the cancellation of the double poles, we find
\begin{equation}
    \mathcal{M}_{\phi\phi}
=\frac{i\kappa ^4 M }{480 (16\pi^2 \epsilon) }(193\omega^2-3 p_i \cdot p_f)-\frac{i \kappa ^4 M }{8 (16\pi ^2 \epsilon) }\omega ^2\log \left(\frac{p_i\cdot p_f}{2\omega^2}\right) \ .
\label{eq:AmpUVIRdiv}
\end{equation}
The remaining single pole contains both ultraviolet and infrared contributions because dimensional regularization uses the same parameter $\epsilon$ for both. The nonanalytic term $\log(p_i\cdot p_f/2\omega^2)$ cannot be canceled by a local counterterm and must be removed with the infrared contribution before extracting the tidal counterterms.

We determine the real infrared contribution from soft-graviton emission and subtract it from Eq.~\eqref{eq:AmpUVIRdiv} to isolate the ultraviolet pole.

\subsubsection{Soft emissions in worldline theory}
\begin{figure}[h]
    \centering
    \begin{subfigure}{0.32\textwidth}
        \centering
        \begin{tikzpicture}[baseline=(current bounding box.center)]
            \begin{feynman}
                \vertex (i)  at ( 0,   -1.5);
                \vertex (a)  at ( 0,    0);
                \vertex (f)  at ( 0,    1.5);
                \vertex (b)  at (-1.3,  0);
                \vertex (f1) at (-2.2,  1.6);
                \vertex (c)  at (-2.0, -0.75);
                \vertex (f3) at (-3.4, -1.2);
                \vertex (f2) at (-2.6,  0.4);
                \diagram*{
                    (i)  -- (f),
                    (a)  -- [graviton] (b),
                    (f1) -- [scalar] (b),
                    (b)  -- [scalar] (c),
                    (c)  -- [scalar] (f3),
                    (c)  -- [graviton] (f2),
                };
            \end{feynman}
        \end{tikzpicture}
    \end{subfigure}%
    \begin{subfigure}{0.32\textwidth}
        \centering
        \begin{tikzpicture}[baseline=(current bounding box.center)]
            \begin{feynman}
                \vertex (i)  at ( 0,    1.5);
                \vertex (a)  at ( 0,    0);
                \vertex (f)  at ( 0,   -1.5);
                \vertex (b)  at (-1.3,  0);
                \vertex (f1) at (-2.2, -1.6);
                \vertex (c)  at (-2.0,  0.75);
                \vertex (f3) at (-3.4,  1.2);
                \vertex (f2) at (-2.6,  1.9);
                \diagram*{
                    (i)  -- (f),
                    (a)  -- [graviton] (b),
                    (f1) -- [scalar] (b),
                    (b)  -- [scalar] (c),
                    (c)  -- [scalar] (f3),
                    (c)  -- [graviton] (f2),
                };
            \end{feynman}
        \end{tikzpicture}
    \end{subfigure}%
    \begin{subfigure}{0.32\textwidth}
        \centering
        \begin{tikzpicture}[baseline=(current bounding box.center)]
            \begin{feynman}
                \vertex (i)  at ( 0,    1.5);
                \vertex (m)  at ( 0,    0.5);
                \vertex (a)  at ( 0,   -0.5);
                \vertex (f)  at ( 0,   -1.5);
                \vertex (b)  at (-1.3, -0.5);
                \vertex (f1) at (-2.8, -1.6);
                \vertex (f3) at (-2.8,  0.6);
                \vertex (g)  at (-1.9,  1.5);
                \diagram*{
                    (i)  -- (m),
                    (m)  -- [pertworld] (a),
                    (a)  -- (f),
                    (a)  -- [graviton] (b),
                    (f1) -- [scalar] (b),
                    (b)  -- [scalar] (f3),
                    (m)  -- [graviton] (g),
                };
            \end{feynman}
        \end{tikzpicture}
    \end{subfigure}
    \caption{Diagrams with soft-graviton emission from the two external scalar legs and from the worldline. Time flows upwards.}
    \label{fig:SoftEmissions}
\end{figure}
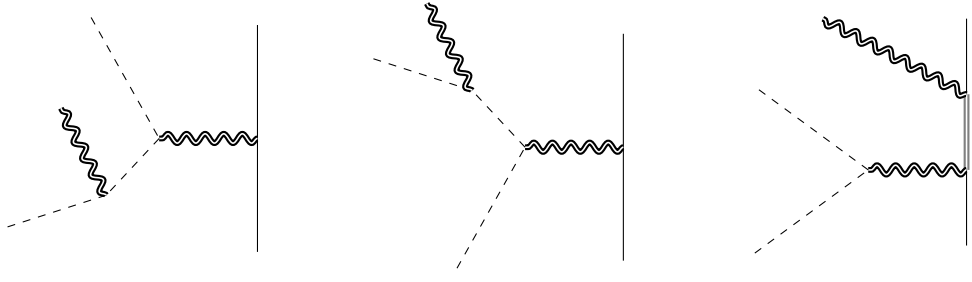

Throughout this section $k^\mu$ denotes the momentum of the emitted graviton, taken outgoing, so that $k^{0} > 0$ and momentum conservation at the hard vertex reads $p_i = p_f + k$. In the soft limit $k \to 0$ the radiative amplitude factorizes as
\begin{equation}
    i\mc{M}_{\phi\to\phi h}
    = i\mc{M}^{(0)}_{\phi\phi}\, S(k) + \mathcal{O}(k^{0}) \ ,
    \label{eq:SoftFactorisation}
\end{equation}
where $\mc{M}^{(0)}_{\phi\phi}$ is the tree-level elastic amplitude for the scalar scattering off the worldline, generated by single-graviton exchange with the static source,
\begin{equation}
    \mc{M}^{(0)}_{\phi\phi}
    = \frac{\kappa^{2}M\,(p_i\!\cdot\!u)(p_f\!\cdot\!u)}{4\,(p_i\!\cdot\!p_f)}
    = \frac{\kappa^{2}M\,\omega^{2}}{4\,(p_i\!\cdot\!p_f)} \ ,
\end{equation}
and $S(k)$ is the leading soft factor. Summing emission from the incoming scalar, the outgoing scalar and the worldline, i.e., the diagrams in Fig.~\ref{fig:SoftEmissions}, and stripping the polarization of the outgoing graviton by writing $S(k) = \tfrac{\kappa}{2}\,\varepsilon^{*}_{\mu}(k)\varepsilon^{*}_{\nu}(k)\, \mathcal{J}^{\mu\nu}(k)$, the soft current $\mathcal{J}^{\mu \nu}$ reads
\begin{equation}
\label{eq:SoftCurrent}
    \mathcal{J}^{\mu\nu}(k)
    = \frac{p_f^\mu p_f^\nu}{k\cdot p_f}
      - \frac{p_i^\mu p_i^\nu}{k\cdot p_i}
    +
    \frac{k\cdot q}{(k\cdot u)^{2}}\,u^\mu u^\nu
      + \frac{q^\mu u^\nu + u^\mu q^\nu}{k\cdot u} \ .
\end{equation}
Here $q \equiv p_i - p_f$ is the momentum transfer of the underlying elastic process, with $q\cdot u = 0$ by the staticity of the source. It is held fixed and independent of $k$ in the factorized leading-soft kinematics.

We square Eq.~\eqref{eq:SoftFactorisation}, sum over graviton polarizations using the $D$-dimensional de Donder sum, and retain the leading soft divergences. Define the rescaled graviton momentum and the total and transferred scalar momenta by
\begin{equation}
\label{eq:SoftVariables}
    n^{\mu} \equiv \frac{k^{\mu}}{k\cdot u} \ , \qquad
    P^{\mu} \equiv p_i^{\mu} + p_f^{\mu} \ , \qquad
    q^{\mu} \equiv p_i^{\mu} - p_f^{\mu} \ .
\end{equation}
Imposing the kinematic conditions $p_i^{2} = p_f^{2} = k^{2} = 0$, $u^{2} = 1$ and $p_i\!\cdot\!u = p_f\!\cdot\!u = \omega$, the squared soft factor (defined below) becomes
\begin{equation}
\label{eq:SoftSquared}
    \mathcal{S}(k) \simeq \frac{\kappa^{2}}{4\,(k\cdot u)^{2}}\left[
      -\,\frac{2\big(p_i\!\cdot\!p_f-\omega\,n\!\cdot\!P\big)^{2}}{(n\!\cdot\!p_i)(n\!\cdot\!p_f)}
      \;+\;4\big(2\omega^{2}-p_i\!\cdot\!p_f\big)
      \;+\;\frac{D-3}{D-2}\,(n\!\cdot\!q)^{2}
    \right] \ .
\end{equation}
To obtain the soft radiation accompanying the elastic process, we define the phase space of the outgoing graviton, with its energy in the rest frame of the body, $k\cdot u$, restricted to lie below a detector resolution $k_{\rm c}$:
\begin{equation}
\label{eq:SoftPhaseSpace}
    \int_{\Pi(k)} \equiv \mu^{2\epsilon}\!\!\int_{k\cdot u < k_{\rm c}}\!\!
    \frac{\diff^{D-1}\vec{k}}{(2\pi)^{D-1}\,2k^{0}} \ .
\end{equation}
Integrating Eq.~\eqref{eq:SoftSquared} over this region defines the soft factor
\begin{equation}
\label{eq:SoftIntegral}
    \mathcal{I}_{\rm soft} \equiv \int_{\Pi(k)} \mathcal{S}(k) \ ,
\end{equation}
which multiplies the elastic rate,
\begin{equation}
    \sum_\lambda\!\int\!\diff\Phi_h\, |\mc{M}_{\phi\to\phi h}|^{2}
    = |\mc{M}^{(0)}_{\phi\phi}|^{2}\,\mathcal{I}_{\rm soft} \ ,
\end{equation}
where the sum is over the helicities of the outgoing graviton.

Since $\mathcal{S}(k)$ carries an overall $(k\cdot u)^{-2}$, the measure factorizes into a radial
integral over the graviton energy $k\cdot u$ and a spherical average over the null direction
$n^{\mu}$ as follows
\begin{equation}
\label{eq:factorised}
    \int_{\Pi(k)}\frac{f(n)}{(k\cdot u)^{2}} = \mathcal{N}\,\big\langle f\big\rangle_{\mathbb{S}^{D-2}} \ , \quad
    \mathcal{N} \equiv \int_{\Pi(k)}\frac{1}{(k\cdot u)^{2}}
    = -\frac{1}{8\pi^{2}\epsilon}
      \left(\frac{\pi\mu^{2}}{k_{\rm c}^{2}}\right)^{\!\epsilon}
      \frac{\Gamma(1-\epsilon)}{\Gamma(2-2\epsilon)} \ .
\end{equation}
Angle brackets denote the normalized angular average over $\mathbb{S}^{D-2}$.\footnote{Explicitly, $\langle f\rangle_{\mathbb{S}^{D-2}}=\frac{\int\diff\Omega_{D-2}\,f(n)}{\int\diff\Omega_{D-2}}$.} Note that all the terms in Eq.~\eqref{eq:SoftSquared} are simple integral powers of $n \cdot p_i$ and $n \cdot p_f$. Therefore, the master integral
\begin{equation}
\label{eq:Jab}
    J_{ab} \equiv \left\langle \frac{1}{ (n\cdot\ell_i)^{a}(n\cdot\ell_f)^{b}}\right\rangle_{\mathbb{S}^{D-2}}
    =\frac{B(1-a-\epsilon,\,1-b-\epsilon)}{2^{a+b}B(1-\epsilon,1-\epsilon)}\;
      {}_2F_1\!\left(a,b;1-\epsilon;1-\frac{p_i\!\cdot\!p_f}{2\omega^{2}}\right) \ ,
\end{equation}
where $\ell_{i,f} \equiv p_{i,f}/\omega$, gives all the integrals we require. Substituting the above master, we find
\begin{equation}
\label{eq:assembled}
    \mathcal{I}_{\rm soft}
    = \frac{\kappa^{2}\omega^{2}}{4}\,\mathcal{N}\left[
      -8\bar z^{2}J_{11} + 16\bar z\,J_{10} - 4J_{1,-1} + 4\big(1-2\bar z\big)
      + \frac{1-2\epsilon}{1-\epsilon}\big(J_{-2,0}-J_{-1,-1}\big)
    \right] \ ,
\end{equation}
where $\bar z \equiv \frac{p_i\cdot p_f}{2\omega^{2}}$.
The collinear poles cancel among the first three terms, while the last two terms are finite as $\epsilon\to0$. The remaining divergence comes from the radial factor $\mathcal N$. The real infrared contribution to the one-loop amplitude is
\begin{equation}
\label{eq:WLsoftdiv}
    -\frac{i}{2}\,\mc{M}^{(0)}_{\phi\phi}\left.\mathcal{I}_{\rm soft}\right|_{\rm div}
    = -\frac{i\kappa^{4}M\omega^{2}}{128\pi^{2}\epsilon}
      \left[\log\!\left(\frac{p_i\!\cdot\!p_f}{2\omega^{2}}\right) - \frac{1}{6}\right] \ .
\end{equation}

\subsubsection{UV divergences and Running of Scalar Love Numbers}

The logarithmic coefficient agrees with the independent soft-emission calculation. Subtracting Eq.~\eqref{eq:WLsoftdiv} gives the ultraviolet pole:

\begin{equation}
\label{eq:AmpUVdiv}
    i\mathcal{M}^{\rm UV}_{\phi\phi}
    = \frac{i\kappa^{4}M}{(4\pi)^{2}\epsilon}\,
      \frac{60\omega^{2} - (p_i\!\cdot\!p_f-\omega^2)}{160} \ .
\end{equation}
The residue is polynomial in the external kinematics. The Love-number counterterms in the minimal subtraction scheme follow by comparing Eq.~\eqref{eq:AmpUVdiv} with the tree-level amplitude generated by the tidal operators of Eq.~\eqref{eq:ActionExpanded}.

We first check whether the divergence in Eq.~\eqref{eq:AmpUVdiv} can be absorbed by any of the five lower-order counterterms.

The first is the graviton two-point function. At zeroth order in $GM$, no diagram contains a worldline insertion. The only contribution is then the one-loop graviton self-energy in flat space. With the external momentum on shell, $k^{2}=0$, this loop integral is scaleless and thus vanishes in dimensional regularization. The second and third are the wavefunction renormalizations of the scalar and of the worldline fluctuation. The scalar self-energy is again scaleless, while $\delta z^{\mu}$ requires no renormalization at all, since it does not appear in the asymptotic states. The fourth is the vertex joining two massless scalars to a graviton, which is likewise unrenormalized. Every one-loop diagram contributing to it vanishes on shell by scalelessness.

The remaining possibility is the worldline mass. At one loop the graviton one-point function on the worldline is a single triangle built from two graviton propagators and one worldline fluctuation. Its numerator is proportional to the square of the frequency of the emitted graviton, and frequency conservation at the static source forces that frequency to vanish on shell. The diagram therefore vanishes, and $M$ is not renormalized. Thus, we conclude that the divergence in Eq.~\eqref{eq:AmpUVdiv} can be absorbed only into $\lambda^{(n)}_{\ell}$.

The tidal operators in Eq.~\eqref{eq:ActionExpanded} generate a two-point vertex on the worldline at which the scalar is absorbed and re-emitted. For an incoming scalar of momentum $p_i$ and an outgoing one of momentum $p_f$, each of the derivative operators is replaced by a suitable projected momentum:
\begin{equation}
    \bar\partial_\parallel \;\to\; i\,(p\!\cdot\!u) \ ,
    \qquad
    \bar\partial_\perp^{\mu} \;\to\; i\,\bar{\mathbb{P}}^{\mu}{}_{\nu}\,p^{\nu} \ .
\end{equation}
The leading tidal operators therefore contribute
\begin{equation}
\begin{split}
    \label{eq:TidalVertex}
    i\mathcal{M}^{\rm tidal}_{\phi\phi}
    &\supset i M \kappa^2 r_{\rm h}^{2} \left[
      \hat\lambda^{(1)}_{0}\,(p_i\!\cdot\!u)(p_f\!\cdot\!u)
      + \hat\lambda^{(0)}_{1}\,
        \bar{\mathbb{P}}_{\mu\nu}\,p_i^{\mu} p_f^{\nu}
    \right]\\
    &= i M \kappa^2 r_{\rm h}^{2} \left[
      \hat\lambda^{(1)}_{0}\,\omega^{2}
      + \hat\lambda^{(0)}_{1}\,(p_i\!\cdot\!p_f - \omega^{2})
    \right] ,
\end{split}
\end{equation}
where in the second equality we used
$\bar{\mathbb{P}}_{\mu\nu} = \eta_{\mu\nu} - u_\mu u_\nu$ together with $p_i\!\cdot\!u = p_f\!\cdot\!u = \omega$. 

Demanding that the counterterms cancel the divergence in Eq.~\eqref{eq:AmpUVdiv}, the tidal couplings must be renormalized as
\begin{equation}
    \delta \hat{\lambda}^{(1)}_{0}
    = -\frac{3}{16 \pi}\,\frac{1}{\epsilon}\frac{M_{\rm pl}^2}{M^2} \ ,
    \qquad
    \delta \hat{\lambda}^{(0)}_{1}
    = \frac{1}{320\pi}\,\frac{1}{\epsilon}\frac{M_{\rm pl}^2}{M^2} \ ,
\end{equation}
where we have written the bare couplings as $\hat\lambda_{\rm bare} = \mu^{2\epsilon} \big(\hat\lambda(\mu) + \delta\hat\lambda\big)$. We have also used 
\begin{equation}
    \frac{\kappa^2}{4\pi r_{\rm h}^2} = 2 \frac{M_{\rm pl}^2}{M^2} \ .
\end{equation}
Requiring $\hat\lambda_{\rm bare}$ to be independent of $\mu$, the residues fix the beta functions
\begin{equation}
\label{eq:BetaFunctions}
\begin{split}
    \beta_{\hat\lambda^{(1)}_{0}}
    &\equiv \frac{\diff \hat\lambda^{(1)}_{0}}{\diff\log\mu}
    = \frac{3}{8\pi}\frac{M_{\rm pl}^2}{M^2} = \frac{3}{2\pi} \frac{\ell_{\rm pl}^2}{r_{\rm h}^2}\ ,\\
    \beta_{\hat\lambda^{(0)}_{1}}&\equiv \frac{\diff \hat\lambda^{(0)}_{1}}{\diff\log\mu}
    = -\frac{1}{160\pi}\frac{M_{\rm pl}^2}{M^2} =-\frac{1}{40\pi}\frac{\ell_{\rm pl}^2}{r_{\rm h}^2}\ .
\end{split}
\end{equation}

At this order, the beta functions are generated by loops of the minimal point-particle theory and are independent of the tidal couplings. The running is therefore $\hat\lambda(\mu)=\hat\lambda(\mu_0)+\beta_{\hat\lambda}\log(\mu/\mu_0)$. A coupling set to zero at $\mu_0$ is nonzero at other scales. Determining its finite value at the reference scale requires matching to the black-hole calculation.

\subsection{Electromagnetic Love Numbers}\label{sec:EMLove}

We next compute the one-loop Compton amplitude $i\mathcal M_{\gamma\gamma}$ for photon scattering from an electrically neutral worldline. A neutral body can still have induced electric and magnetic multipoles~\cite{Hui:2020xxx,Charalambous:2021mea,Pereniguez:2021xcj,Rai:2024lho}. Its ultraviolet pole renormalizes the electric and magnetic polarizabilities. The calculation follows the scalar case, so we give the additional ingredients and results. We supplement Eq.~\eqref{eq:ActionExpanded} with the Maxwell and gauge-fixing terms:
\[
S_{\mathrm{M}} & =-\frac{1}{4} \int \diff^4 x \sqrt{-g} F_{\mu \nu} F^{\mu \nu}\ , \\
S_{\mathrm{gf}}^{(A)} & =-\frac{1}{2 \zeta} \int \diff^4 x \sqrt{-g}\left(\nabla_\mu A^\mu\right)^2\ ,
\]
where $F_{\mu \nu} = \partial_\mu A_\nu -\partial_\nu A_\mu$ and $\nabla_\mu$ is the covariant derivative with respect to $g_{\mu\nu}$. Gauge fixing also gives the ghost action
\[
S_{\mathrm{gh}}^{(A)}=-\int \diff^4 x \sqrt{-g} \; \bar{c} \, \square_g c= \int \diff^4 x\sqrt{-g}\,g^{\mu\nu}
 \partial_\mu\bar c\,\partial_\nu c \ .
\]
The photon ghosts $c$ and $\bar c$ are Grassmann scalar fields, whereas the graviton ghosts in Eq.~\eqref{eq:GhostAction} are vectors. Although these fields do not couple to the photon itself, they contribute to the amplitude through their coupling to gravitons.

The UV divergences of the one-loop amplitude $i\mathcal{M}_{\gamma \gamma}$ require the addition of counterterm vertices. As before, the bulk counterterms can be absorbed into field redefinitions leaving only the worldline counterterms corresponding to the electromagnetic Love numbers. To construct the worldline counterterms, define
\[
\Ecal_\mu\equiv F_{\mu\nu}u^\nu\ ,
 \qquad
 \Bcal_\mu\equiv \widetilde F_{\mu\nu}u^\nu \ ,
 \qquad
 \widetilde F_{\mu\nu}\equiv
 \frac12\epsilon_{\mu\nu\rho\sigma}F^{\rho\sigma}\ .
\]
Here, $\Ecal_\mu$ and $\Bcal_\mu$ describe the electric and magnetic components of the field strength tensor in the rest frame of the worldline. The Love counterterm basis can be expressed by contracting STF tensors composed of these operators and their derivatives:
\[\label{eq:PhotonLoveTower}
S_{\rm tidal}^{(A)}
 =\frac{M\kappa^2}{2}
 \sum_{\ell=1}^{\infty}\sum_{n=0}^{\infty}
 \rh^{\,2(\ell+n)}
 \int\dd\tau\,
 \bigg[
 &\hat\lambda_{E,\ell}^{(n)}
 \Ecal_{\langle\mu_1\cdots\mu_\ell\rangle}^{(n)}
 \Ecal^{(n)\langle\mu_1\cdots\mu_\ell\rangle}
+
 \hat\lambda_{B,\ell}^{(n)}
 \Bcal_{\langle\mu_1\cdots\mu_\ell\rangle}^{(n)}
 \Bcal^{(n)\langle\mu_1\cdots\mu_\ell\rangle}
 \bigg]\ ,
\]
where we have adopted the notation in Eq.~\eqref{eq:STFnotationu} and defined
\[
\Ecal_{\langle\mu_1\cdots\mu_\ell\rangle}^{(n)}
 =
\bar{\partial}_\parallel^n
\left(\bar{\partial}_{\perp}\Ecal\right)_{\langle \mu \rangle_\ell}\ ,\quad
 \\ \left(\bar{\partial}_{\perp}\Ecal\right)_{\langle \mu \rangle_\ell} \equiv {\rm STF} \left[\prod_{i=1}^{\ell-1} \bar{\mathbb{P}}_{\mu_i}^{\ \nu_i} \partial_{\nu_i} \Ecal_{\mu_\ell}\right] \ .
\]

Working at one loop and retaining the leading power in $M$, our UV pole must take the form
\[
i \mathcal{M}_{\gamma \gamma}^{\rm UV} = \frac{i \kappa^4 M}{\epsilon} \mathcal{P}(p_i,p_f,\varepsilon_i,\varepsilon_f,u)\ .
\]
Dimensional analysis implies that $\mathcal{P}(p_i,p_f,\varepsilon_i,\varepsilon_f,u)$ must be quadratic in the photon momenta $p_i$ and $p_f$. We therefore expect the UV pole to be fully captured by the $\ell=1$ sector of the expansion in Eq.~\eqref{eq:PhotonLoveTower}. Specifically, the two counterterms
\[
 S_{\rm tidal}^{(A)}
 \supset
 \frac{M\kappa^2\rh^2}{2}
 \int\dd\tau\left[
 \hat\lambda_E\,\Ecal_\mu\Ecal^\mu
 +\hat\lambda_B\,\Bcal_\mu\Bcal^\mu
 \right]\ ,
\]
suffice to subtract all UV divergences at one loop. 
We use the photon gauge freedom to choose polarization vectors orthogonal to the worldline four-velocity $u$:
\[
\varepsilon_i \cdot u = \varepsilon_f \cdot u=0 \ .
\]
In this gauge, the tidal operators contribute the following Compton amplitude:
\[\label{eq:MphotonLove}
 i \mathcal{M}_{\gamma\gamma}^{\rm tidal}
 =
 i M\kappa^2\rh^2\bigg[
 &-(\hat\lambda_E+\hat\lambda_B)
 \omega^2(\varepsilon_i\cdot\varepsilon_f^*)
 +\hat\lambda_B\left\{
 (p_i\cdot p_f)(\varepsilon_i\cdot\varepsilon_f^*)
 +(\varepsilon_i\cdot q)(\varepsilon_f^*\cdot q)
 \right\}
 \bigg]\ .
\]

\subsubsection{UV divergences and Running of Electromagnetic Love Numbers}

The photon calculation uses the scalar kinematics, with incoming and outgoing momenta $p_i$ and $p_f$:
\[ \label{eq:kinematicsEM}
 q\equiv p_i-p_f\ ,
 \qquad
 p_i^2=p_f^2=0\ ,
 \qquad
 q\cdot u=0\ , \qquad
 p_i\cdot u=p_f\cdot u\equiv\omega\ .
\]
The physical polarizations obey
\[
 \varepsilon_i\cdot p_i=0\ ,
 \qquad
 \varepsilon_f^*\cdot p_f=0\ .
 \label{eq:transversality}
\]

The contributing topologies are the analog to the nine diagrams for the scalar and a photon-ghost vacuum-polarization diagram. The corresponding integrals reduce to the same family $J_{n_1n_2n_3n_4}$ in Eq.~\eqref{eq:DefMaster}. The leading soft-graviton factor is the same as in the scalar case, so the infrared contribution factorizes as follows:
\[
i \mathcal{M}_{\gamma\gamma}^{\mathrm{IR}}=  i \mathcal{M}_{\gamma\gamma}^{(0)} \mathcal{I}_{\mathrm{IR}}\ ,
\]
where 
\[
&\mathcal{M}_{\gamma\gamma}^{(0)}   =
-\frac{i\kappa^2 M}{16 p_i \cdot p_f}
\left[
(4\omega^2-2 p_i \cdot p_f)
(\varepsilon_i\cdot\varepsilon_f^{*})
-2(\varepsilon_i\cdot q)
  (\varepsilon_f^{*}\cdot q)
\right], \\
&\mathcal{I}_{\mathrm{IR}}
=
\frac{\kappa^2\, p_i \cdot p_f}
     {32\pi^2\epsilon}
\left[
\frac{1}{6}
-\log\left(\frac{p_i \cdot p_f}{2\omega^2}\right)
\right] \ ,
\label{eq:photon-tree-amplitude}
\]
are the tree-level amplitude and soft-graviton IR function respectively. The external photons and worldline are electrically neutral, so this calculation contains no soft-photon exchange between charged on-shell external legs.

We can now directly extract the UV pole by subtracting the IR contribution from the full $\mathcal{M}_{\gamma\gamma}$ amplitude, giving
\[
 i\mathcal{M}_{\gamma\gamma}^{\rm UV}
 =
 \frac{i \kappa^4M}{(4\pi)^2\epsilon}
 \frac{39}{160}
 \left[
 \bigl(p_i\cdot p_f-2\omega^2\bigr)
 (\varepsilon_i\cdot\varepsilon_f^*)
 +(\varepsilon_i\cdot q)(\varepsilon_f^*\cdot q)
 \right].
 \label{eq:UVpole}
\]
Recalling that there are no lower-order bulk counterterms, we can match this result directly to the counterterm contribution in Eq.~\eqref{eq:MphotonLove} to find
\[
 \delta\hat\lambda_E = \delta\hat\lambda_B = -\frac{39}{320\pi\epsilon}
 \frac{M_{\rm pl}^2}{M^2}.
\]
These counterterms give the beta functions below.
\begin{equation}
    \begin{split}
        \beta_{\hat{\lambda}_E}&=\frac{\diff \hat{\lambda}_E}{\diff \log \mu}=\frac{39}{160 \pi}\frac{M_{\rm pl}^2}{M^2}=\frac{39}{40 \pi}\frac{\ell_{\rm pl}^2}{r_{\rm h}^2}\ ,\\
\beta_{\hat{\lambda}_B}&=\frac{\diff \hat{\lambda}_B}{\diff \log \mu}=\frac{39}{160 \pi}\frac{M_{\rm pl}^2}{M^2} = \frac{39}{40 \pi}\frac{\ell_{\rm pl}^2}{r_{\rm h}^2}\ .
    \end{split}
\end{equation}

\subsubsection{Comments on Electric-Magnetic Duality}
The electric and magnetic beta functions are equal, which implies
\begin{equation}
    \frac{\diff}{\diff\log\mu}\left(\hat\lambda_E-\hat\lambda_B\right)=0
\end{equation}
at the order considered, consistent with source-free electric--magnetic duality.
To identify the operator that duality protects, decompose the field strength into its self-dual and anti-self-dual parts,
\[
 F^{\pm}_{\mu\nu}\equiv\frac12\left(F_{\mu\nu}\mp i\widetilde F_{\mu\nu}\right)\ ,
 \qquad
 \widetilde F^{\pm}_{\mu\nu}=\pm i\,F^{\pm}_{\mu\nu}\ ,
 \qquad
 \Ecal^{\pm}_\mu\equiv F^{\pm}_{\mu\nu}u^\nu=\frac12\left(\Ecal_\mu\mp i\Bcal_\mu\right)\ .
\]
The $\ell=1$ counterterms then split into duality-even and duality-odd combinations,
\[
 \hat\lambda_E\,\Ecal_\mu\Ecal^\mu+\hat\lambda_B\,\Bcal_\mu\Bcal^\mu
 &=
 \frac{\hat\lambda_E+\hat\lambda_B}{2}
 \left(\Ecal_\mu\Ecal^\mu+\Bcal_\mu\Bcal^\mu\right)
 +
 \frac{\hat\lambda_E-\hat\lambda_B}{2}
 \left(\Ecal_\mu\Ecal^\mu-\Bcal_\mu\Bcal^\mu\right)
 \\
 &=
 2\left(\hat\lambda_E+\hat\lambda_B\right)\Ecal^{+}_\mu\Ecal^{-\mu}
 +
 \left(\hat\lambda_E-\hat\lambda_B\right)
 \left(\Ecal^{+}_\mu\Ecal^{+\mu}+\Ecal^{-}_\mu\Ecal^{-\mu}\right)\ .
 \label{eq:DualityBasis}
\]
Electric--magnetic duality, $F_{\mu\nu}\to\widetilde F_{\mu\nu}$, acts as $\Ecal^{\pm}_\mu\to\pm i\,\Ecal^{\pm}_\mu$. The operator $\Ecal^{+}_\mu\Ecal^{-\mu}$ is therefore duality even, whereas $\Ecal^{\pm}_\mu\Ecal^{\pm\mu}$ are duality odd.
The same decomposition fixes the helicity content of the two operators. The linearized field strength of a photon of definite helicity is self-dual or anti-self-dual, and the conjugate polarization of an outgoing photon carries the opposite duality eigenvalue to that of an incoming photon of the same helicity. Hence $\Ecal^{+}_\mu\Ecal^{-\mu}$ contributes only to the helicity-preserving Compton amplitude, and $\Ecal^{\pm}_\mu\Ecal^{\pm\mu}$ only to the helicity-flip amplitude. This is the origin of the structure of Eq.~\eqref{eq:MphotonLove}, where the helicity-preserving contribution depends on $\hat\lambda_E+\hat\lambda_B$ while the helicity-flip contribution is proportional to $\hat\lambda_E-\hat\lambda_B$.

The one-loop divergence renormalizes only the duality-invariant, helicity-preserving operator $\Ecal^{+}_\mu\Ecal^{-\mu}$, whereas the duality-odd, helicity-flip operator is not renormalized. Thus the running generated by the minimal worldline theory preserves the duality-invariant relation $\hat\lambda_E=\hat\lambda_B$. This statement concerns the logarithmic running and does not determine finite matching contributions.

This result admits a simple four-dimensional unitarity explanation~\cite{Bern:1994zx}. The tree amplitudes of the minimally coupled, electrically neutral theory obey the electromagnetic-duality helicity selection rule (i.e., the helicity-flip amplitude vanishes). Sewing these amplitudes therefore gives vanishing unitarity cuts in the helicity-flip channel, consistent with the absence of the corresponding one-loop ultraviolet logarithm. This argument does not exclude finite rational contributions, which do not affect the running.

\subsection{Comments on Gravitational Love numbers}

\begin{figure}[h]
    \centering
    \begin{subfigure}{0.49\textwidth}
        \centering
        \begin{tikzpicture}
  \begin{feynman}
    \vertex (p1) at (-1,  1.5);
    \vertex (p2) at (-1, -1.5);
    \vertex (a1) at ( 0,  0.9);
    \vertex (a2) at ( 0, -0.9);
    \vertex (c1) at ( 1.75,  0.9);
    \vertex (c2) at ( 1.75, -0.9);
    \vertex (t1) at ( 3.5,  0.9);
    \vertex (t0) at ( 3.5, -0.9);
    \vertex (y)  at ( 3.5,  2.0);
    \vertex (x)  at ( 3.5, -2.0);
    \diagram* {
      (p1) -- [graviton] (a1)  -- [graviton] (c1),
      (p2) -- [graviton] (a2)  -- [graviton] (c2),
      (a1) -- [graviton] (a2),
      (c1) -- [graviton] (c2),
      (c1) -- [graviton] (t1),
      (c2) -- [graviton] (t0),
      (t0) -- [pertworld] (t1),
      (y) -- (t1),
      (t0) -- (x),
    };
  \end{feynman}
\end{tikzpicture}
\caption{Two quantum loops}
\label{fig:GravitonLoveTwoLoop}
\end{subfigure}
    \begin{subfigure}{0.49\textwidth}
    \centering
    \begin{tikzpicture}
  \begin{feynman}
    \vertex (p1) at (-1,  1.5);
    \vertex (p2) at (-1, -1.5);
    \vertex (a1) at ( 0,  0.9);
    \vertex (a2) at ( 0, -0.9);
    \vertex (c1) at ( 1.75,  0.9);
    \vertex (cm) at ( 1.75,  0);
    \vertex (c2) at ( 1.75, -0.9);
    \vertex (t1) at ( 3.5,  0.9);
    \vertex (tm) at ( 3.5,  0);
    \vertex (t0) at ( 3.5, -0.9);
    \vertex (y)  at ( 3.5,  2.0);
    \vertex (x)  at ( 3.5, -2.0);
    \diagram* {
      (p1) -- [graviton] (a1)  -- [graviton] (c1),
      (p2) -- [graviton] (a2)  -- [graviton] (c2),
      (a1) -- [graviton] (a2),
      (c1) -- [graviton] (cm) -- [graviton] (c2),
      (c1) -- [graviton] (t1),
      (cm) -- [graviton] (tm),
      (c2) -- [graviton] (t0),
      (t0) -- (tm) -- (t1),
      (y) -- (t1),
      (t0) -- (x),
    };
  \end{feynman}
\end{tikzpicture}

    \caption{One quantum loop and two classical loops}
    \label{subfig:BoxSplit}
    \end{subfigure}
    \caption{Diagrams contributing to the scattering of gravitons against a worldline that are relevant to the gravitational Love number.}
    \label{fig:GravitonLove}
\end{figure}
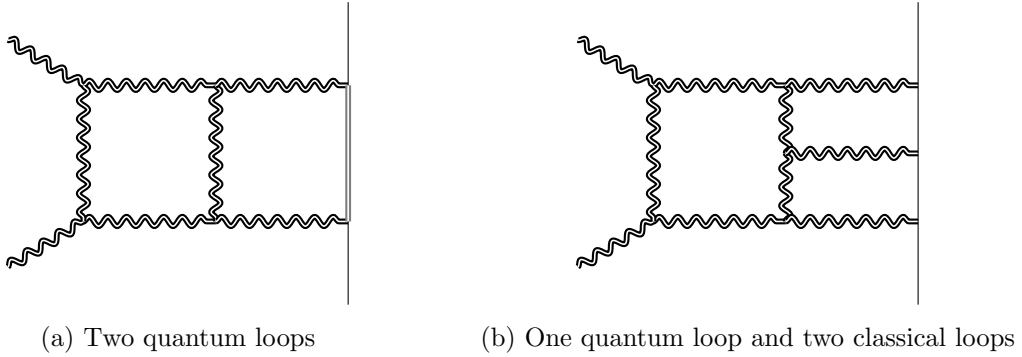

Gravitational tidal couplings begin at quadrupole order, $\ell=2$. Mass and momentum conservation exclude independent monopole and dipole tidal responses; corresponding finite-size operators are forbidden by gauge invariance or removable by field redefinitions~\cite{Goldberger:2004jt,Bern:2020uwk}. The leading tidal action is
\begin{equation}
    S_{\rm tidal}^{(h)}
\sim 
 \int\dd\tau\left[
 \lambda_E\,\Ecal_{\mu\nu}\Ecal^{\mu\nu}
 +\lambda_B\,\Bcal_{\mu\nu}\Bcal^{\mu\nu}
 \right] \ ,
\end{equation}
where $\Ecal_{\mu\nu}=R_{\mu\alpha\nu\beta}u^\alpha u^\beta$, and $\Bcal_{\mu\nu}$ is analogously defined using the dual of the Riemann curvature tensor.
However, since the electric and magnetic tidal fields have dimensions of $[\text{length}^{-2}]$, the associated coefficients in the action must scale as $[\text{mass}\times  \text{length}^{4}]$ for consistency. This means that we can either have $M r_{\rm h}^2 \ell_{\rm pl}^2$, or $M \ell_{\rm pl}^4$. The term $M r_{\rm h}^2\ell_{\rm pl}^2$ is first order in the quantum expansion and occurs at 3PM, with two classical loops and one quantum loop. The term $M\ell_{\rm pl}^4$ is linear in $M$ but second order in the quantum expansion, with two quantum loops. One representative diagram for each scaling is shown in Fig.~\ref{fig:GravitonLove}.

\section{Approaching the Worldline from a Massive Scalar}\label{sec:WorldlineVsQFT}

We compare the worldline calculation with the soft limit of a relativistic massive scalar $\Phi$ coupled to gravity,
\begin{equation}\label{eq:Scl}
  S = \int\!\diff^{4}x\,\sqrt{-g}\,
  \left[-\frac{2}{\kappa^{2}}\,R
    +\frac{1}{2}(\partial\phi)^{2}
    +\frac{1}{2}(\partial\Phi)^{2}
    -\frac{1}{2}M^{2}\Phi^{2}\right] \ ,
\end{equation}
and compute the one-loop correction to $\phi\Phi\to\phi\Phi$ scattering, with no post-Minkowskian expansion~\cite{Donoghue:1994dn,Bjerrum-Bohr:2002gqz,Bjerrum-Bohr:2014zsa,Bjerrum-Bohr:2016hpa}. The Love number coupling now corresponds to operators built from $\phi^{2}$, $\Phi^{2}$ and derivatives, and the question is whether such couplings are generated by the renormalization-group flow.

\begin{figure}[h]
    \centering
    \begin{subfigure}{0.32\textwidth}
        \centering
        \feynmandiagram [small, horizontal=b to c] {
            a -- [graviton] b -- [graviton, half left] c -- [graviton, half left] b,
            c -- [graviton] d,
            e -- [scalar] a, f -- [scalar] a,
            g -- [plain, thick] d, h -- [plain, thick] d,
        };
    \end{subfigure}%
    \begin{subfigure}{0.32\textwidth}
        \centering
        \feynmandiagram [small, horizontal=a to d] {
            a -- [graviton] b -- [ghost, half left] c -- [ghost, half left] b,
            c -- [graviton] d,
            e -- [scalar] a, f -- [scalar] a,
            g -- [plain, thick] d, h -- [plain, thick] d,
        };
    \end{subfigure}%
    \begin{subfigure}{0.32\textwidth}
        \centering
        \feynmandiagram [small, horizontal=b to c] {
            a -- [graviton] b -- [scalar, half left] c -- [scalar, half left] b,
            c -- [graviton] d,
            e -- [scalar] a, f -- [scalar] a,
            g -- [plain, thick] d, h -- [plain, thick] d,
        };
    \end{subfigure}\\[6pt]
    \begin{subfigure}{0.32\textwidth}
        \centering
        \feynmandiagram [small, horizontal=a to d] {
            a -- [graviton] b -- [graviton] d,
            b -- [plain, thick, out=135, in=45, loop, min distance=2cm] b,
            e -- [scalar] a, f -- [scalar] a,
            g -- [plain, thick] d, h -- [plain, thick] d,
        };
    \end{subfigure}%
    \begin{subfigure}{0.32\textwidth}
        \centering
        \feynmandiagram [small, horizontal=a to d] {
            a -- [graviton] b -- [plain, thick, half left] c
              -- [plain, thick, half left] b,
            c -- [graviton] d,
            e -- [scalar] a, f -- [scalar] a,
            g -- [plain, thick] d, h -- [plain, thick] d,
        };
    \end{subfigure}
    \caption{Corrections to the exchanged graviton.}
    \label{fig:GravitonBubbles}
\end{figure}
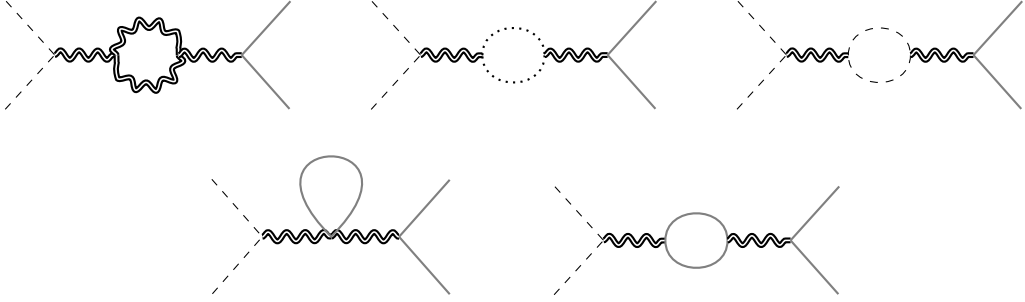

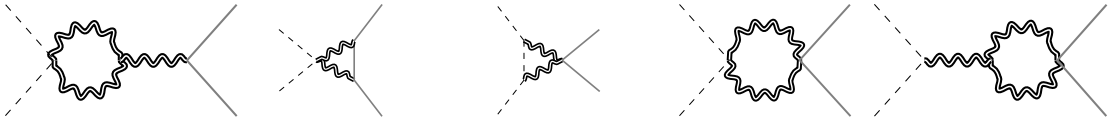
\begin{figure}[h]
    \centering
    \begin{subfigure}{0.19\textwidth}
        \centering
        \resizebox{!}{1.5cm}{%
        \feynmandiagram [small, horizontal=c to d] {
            b -- [graviton, half left] c -- [graviton, half left] b,
            c -- [graviton] d,
            e -- [scalar] b, f -- [scalar] b,
            g -- [plain, thick] d, h -- [plain, thick] d,
        };}
    \end{subfigure}%
    \begin{subfigure}{0.19\textwidth}
        \centering
        \resizebox{!}{1.5cm}{%
        \feynmandiagram [small, vertical=t3 to t2] {
            t1 -- [graviton] t2 -- [plain, thick] t3 -- [graviton] t1,
            t2 -- [plain, thick] p1,
            t3 -- [plain, thick] p2,
            g -- [scalar] t1, h -- [scalar] t1,
        };}
    \end{subfigure}%
    \begin{subfigure}{0.19\textwidth}
        \centering
        \resizebox{!}{1.5cm}{%
        \feynmandiagram [small, vertical=t2 to t3] {
            t1 -- [graviton] t2 -- [scalar] t3 -- [graviton] t1,
            t2 -- [scalar] p1, t3 -- [scalar] p2,
            g -- [plain, thick] t1, h -- [plain, thick] t1,
        };}
    \end{subfigure}%
    \begin{subfigure}{0.19\textwidth}
        \centering
        \resizebox{!}{1.5cm}{%
        \feynmandiagram [small, vertical=e to f] {
            b -- [graviton, half right] c -- [graviton, half right] b,
            e -- [scalar] b, f -- [scalar] b,
            g -- [plain, thick] c, h -- [plain, thick] c,
        };}
    \end{subfigure}%
    \begin{subfigure}{0.19\textwidth}
        \centering
        \resizebox{!}{1.5cm}{%
        \feynmandiagram [small, horizontal=d to c] {
            b -- [graviton, half left] c -- [graviton, half left] b,
            c -- [graviton] d,
            e -- [plain, thick] b, f -- [plain, thick] b,
            g -- [scalar] d, h -- [scalar] d,
        };}
    \end{subfigure}
    \caption{One-loop corrections to the scalar--scalar--graviton--graviton vertices.}
    \label{fig:VertexCorrections}
\end{figure}

\begin{figure}[h]
    \centering
    \begin{subfigure}{0.24\textwidth}\centering
        \begin{tikzpicture}[baseline=(a.base), scale=0.7]\begin{feynman}
            \gravskeletonpp
            \diagram* {
                (a) -- [graviton] (d),
                (e) -- [scalar] (x) -- [scalar] (a),
                (x) -- [graviton, half left, min distance=0.7cm] (a),
                (f) -- [scalar] (a),
                (g) -- [plain, thick] (d), (h) -- [plain, thick] (d),
            };
        \end{feynman}\end{tikzpicture}
    \end{subfigure}%
    \begin{subfigure}{0.24\textwidth}\centering
        \begin{tikzpicture}[baseline=(a.base), scale=0.7]\begin{feynman}
            \gravskeletonpp
            \diagram* {
                (a) -- [graviton] (d),
                (e) -- [scalar] (a),
                (f) -- [scalar] (y) -- [scalar] (a),
                (y) -- [graviton, half right, min distance=0.7cm] (a),
                (g) -- [plain, thick] (d), (h) -- [plain, thick] (d),
            };
        \end{feynman}\end{tikzpicture}
    \end{subfigure}%
    \begin{subfigure}{0.24\textwidth}\centering
        \begin{tikzpicture}[baseline=(a.base), scale=0.7]\begin{feynman}
            \gravskeletonpp
            \diagram* {
                (a) -- [graviton] (d),
                (e) -- [scalar] (a), (f) -- [scalar] (a),
                (g) -- [plain, thick] (z) -- [plain, thick] (d),
                (z) -- [graviton, half right, min distance=0.7cm] (d),
                (h) -- [plain, thick] (d),
            };
        \end{feynman}\end{tikzpicture}
    \end{subfigure}%
    \begin{subfigure}{0.24\textwidth}\centering
        \begin{tikzpicture}[baseline=(a.base), scale=0.7]\begin{feynman}
            \gravskeletonpp
            \diagram* {
                (a) -- [graviton] (d),
                (e) -- [scalar] (a), (f) -- [scalar] (a),
                (g) -- [plain, thick] (d),
                (h) -- [plain, thick] (w) -- [plain, thick] (d),
                (w) -- [graviton, half left, min distance=0.7cm] (d),
            };
        \end{feynman}\end{tikzpicture}
    \end{subfigure}\\
    \begin{subfigure}{0.24\textwidth}
        \centering
        \resizebox{0.86\linewidth}{!}{%
        \feynmandiagram [small, vertical=t3 to t2] {
            a -- [graviton] t1 -- [graviton] t2 -- [plain, thick] t3 -- [graviton] t1,
            t2 -- [plain, thick] p1,
            t3 -- [plain, thick] p2,
            g -- [scalar] a, h -- [scalar] a,
        };}
    \end{subfigure}%
    \begin{subfigure}{0.24\textwidth}
        \centering
        \resizebox{0.86\linewidth}{!}{%
        \feynmandiagram [small, vertical=t2 to t3] {
            a -- [graviton] t1 -- [graviton] t2 -- [scalar] t3 -- [graviton] t1,
            t2 -- [scalar] p1, t3 -- [scalar] p2,
            g -- [plain, thick] a, h -- [plain, thick] a,
        };}
    \end{subfigure}%
    \begin{subfigure}{0.24\textwidth}
        \centering
        \resizebox{0.86\linewidth}{!}{%
        \feynmandiagram [small, vertical=c to b] {
            a -- [plain, thick] c -- [plain, thick] e -- [plain, thick] b
              -- [plain, thick] d,
            b -- [graviton] c,
            e -- [graviton] f,
            g -- [scalar] f, h -- [scalar] f,
        };}
    \end{subfigure}%
    \begin{subfigure}{0.24\textwidth}
        \centering
        \resizebox{0.86\linewidth}{!}{%
        \feynmandiagram [small, vertical=b to c] {
            a -- [scalar] c -- [scalar] e -- [scalar] b -- [scalar] d,
            b -- [graviton] c,
            e -- [graviton] f,
            g -- [plain, thick] f, h -- [plain, thick] f,
        };}
    \end{subfigure}
    \caption{Insertions on external legs and triangle topologies.}
    \label{fig:TrianglesCounterterms}
\end{figure}
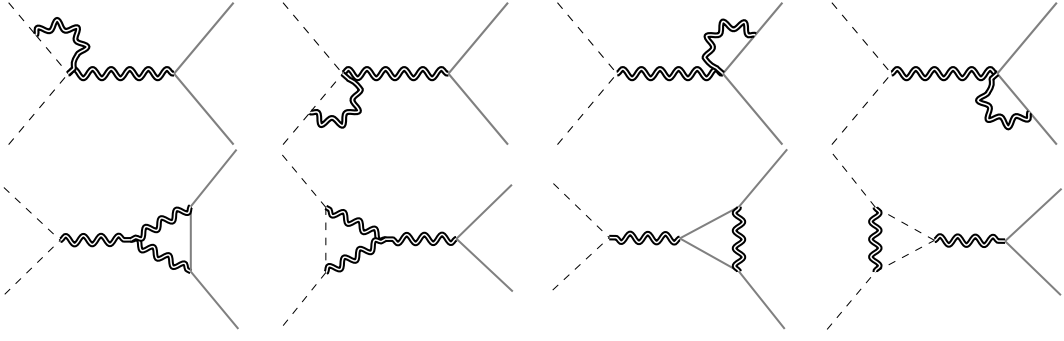

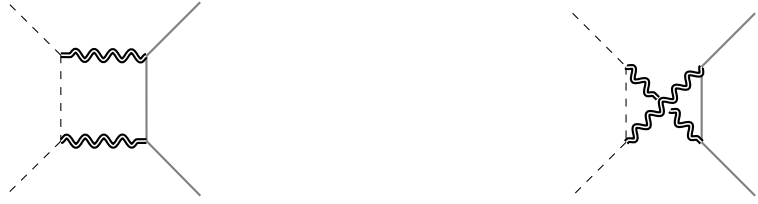
\begin{figure}[h]
    \centering
    \begin{subfigure}{0.49\textwidth}
        \centering
        \feynmandiagram [small, vertical=t2 to t3] {
            t0 -- [plain, thick] t1,
            t0 -- [graviton] t2,
            t2 -- [scalar] t3,
            t3 -- [graviton] t1,
            t2 -- [scalar] p1,
            t3 -- [scalar] p2,
            t0 -- [plain, thick] p4,
            t1 -- [plain, thick] p3,
        };
    \end{subfigure}%
    \begin{subfigure}{0.49\textwidth}
        \centering
        \begin{tikzpicture}[scale=0.5]
        \begin{feynman}
            \vertex (t0) at (0, 1);   \vertex (t2) at (2, 1);
            \vertex (t1) at (0,-1);   \vertex (t3) at (2,-1);
            \vertex[above left=1cm of t0] (p4);
            \vertex[above right=1cm of t2] (p1);
            \vertex[below left=1cm of t1] (p3);
            \vertex[below right=1cm of t3] (p2);
            \vertex (g1) at (0.84, 0.16);
            \vertex (g2) at (1.16,-0.16);
            \diagram*{
                (p4) -- [scalar] (t0), (p1) -- [plain, thick] (t2),
                (t1) -- [scalar] (p3), (t3) -- [plain, thick] (p2),
                (t0) -- [scalar] (t1), (t2) -- [plain, thick] (t3),
                (t2) -- [graviton] (t1),
                (t0) -- [graviton] (g1), (g2) -- [graviton] (t3),
            };
        \end{feynman}
        \end{tikzpicture}
    \end{subfigure}
    \caption{Box topologies.}
    \label{fig:Boxes}
\end{figure}

To answer this, we compute the one-loop amplitude from the diagrams collected in Figs.~\ref{fig:GravitonBubbles}, \ref{fig:VertexCorrections}, \ref{fig:TrianglesCounterterms} and \ref{fig:Boxes}, and absorb the resulting divergences into the counterterm action~\cite{tHooft:1974toh}
\begin{equation}
\label{eq:LagrangianCounterterm}
\begin{split}
    S_{\rm ct} = \int\!\diff^{4}x\,\sqrt{-g}\,\Bigg[
    &-\frac{2}{\kappa^{2}}\big(\delta c_{\rm EH}R-2\delta\Lambda_{\rm cc}\big)
    +\frac{\delta Z_{\phi}}{2}(\partial\phi)^{2}
    +\frac{\delta Z_{\Phi}}{2}\Big((\partial\Phi)^{2}-\delta_{M}M^{2}\Phi^{2}\Big)
    \\
    &+\frac{\lambda_{\phi}}{4!}(\partial\phi)^{4}
    +\frac{\lambda_{\Phi}}{4!}(\partial\Phi)^{4}
    +\frac{\lambda^{M}_{\Phi}}{4!}\,M^{4}\Phi^{4}
    \\
    &+\kappa^{2}\frac{\lambda_{\phi\Phi}}{2!\,2!}(\partial\phi)^{2}(\partial\Phi)^{2}
    +\kappa^{2}\frac{\tilde\lambda_{\phi\Phi}}{2!\,2!}
      \big(g^{\mu\nu}\partial_{\mu}\phi\partial_{\nu}\Phi\big)^{2}
    +\kappa^{2}\frac{\lambda^{M}_{\phi\Phi}}{2!\,2!}\,M^{2}\Phi^{2}(\partial\phi)^{2}
    \Bigg] \ ,
\end{split}
\end{equation}
where $(\partial X)^{2}\equiv g^{\mu\nu}\partial_{\mu}X\partial_{\nu}X$. To extract the worldline response, we must isolate the soft contribution in the heavy-mass expansion. We first examine the naive limit of the integrated relativistic result to show why it gives different beta functions.

We use a basis without the non-minimal operators $R(\partial\phi)^2$, $R^{\mu\nu}\partial_\mu\phi\partial_\nu\phi$, $R\Phi^2$, $R^2$, and $R_{\mu\nu}R^{\mu\nu}$. These operators can be removed by field redefinitions using the Einstein equation, which relates the Ricci tensor to the scalar stress tensor:
\begin{equation}
  R_{\mu\nu} \sim \kappa^{2}\Big(\partial_{\mu}\phi\,\partial_{\nu}\phi
  + \partial_{\mu}\Phi\,\partial_{\nu}\Phi
  + g_{\mu\nu}\,M^{2}\Phi^{2}\Big) \ ,
\end{equation}
so that, for instance, $R^{\mu\nu}\partial_{\mu}\phi\partial_{\nu}\phi \to \kappa^{2}\big[(\partial\phi)^{4} + (\partial\phi\!\cdot\!\partial\Phi)^{2} + \dots\big]$. The one curvature-squared structure not of Ricci type, $R_{\mu\nu\rho\sigma}R^{\mu\nu\rho\sigma}$, is removed instead by the Gauss--Bonnet identity, which renders it topological in four dimensions. These basis reductions leave the quartic scalar operators listed in the counterterm action.

Isolating the cross couplings $\lambda_{\phi\Phi}$,
$\tilde\lambda_{\phi\Phi}$ and $\lambda^{M}_{\phi\Phi}$ requires first fixing every lower-order
counterterm and including the corresponding diagrams. This is done in
Appendix~\ref{app:counterterms}. The infrared contributions are determined from soft emission and subtracted as described in Appendix~\ref{app:IR}. The UV divergence, after the IR poles are subtracted, reads
\begin{equation}
\label{eq:crosscouplings}
\begin{alignedat}{2}
\lambda_{\phi\Phi}
&= \frac{49}{120}\frac{\kappa^{2}}{16\pi^{2}\epsilon} \ , \quad
\tilde\lambda_{\phi\Phi}
= -\frac{101}{60}\frac{\kappa^{2}}{16\pi^{2}\epsilon} \ ,\quad
\lambda^{M}_{\phi\Phi}
&= \frac{7}{20}\frac{\kappa^{2}}{16\pi^{2}\epsilon} \ ,
\end{alignedat}
\end{equation}
from which the one-loop beta functions follow as twice the residues,
\begin{equation}
\label{eq:betafunctions}
\beta_{\lambda_{\phi\Phi}} = -\frac{49}{60}\frac{\kappa^{2}}{16\pi^{2}} \ ,
\qquad
\beta_{\tilde\lambda_{\phi\Phi}} = \frac{101}{30}\frac{\kappa^{2}}{16\pi^{2}} \ ,
\qquad
\beta_{\lambda^{M}_{\phi\Phi}} = -\frac{7}{10}\frac{\kappa^{2}}{16\pi^{2}} \ .
\end{equation}
All three cross couplings therefore run, and the Love number operators are generated by the flow, as in
the worldline theory.

We match the mixed counterterms to the worldline response coefficients. Consider the scattering process
$\Phi(P_i)\phi(p_i)\to\Phi(P_f)\phi(p_f)$, with $P_i^2=P_f^2=M^2$ and $p_i^2=p_f^2=0$. The mixed
operators of Eq.~\eqref{eq:LagrangianCounterterm} generate the contact amplitude
\begin{equation}
\label{eq:QFTtidalCTamplitude}
\begin{split}
\mathcal{M}_{\rm ct}^{\rm QFT}
=i\kappa^{2}M^2\left[\lambda_{\phi\Phi}\, p_i\!\cdot\!p_f \frac{P_i\!\cdot\!P_f}{M^2}
   + \lambda_{\phi\Phi}^{M}\, p_i\!\cdot\!p_f+\frac{\widetilde\lambda_{\phi\Phi}}{2}
 \Big[\frac{p_i\!\cdot\!P_i\ p_f\!\cdot\!P_f}{M^2}+\frac{p_i\!\cdot\!P_f\ p_f\!\cdot\!P_i}{M^2}\Big]\right] \ .
\end{split}
\end{equation}
To compare with the worldline normalization we divide by $2M$ and retain the leading term in the
heavy-mass expansion. Setting $P_i^{\mu}\simeq P_f^{\mu}\simeq M\,u^{\mu}$ and $u\cdot p_i=u\cdot p_f=\omega$, so that
$P_i\!\cdot\!P_f\to M^{2}$ and $p\!\cdot\!P\to M\omega$ for each of the four mixed products,
\begin{equation}
\label{eq:QFTtidalCTheavy}
\left.\frac{\mathcal{M}_{\rm ct}^{\rm QFT}}{2M}\right|_{\rm heavy}
= \frac{iM\kappa^{2}}{2}
  \Big[\big(\lambda_{\phi\Phi}+\lambda_{\phi\Phi}^{M}\big)(p_i\!\cdot\!p_f)
       +\widetilde\lambda_{\phi\Phi}\,\omega^{2}\Big] \ .
\end{equation}
At leading order in the heavy-mass expansion, the operators multiplying $\lambda_{\phi\Phi}$ and $\lambda_{\phi\Phi}^{M}$ both contribute through $M^2(p_i\cdot p_f)$. Only their sum enters, leaving two independent worldline momentum structures. The corresponding worldline amplitude is
\begin{equation}
\label{eq:WLtidalCTamplitude}
\mathcal{M}_{\rm ct}^{\rm WL}
= iM\kappa^{2}r_{\rm h}^{2}
  \Big[\hat\lambda_{1}^{(0)}\big(p_i\!\cdot\!p_f-\omega^{2}\big)+\hat\lambda_{0}^{(1)}\,\omega^{2}\Big] \ .
\end{equation}
Matching the coefficients of $p_i\!\cdot\!p_f$ and of $\omega^{2}$ between
Eqs.~\eqref{eq:QFTtidalCTheavy} and \eqref{eq:WLtidalCTamplitude},
\begin{equation}
\label{eq:QFTtoWLcouplings}
r_{\rm h}^{2}\hat\lambda_{1}^{(0)}=\frac{1}{2}\big(\lambda_{\phi\Phi}+\lambda_{\phi\Phi}^{M}\big) \ ,
\qquad
r_{\rm h}^{2}\hat\lambda_{0}^{(1)}=\frac{1}{2}
  \big(\lambda_{\phi\Phi}+\lambda_{\phi\Phi}^{M}+\widetilde\lambda_{\phi\Phi}\big) \ .
\end{equation}

Inserting the ultraviolet poles of Eq.~\eqref{eq:crosscouplings} into
Eq.~\eqref{eq:QFTtoWLcouplings}, with $r_{\rm h}=2GM$,
naively, we obtain the beta functions,
\begin{equation}
\label{eq:WLbetafunctions}
\beta_{\hat\lambda_{1}^{(0)}}^{\rm \small naive} = -\frac{91}{240\pi}\frac{M_{\rm pl}^{2}}{M^{2}} \ ,
\qquad
\beta_{\hat\lambda_{0}^{(1)}}^{\rm \small naive} = \frac{37}{80\pi}\frac{M_{\rm pl}^{2}}{M^{2}} \ ,
\end{equation}
in the worldline limit. These disagree with the beta functions obtained directly in the worldline theory,
Eq.~\eqref{eq:BetaFunctions}.

The poles in Eq.~\eqref{eq:crosscouplings} were obtained by integrating before expanding in $M$, and therefore include contributions from the hard region, where the loop momentum is of order $M$~\cite{Beneke:1997zp}. The worldline calculation contains only the soft contribution. The relativistic theory also includes closed loops of $\Phi$, which have no counterpart in the fixed-worldline description. These contributions must be distinguished when extracting the worldline beta functions.

\begin{figure}[h]
    \centering
    \newcommand{\wlscale}{0.62}
    \newcommand{\dgm}[1]{\raisebox{-0.5\height}{#1}}
    \newcommand{\pl}{\raisebox{-0.5\height}{$+$}}
    \newcommand{\desc}[1]{\dgm{\resizebox{!}{2.0cm}{#1}}}

    \begin{tabular}{@{}c@{\hspace{6pt}}c@{\hspace{6pt}}c@{\hspace{6pt}}c@{\hspace{6pt}}c@{\hspace{10pt}}c@{\hspace{10pt}}c@{}}

    \desc{\begin{tikzpicture}
    \begin{feynman}
      \vertex (t0) at (0, 0.75);   \vertex (t2) at (1.5, 0.75);
      \vertex (t1) at (0,-0.75);   \vertex (t3) at (1.5,-0.75);
      \vertex[above left=1.3cm of t0] (p4);
      \vertex[below left=1.3cm of t1] (p3);
      \vertex[above right=1.3cm of t2] (p1);
      \vertex[below right=1.3cm of t3] (p2);
      \diagram*{
        (p4) -- [scalar] (t0), (t1) -- [scalar] (p3),
        (t0) -- [scalar] (t1),
        (t2) -- [plain, thick] (p1), (t3) -- [plain, thick] (p2),
        (t2) -- [plain, thick] (t3),
        (t0) -- [graviton] (t2), (t1) -- [graviton] (t3),
      };
    \end{feynman}
  \end{tikzpicture}}
    & \pl &
    \desc{\begin{tikzpicture}
    \begin{feynman}
      \vertex (t0) at (0, 0.75);   \vertex (t2) at (1.5, 0.75);
      \vertex (t1) at (0,-0.75);   \vertex (t3) at (1.5,-0.75);
      \vertex[above left=1.3cm of t0] (p4);
      \vertex[below left=1.3cm of t1] (p3);
      \vertex[above right=1.3cm of t2] (p1);
      \vertex[below right=1.3cm of t3] (p2);
      \vertex (g1) at (0.63, 0.12);  \vertex (g2) at (0.87,-0.12);
      \diagram*{
        (p4) -- [scalar] (t0), (t1) -- [scalar] (p3),
        (t0) -- [scalar] (t1),
        (t2) -- [plain, thick] (p1), (t3) -- [plain, thick] (p2),
        (t2) -- [plain, thick] (t3),
        (t2) -- [graviton] (t1),
        (t0) -- [graviton] (g1), (g2) -- [graviton] (t3),
      };
    \end{feynman}
  \end{tikzpicture}}
    & \pl &
    \desc{\begin{tikzpicture}
    \begin{feynman}
      \vertex (e) at (-1.5,  1.5);   \vertex (f) at (-1.5, -1.5);
      \vertex (t2) at (-0.3, 0.85);  \vertex (t3) at (-0.3, -0.85);
      \vertex (b) at ( 0.9, 0);
      \vertex (g) at ( 2.1, 1.1);    \vertex (h) at ( 2.1, -1.1);
      \diagram* {
        (e) -- [scalar] (t2), (f) -- [scalar] (t3),
        (t2) -- [scalar] (t3),
        (t2) -- [graviton] (b), (t3) -- [graviton] (b),
        (b) -- [plain, thick] (g), (b) -- [plain, thick] (h),
      };
    \end{feynman}
  \end{tikzpicture}}
    &
    $\longrightarrow$
    &
    \dgm{\begin{tikzpicture}[scale=\wlscale]
      \begin{feynman}
        \vertex (p1) at (-1.0,  1.2); \vertex (p2) at (-1.0, -1.2);
        \vertex (t2) at ( 0.5, 0.9);  \vertex (t3) at ( 0.5,-0.9);
        \vertex (t1) at ( 2,   0.9);  \vertex (t0) at ( 2,  -0.9);
        \vertex (y)  at ( 2,   1.8);  \vertex (x)  at ( 2,  -1.8);
        \diagram* {
          (p1) -- [scalar] (t2) -- [scalar] (t3) -- [scalar] (p2),
          (t2) -- [graviton] (t1),
          (t0) -- [pertworld] (t1),
          (t3) -- [graviton] (t0),
          (y) -- (t1), (t0) -- (x),
        };
      \end{feynman}
      \path (-2,-1.8) rectangle (2.1,1.8);
    \end{tikzpicture}}
    \\[3ex]

    \desc{\begin{tikzpicture}
    \begin{feynman}
      \vertex (e) at (-2.4,  1.1);  \vertex (f) at (-2.4, -1.1);
      \vertex (b) at (-1.2, 0);
      \vertex (c) at ( 0.0, 0);
      \vertex (d) at ( 0.85, 0);
      \vertex (g) at ( 2.05, 1.1);  \vertex (h) at ( 2.05, -1.1);
      \diagram* {
        (e) -- [scalar] (b), (f) -- [scalar] (b),
        (b) -- [graviton] (c),
        (c) -- [graviton, out=55, in=125, looseness=1.5] (d),
        (c) -- [graviton, out=-55, in=-125, looseness=1.5] (d),
        (d) -- [plain, thick] (g), (d) -- [plain, thick] (h),
      };
    \end{feynman}
  \end{tikzpicture}}
    & \pl &
    \desc{\begin{tikzpicture}
        \begin{feynman}
          \vertex (e) at (-2.2,  1.1);  \vertex (f) at (-2.2, -1.1);
          \vertex (b) at (-1.1, 0);
          \vertex (t1) at ( 0.0, 0);
          \vertex (t2) at ( 1.0, 0.85); \vertex (t3) at ( 1.0, -0.85);
          \vertex (g) at ( 2.1, 1.5);   \vertex (h) at ( 2.1, -1.5);
          \diagram* {
            (e) -- [scalar] (b), (f) -- [scalar] (b),
            (b) -- [graviton] (t1),
            (t1) -- [graviton] (t2), (t1) -- [graviton] (t3),
            (t2) -- [plain, thick] (t3),
            (t2) -- [plain, thick] (g), (t3) -- [plain, thick] (h),
          };
        \end{feynman}
      \end{tikzpicture}}
    & &
    &
    $\longrightarrow$
    &
    \dgm{\begin{tikzpicture}[scale=\wlscale]
      \begin{feynman}
        \vertex (e) at (-2,  1);   \vertex (f) at (-2, -1);
        \vertex (a) at (-1,  0);   \vertex (b) at (0.5, 0);
        \vertex (c) at ( 2,  0.9); \vertex (d) at ( 2, -0.9);
        \vertex (y) at ( 2,  1.8); \vertex (x) at ( 2, -1.8);
        \diagram* {
          (e) -- [scalar] (a) -- [scalar] (f),
          (a) -- [graviton] (b),
          (b) -- [graviton] (c),
          (c) -- [pertworld] (d),
          (d) -- [graviton] (b),
          (y) -- (c), (d) -- (x),
        };
      \end{feynman}
      \path (-2,-1.8) rectangle (2.1,1.8);
    \end{tikzpicture}}
    \\[3ex]

    \desc{\begin{tikzpicture}
        \begin{feynman}
          \vertex (e) at (-2.4,  1.1);  \vertex (f) at (-2.4, -1.1);
          \vertex (b) at (-1.2, 0);
          \vertex (c) at (-0.35, 0);
          \vertex (g) at ( 0.85, 1.1);  \vertex (h) at ( 0.85, -1.1);
          \diagram* {
            (e) -- [scalar] (b), (f) -- [scalar] (b),
            (b) -- [graviton, out=55, in=125, looseness=1.5] (c),
            (b) -- [graviton, out=-55, in=-125, looseness=1.5] (c),
            (c) -- [plain, thick] (g), (c) -- [plain, thick] (h),
          };
        \end{feynman}
      \end{tikzpicture}}
    & \pl &
    \desc{\begin{tikzpicture}
        \begin{feynman}
          \vertex (e) at (-2.2,  1.1);  \vertex (f) at (-2.2, -1.1);
          \vertex (t1) at (-1.1, 0);
          \vertex (t2) at ( 0.1, 0.85); \vertex (t3) at ( 0.1, -0.85);
          \vertex (g) at ( 1.3, 1.5);   \vertex (h) at ( 1.3, -1.5);
          \diagram* {
            (e) -- [scalar] (t1), (f) -- [scalar] (t1),
            (t1) -- [graviton] (t2), (t1) -- [graviton] (t3),
            (t2) -- [plain, thick] (t3),
            (t2) -- [plain, thick] (g), (t3) -- [plain, thick] (h),
          };
        \end{feynman}
      \end{tikzpicture}}
    & &
    &
    $\longrightarrow$
    &
    \dgm{\begin{tikzpicture}[scale=\wlscale]
      \begin{feynman}
        \vertex (e) at (-1.5,  1.5); \vertex (f) at (-1.5, -1.5);
        \vertex (b) at (0.2, 0);
        \vertex (c) at ( 2,  0.9);   \vertex (d) at ( 2, -0.9);
        \vertex (y) at ( 2,  1.8);   \vertex (x) at ( 2, -1.8);
        \diagram* {
          (e) -- [scalar] (b) -- [scalar] (f),
          (b) -- [graviton] (d),
          (c) -- [pertworld] (d),
          (c) -- [graviton] (b),
          (y) -- (c), (d) -- (x),
        };
      \end{feynman}
      \path (-2,-1.8) rectangle (2.1,1.8);
    \end{tikzpicture}}
    \end{tabular}

    \caption{Scalar one-loop diagrams (left of each arrow) together with the worldline one-loop diagram they reduce to in the heavy-mass limit (right of each arrow).}
    \label{fig:RecoilDescendants}
\end{figure}
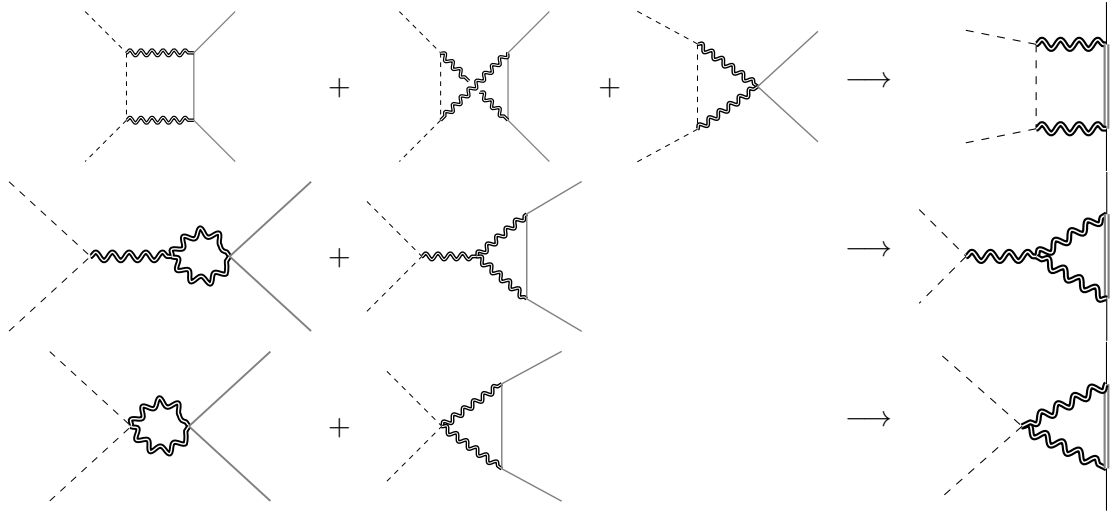

We expand the integrands in the soft region before integration~\cite{Beneke:1997zp,Damgaard:2019lfh,Aoude:2020onz,Brandhuber:2021eyq,Mogull:2020sak}. Writing the loop momentum as $l^{\mu}=M u^{\mu}+\ell^{\mu}$
with $u^{\mu}$ the worldline four-velocity and expanding the integrand in the soft region
$\ell\ll M$, the massive propagators linearize onto the worldline recoil propagators. One then retains the term linear in $M$ after the relativistic $1/(2M)$ normalization. Under this expansion most diagrams with massive propagators in the loop become scaleless and vanish in dimensional regularization. The nonvanishing diagrams are shown in Fig.~\ref{fig:RecoilDescendants} together with the worldline diagrams they reduce to. Diagrams with no massive propagator inside the loop match directly, both at the integrand level and after integration.

For a single heavy propagator, write the momentum of the heavy line as $P^{\mu}=Mu^{\mu}$ with $u^{2}=1$, and let $\ell^{\mu}$ be the momentum flowing in from the loop. The massive propagator is then
\begin{equation}
\label{eq:MassivePropExpansion}
  \frac{i}{(P+\ell)^{2}-M^{2}+i0^+}
  = \frac{i}{2M\,u\!\cdot\!\ell+\ell^{2}+i0^+}
  = \frac{i}{2M}\,\frac{1}{u\!\cdot\!\ell+i0^+}
    \left[1-\frac{\ell^{2}}{2M\,u\!\cdot\!\ell}+\mathcal{O}\!\left(\frac{\ell^{2}}{M^{2}}\right)\right] \ ,
\end{equation}
where the expansion is in the soft region $\ell\ll M$. The leading term is the Fourier transform of the ordering of vertex insertions along the trajectory, $\int\!\diff\tau\,e^{i\,u\cdot\ell\,\tau}\theta(\tau)=i/(u\!\cdot\!\ell+i0)$, and localizes the graviton couplings on the worldline. The sub-leading term is the propagator of the worldline fluctuation.

Carrying out the calculation this way, we find
\begin{equation}
\begin{split}
    \beta^{\rm QFT \to WL}_{\hat\lambda^{(1)}_{0}}
    &= \frac{3}{2\pi} \frac{\ell_{\rm pl}^2}{r_{\rm h}^2}\ ,\quad \beta^{\rm QFT \to WL}_{\hat\lambda^{(0)}_{1}}=-\frac{1}{40\pi}\frac{\ell_{\rm pl}^2}{r_{\rm h}^2}\ .
\end{split}
\end{equation}
These agree with the worldline beta functions in Eq.~\eqref{eq:BetaFunctions}. The soft-region expansion reproduces the worldline result, while the full relativistic integrals also contain contributions from loop momenta of order $M$.

The relation $r_{\rm h}\lambda_{\rm C}=2\ell_{\rm pl}^2$, with $\lambda_{\rm C}=1/M$ and $r_{\rm h}=2GM$, distinguishes the parametrically separated regimes. For $M\gg M_{\rm pl}$, the ordering is $\lambda_{\rm C}\ll\ell_{\rm pl}\ll r_{\rm h}$. The agreement of the beta functions establishes agreement of the soft contributions; it does not determine the short-distance matching coefficients of a black hole.

\section{Discussion}

We have computed the one-loop, first post-Minkowskian amplitude for a massless scalar and a photon scattering off a Schwarzschild worldline, and extracted from its ultraviolet divergence the renormalization of the leading tidal operators. We find that both the leading static and the leading dynamical Love numbers acquire a non-trivial renormalization-group flow from quantum corrections. The flow is inhomogeneous, i.e., independent of the tidal couplings themselves: the beta functions are nonzero even when the tidal couplings vanish. The classical statement that $\hat\lambda^{(0)}_{\ell} = 0$ is therefore not preserved by the flow and cannot be imposed at more than one scale.

The EFT fixes all logarithms associated with this running at the order considered. Their coefficients are determined by the low-energy theory, independently of the finite matching contributions from black-hole structure, so no choice of finite matching constants can make the Love numbers vanish at all scales. The finite values at a reference scale require one-loop matching to black-hole perturbation theory, which we leave for future work.

\section*{Acknowledgments}

We thank Austin Joyce, R. Loganayagam, Donal O'Connell, Ashoke Sen, and Jan Steinhoff for helpful discussions. We acknowledge the use of AI assistants (Claude and ChatGPT) for evaluating integrals and for writing Mathematica code. G.M. thanks IHES for hospitality during a brief visit, which led to the discussions that initiated this project.  The research of M.V.S.S. is supported by the
National Post-Doctoral Fellowship (PDF/2025/004764),
ANRF, Government of India. G.M. and M.V.S.S. acknowledge support from the Department of Atomic Energy, Government of India, under project no. RTI4019. The work of J.P.-M. was supported by the European Research
Council (ERC) under the European Union’s Horizon 2020 research and innovation programme, grant agreement ERC-StG-101221094 GravitaS, PI Julio Parra-Martinez.

\appendix

\section{Evaluation of Master Integrals}\label{sec:AppMasters}

In this appendix we collect the master integrals required in Sec.~\ref{sec:LoveRunning}, together with their derivations and $\epsilon$ expansions. After integration-by-parts (IBP) reduction~\cite{Chetyrkin:1981qh}, every one-loop integral arising in the worldline calculation can be written in terms of four members of the family in Eq.~\eqref{eq:DefMaster}: $J_{0011}$, $J_{1100}$, $J_{1110}$, and $J_{1111}$.

After IBP reduction, eight of the nine diagrams reduce to $J_{1100}$. The recoil box in Fig.~\ref{subfig:RecoilTri} requires all four master integrals. Masters with $n_3\ne0$ arise only from diagrams containing worldline fluctuation propagators.

\subsection*{Evaluation of $J_{1100}$}
Feynman parametrization and Wick rotation give the massless bubble integral:
\begin{equation}
    J_{1100} = \frac{i}{(4 \pi )^{\frac{D}{2}} }(2 p_f\cdot p_i)^{\frac{D}{2}-2}  \Gamma \left(2-\frac{D}{2}\right) \frac{\Gamma \left(\frac{D}{2}-1\right)^2}{\Gamma (D-2)}\ .
\end{equation}

\subsection*{Evaluation of $J_{0011}$}
Shifting $l \to l + p_f$ removes the external momentum from the quadratic propagator, leaving
\begin{equation}
    J_{0011} = \int\!\frac{\diff^D l}{(2\pi)^D}\,
    \frac{1}{l^{2}\,(l\cdot u + \omega)} \ ,
    \qquad \omega \equiv p_f\!\cdot\!u \ ,
\end{equation}
so that the only scale is the frequency $\omega$. Working in the rest frame $u^\mu = (1,\vec{0})$ and splitting $l^\mu = (l^0, \vec{l}\,)$,
\begin{equation}
    J_{0011} = \int\!\frac{\diff l^{0}}{2\pi}
    \int\!\frac{\diff^{D-1}\vec{l}}{(2\pi)^{D-1}}\,
    \frac{1}{\big[(l^{0})^{2} - \vec{l}^{\,2}\big]\,(l^{0}+\omega)} \ .
\end{equation}
Performing the $l^0$ integral by Wick rotation and the remaining spatial integral in $D-1$ dimensions gives
\begin{equation}
\label{eq:J0011}
    J_{0011}
    = \frac{2i \pi^{\frac{3}{2}}}{(4\pi)^{\frac{D}{2}}\,
      \sin(D\pi)\,\Gamma\!\left(\tfrac{D-1}{2}\right)}(-\omega)^{D-3} \ .
\end{equation}
The result has mass dimension $D-3$. Near four dimensions, $\sin(D\pi)=-2\pi\epsilon+\mathcal O(\epsilon^3)$, so $J_{0011}$ has a simple pole.

\subsection*{Evaluation of $J_{1110}$}
Since $q \equiv p_i - p_f$ satisfies
$q\cdot u = 0$ in our kinematics, $q^\mu$ is purely spatial in the rest frame,
and with $\vec{q}^{\,2} = -q^{2} = 2\,p_i\!\cdot\!p_f$ the factors in the denominator read
\begin{equation}
    l^{2} = (l^{0})^{2} - \vec{l}^{\,2} \ , \qquad
    (l+q)^{2} = (l^{0})^{2} - (\vec{l}+\vec{q}\,)^{2} \ , \qquad
    l\cdot u = l^{0} \ ,
\end{equation}
so that
\begin{equation}
    J_{1110} = \int\!\frac{\diff l^{0}}{2\pi}
    \int\!\frac{\diff^{D-1}\vec{l}}{(2\pi)^{D-1}}\,
    \frac{1}{\big[(l^{0})^{2} - E_{1}^{2}\big]
             \big[(l^{0})^{2} - E_{2}^{2}\big]\,(l^{0}+i0^+)} \ ,
\end{equation}
where $E_{1} \equiv |\vec{l}\,|$ and $E_{2} \equiv |\vec{l}+\vec{q}\,|$. The
integrand has five poles in $l^{0}$, at $\pm(E_{1}-i0^+)$, $\pm(E_{2}-i0^+)$ and
$-i0^+$, and falls off as $(l^{0})^{-5}$, so the arc at infinity may be discarded
and the contour closed in either half-plane. Closing above encloses only
$l^{0} = -E_{1}+i0^+$ and $l^{0} = -E_{2}+i0^+$, whose residues combine as
\begin{equation}
    \mathrm{Res}_{-E_{1}} + \mathrm{Res}_{-E_{2}}
    = \frac{1}{2(E_{1}^{2}-E_{2}^{2})}
      \left(\frac{1}{E_{1}^{2}} - \frac{1}{E_{2}^{2}}\right)
    = -\frac{1}{2\,E_{1}^{2}E_{2}^{2}} \ ,
\end{equation}
and therefore
\begin{equation}
    \int\!\frac{\diff l^{0}}{2\pi}\,
    \frac{1}{\big[(l^{0})^{2}-E_{1}^{2}\big]\big[(l^{0})^{2}-E_{2}^{2}\big]
             (l^{0}+i0^+)}
    = -\frac{i}{2\,E_{1}^{2}E_{2}^{2}} \ .
\end{equation}
The energy integral reduces $J_{1110}$ to a spatial Euclidean bubble:
\begin{equation}
    J_{1110} = -\frac{i}{2}
    \int\!\frac{\diff^{D-1}\vec{l}}{(2\pi)^{D-1}}\,
    \frac{1}{\vec{l}^{\,2}\,(\vec{l}+\vec{q}\,)^{2}} \ .
\end{equation}
This is the standard massless bubble in $d = D-1$ dimensions,
\begin{equation}
    \int\!\frac{\diff^{d}\vec{l}}{(2\pi)^{d}}\,
    \frac{1}{\vec{l}^{\,2}(\vec{l}+\vec{q}\,)^{2}}
    = \frac{1}{(4\pi)^{d/2}}\,
      \frac{\Gamma\!\left(\tfrac{d}{2}-1\right)^{2}
            \Gamma\!\left(2-\tfrac{d}{2}\right)}{\Gamma(d-2)}\,
      \big(\vec{q}^{\,2}\big)^{\frac{d}{2}-2} \ ,
\end{equation}
so that
\begin{equation}
\label{eq:J1110}
    J_{1110} = -\frac{i}{2}\,\frac{1}{(4\pi)^{\frac{D-1}{2}}}\,
    \frac{\Gamma\!\left(\tfrac{D-3}{2}\right)^{2}
          \Gamma\!\left(\tfrac{5-D}{2}\right)}{\Gamma(D-3)}\,
    \big(2\,p_i\!\cdot\!p_f\big)^{\frac{D-5}{2}} \ .
\end{equation}
$J_{1110}$ depends on the single scale $2\,p_i\!\cdot\!p_f = -q^{2}$, the same scale that controls $J_{1100}$; every $\Gamma$ function in Eq.~\eqref{eq:J1110} is regular at $d=4$, confirming that this master is finite in both the ultraviolet and the infrared. An alternative derivation uses the Sokhotski--Plemelj identity
\begin{equation}
    \frac{1}{l\cdot u + i0^+}
    = \mathrm{PV}\!\left(\frac{1}{l\cdot u}\right) - i\pi\,\delta(l\cdot u) \ .
\end{equation}
The remaining integrand is even in $l^{0}$, so the principal-value piece
integrates to zero, and the delta function sets $l^{0} = 0$, reproducing
the spatial bubble.

\subsection*{Evaluation of $J_{1111}$}
The master $J_{1111}$ depends on two scales. Feynman parametrization followed by a Mellin--Barnes split~\cite{Smirnov:2012gma} gives a representation in terms of Meijer $G$ functions. Combining the three quadratic propagator denominators gives
\begin{equation}
    \frac{1}{l^{2}(l+q)^{2}(l-p_f)^{2}}
    = 2\int_{0}^{1}\!\diff x\,\diff y\,\diff z\;\delta(1-x-y-z)\,
      \frac{1}{\big[(l+P)^{2}-m_{\perp}^{2}\big]^{3}} \ ,
\end{equation}
where the shift momentum and the induced mass are
\begin{equation}
    P^{\mu} \equiv y\,q^{\mu} - z\,p_f^{\mu} \ ,
    \qquad
    m_{\perp}^{2} \equiv P^{2} - y\,q^{2} = -q^{2}xy = \vec{q}^{\,2}xy \ .
\end{equation}
Here $p_f^2=0$ and $q\cdot p_f=-q^2/2$. For spacelike $q$, $m_\perp^2=-q^2xy$ is nonnegative.

Next, we combine this cubic denominator with the linear propagator using a Feynman-like parametrization of the form
\begin{equation}
    \frac{1}{A^{a}B^{b}} = \frac{\Gamma(a+b)}{\Gamma(a)\Gamma(b)}
    \int_{0}^{\infty}\!\diff\lambda\,
    \frac{2^{b}\lambda^{b-1}}{\big[A+2\lambda B\big]^{a+b}} \ .
\end{equation}
Integrating over the loop momentum gives
\begin{equation}
\label{eq:J1111parametric}
    J_{1111} = \frac{2i\,\Gamma(2+\epsilon)}{(4\pi)^{2-\epsilon}}
    \int_{0}^{1}\!\diff x\,\diff y\,\diff z\;\delta(1-x-y-z)
    \int_{0}^{\infty}\!\diff\lambda\;\Xi^{-2-\epsilon} \ ,
\end{equation}
where
\begin{equation}
\label{eq:Xidef}
    \Xi \equiv \vphantom{\lambda^2}\;m_{\perp}^{2}+\lambda^{2}
              - 2\lambda z\omega - i0^+ \ ,
    \qquad
    \omega \equiv p_f\cdot u \ .
\end{equation}
Here, we are being explicit about the $i0^{+}$ prescription since it affects the result.

The final result depends on the dimensionless ratio $-q^2/(4\omega^2)$, which appears as the argument of the Meijer $G$ functions. To derive this representation, define
\begin{equation}
    \Xi_{\perp} \equiv m_{\perp}^2= \vec{q}^{\,2}xy \ ,
    \qquad
    \Xi_{\rm wl} \equiv \lambda^{2} - 2\lambda z\omega - i0^+ \ ,
    \qquad
    \Xi = \Xi_{\perp} + \Xi_{\rm wl} \ ,
\end{equation}
we separate the worldline contribution by using the Mellin--Barnes representation
of the binomial
\begin{equation}
    \frac{1}{\big[\Xi_{\rm wl} + \Xi_{\perp}\big]^{2+\epsilon}}
    = \frac{1}{\Gamma(2+\epsilon)}\frac{1}{2\pi i}\int\!\diff s\;
      \Gamma(-s)\,\Gamma(2+\epsilon+s)\,
      \frac{\Xi_{\perp}^{\,s}}{\Xi_{\rm wl}^{\,2+\epsilon+s}} \ .
\end{equation}
The contour of integration runs parallel to the imaginary axis and separates the
poles of $\Gamma(2+\epsilon+s)$ from those of $\Gamma(-s)$. In
the full integrand it is confined to the strip $-1<\text{Re}\, s<-1-\epsilon$,
which is nonempty only for $\epsilon<0$. This factorizes the parametric
integrals. The $\lambda$ integral is done by rescaling
$\lambda = 2z\omega\,t$, giving
\begin{equation}
    \int_{0}^{\infty}\!\diff\lambda\;\Xi_{\rm wl}^{-\alpha}
    = (2z\omega)^{1-2\alpha}\int_{0}^{\infty}\!\diff t\;
      t^{-\alpha}\,(t-1-i0^{+})^{-\alpha} \ ,
    \qquad \alpha \equiv 2+\epsilon+s \ ,
\end{equation}
where the $-i0^{+}$ is inherited from the worldline propagator and fixes the branch
on $0<t<1$, where $\Xi_{\rm wl}<0$. Splitting at $t=1$ and evaluating each
segment as a Beta function, we obtain
\begin{equation}
\label{eq:lambdaint}
    \int_{0}^{\infty}\!\diff\lambda\;\Xi_{\rm wl}^{-\alpha}
    = (2z\omega)^{1-2\alpha}\left[
      \frac{\Gamma(1-\alpha)\Gamma(2\alpha-1)}{\Gamma(\alpha)}
      + e^{i\pi\alpha}\,\frac{\Gamma(1-\alpha)^{2}}{\Gamma(2-2\alpha)}
    \right] \ .
\end{equation}
Using the reflection formula on
$\Gamma(2-2\alpha)^{-1}$ and on $\Gamma(1-\alpha)/\Gamma(\alpha)$ in the second term, the bracket collapses to
\begin{equation}
\label{eq:bracketcollapse}
    \Big[\,\cdots\Big]
    = \frac{\Gamma(1-\alpha)\Gamma(2\alpha-1)}{\Gamma(\alpha)}
      \big[1-2e^{i\pi\alpha}\cos\pi\alpha\big]
    = -\,e^{2\pi i\alpha}\,
      \frac{\Gamma(1-\alpha)\Gamma(2\alpha-1)}{\Gamma(\alpha)} \ ,
\end{equation}
so that keeping only the first term would change the overall sign and discard
the phase $e^{2\pi i\alpha}$, which is precisely the Coulomb phase of the
worldline. The simplex integral is Dirichlet,
\begin{equation}
\label{eq:simplexint}
    \int_{0}^{1}\!\diff x\,\diff y\,\diff z\;\delta(1-x-y-z)\,
    (xy)^{s}\,z^{-3-2\epsilon-2s}
    = \frac{\Gamma(1+s)^{2}\,\Gamma(-2-2\epsilon-2s)}{\Gamma(-2\epsilon)} \ .
\end{equation}

Both $\Gamma$ functions carrying $2s$ are now reduced by the Legendre duplication
formula, $\Gamma(2w) = 2^{2w-1}\pi^{-1/2}\,\Gamma(w)\Gamma(w+\tfrac12)$, applied
at $w = \tfrac32+\epsilon+s$ and $w = -1-\epsilon-s$ respectively:
\begin{equation}
\begin{split}
    \Gamma(3+2\epsilon+2s)
    &= \frac{2^{2+2\epsilon+2s}}{\sqrt{\pi}}\,
       \Gamma\!\big(\tfrac{3}{2}+\epsilon+s\big)\,\Gamma(2+\epsilon+s) \ , \\
    \Gamma(-2-2\epsilon-2s)
    &= \frac{2^{-3-2\epsilon-2s}}{\sqrt{\pi}}\,
       \Gamma(-1-\epsilon-s)\,\Gamma\!\big(-\tfrac{1}{2}-\epsilon-s\big) \ .
\end{split}
\end{equation}
The factor $\Gamma(2+\epsilon+s)$ so produced cancels against the denominator
of Eq.~\eqref{eq:bracketcollapse}, leaving an
$s$-independent constant $1/(2\pi)$. What remains of the $s$ dependence outside
the $\Gamma$ functions is a single dimensionless ratio
\begin{equation}
    \big(\vec{q}^{\,2}\big)^{s}\,(2\omega)^{-2s} = \zeta^{\,s} \ ,
    \qquad
    \zeta \equiv \frac{\vec{q}^{\,2}}{4\omega^{2}}
    = -\frac{q^{2}}{4(p_f \cdot u)^{2}} \ .
\end{equation}
In the rest frame $\zeta = \sin^{2}(\theta/2)$ with $\theta$ the scattering angle, so $\zeta \in [0,1]$ throughout the physical
region.

Keeping the two terms in Eq.~\eqref{eq:lambdaint} separate gives a principal-branch representation for each. The first term gives
\begin{equation}
\label{eq:MBfinal}
\begin{split}
    \frac{1}{2\pi}\frac{1}{2\pi i}\int\!\diff s\;
    &\Gamma(-s)\,\Gamma(-1-\epsilon-s)^{2}\,
      \Gamma\!\big(-\tfrac12-\epsilon-s\big) \Gamma(1+s)^{2}\,\Gamma\!\big(\tfrac32+\epsilon+s\big)\,
      \Gamma(2+\epsilon+s)
    \;\zeta^{\,s} \ ,
\end{split}
\end{equation}
in which one factor of $\Gamma(-1-\epsilon-s)$ comes from the $\lambda$
integration and the other from duplicating the Dirichlet $\Gamma$ in
\eqref{eq:simplexint}. The second term of \eqref{eq:lambdaint} has its
$\Gamma(2-2\alpha)$ canceled directly against the numerator
$\Gamma(-2-2\epsilon-2s)$ of \eqref{eq:simplexint}, needs no duplication, and
gives
\begin{equation}
\label{eq:MBsecond}
    e^{i\pi\epsilon}\,\frac{1}{2\pi i}\int\!\diff s\;
    \Gamma(-s)\,\Gamma(-1-\epsilon-s)^{2}\,
    \Gamma(1+s)^{2}\,\Gamma(2+\epsilon+s)\;
    \big(e^{i\pi}\zeta\big)^{s} \ .
\end{equation}
Both are Mellin--Barnes representations of Meijer functions,
\begin{equation}
    G^{m,n}_{p,q}\!\left(\left.\begin{matrix} a_{1},\dots,a_{p}\\
    b_{1},\dots,b_{q}\end{matrix}\right| \zeta\right)
    = \frac{1}{2\pi i}\int\!\diff s\;
    \frac{\prod_{j=1}^{m}\Gamma(b_{j}-s)\;
          \prod_{j=1}^{n}\Gamma(1-a_{j}+s)}
         {\prod_{j=m+1}^{q}\Gamma(1-b_{j}+s)\;
          \prod_{j=n+1}^{p}\Gamma(a_{j}-s)}\;\zeta^{\,s} \ ,
\end{equation}
in the maximally balanced cases $m=n=p=q=4$ and $m=n=p=q=3$, for which the
denominator products are empty. Reading off the $\Gamma(b_{j}-s)$ from the first
lines and the $\Gamma(1-a_{j}+s)$ from the second,
\begin{equation}
\begin{split}
    \{b_{j}\}^{(4)} &= \Big\{0,\;-1-\epsilon,\;-1-\epsilon,\;
        -\tfrac12-\epsilon\Big\} \ , \quad
    \{a_{j}\}^{(4)} = \Big\{0,\;0,\;-\tfrac12-\epsilon,\;
        -1-\epsilon\Big\} \ , \\
    \{b_{j}\}^{(3)} &= \Big\{0,\;-1-\epsilon,\;-1-\epsilon\Big\} \ ,
    \qquad\quad\ \
    \{a_{j}\}^{(3)} = \Big\{0,\;0,\;-1-\epsilon\Big\} \ .
\end{split}
\end{equation}

Restoring the prefactor of Eq.~\eqref{eq:J1111parametric}, the $1/\Gamma(2+\epsilon)$ of the Mellin--Barnes split, the $1/\Gamma(-2\epsilon)$ of Eq.~\eqref{eq:simplexint} and the constant $1/(2\pi)$ from the duplication, we obtain
\begin{equation}
\label{eq:J1111}
\begin{split}
    J_{1111} = \frac{i\,4^{-1-\epsilon}\,(-p_f\cdot u)^{-3-2\epsilon}}
    {(4\pi)^{2-\epsilon}\,\Gamma(-2\epsilon)}
    \Bigg[\;&\frac{1}{2\pi}\,
    G^{4,4}_{4,4}\!\left(\left.
    \begin{matrix}
      0,\;0,\;-\tfrac12-\epsilon,\;-1-\epsilon \\
      0,\;-1-\epsilon,\;-1-\epsilon,\;-\tfrac12-\epsilon
    \end{matrix}
    \;\right|\; \zeta\right)\\
    +\;&e^{i\pi\epsilon}\,
    G^{3,3}_{3,3}\!\left(\left.
    \begin{matrix}
      0,\;0,\;-1-\epsilon \\
      0,\;-1-\epsilon,\;-1-\epsilon
    \end{matrix}
    \;\right|\; -\zeta\right)\Bigg] \ ,
\end{split}
\end{equation}
with $\zeta = -q^{2}/[4(p_f\cdot u)^{2}]$.

The poles in $\epsilon$ arise from the pinching of the contour between the double pole of $\Gamma(-1-\epsilon-s)^{2}$ at $s=-1-\epsilon$ and that of $\Gamma(1+s)^{2}$ at $s=-1$, whose separation is exactly $\epsilon$. Two colliding double poles produce $\epsilon^{-3}$, and since $1/\Gamma(-2\epsilon) = -2\epsilon + \mathcal{O}(\epsilon^{2})$ this leaves the $\epsilon^{-2}$ infrared pole identified in the main text. Evaluating the pinch residue at $s=-1-\epsilon+\sigma$, where the half-integer
$\Gamma$'s collapse by
$\Gamma(\tfrac12-\sigma)\Gamma(\tfrac12+\sigma)=\pi\sec\pi\sigma$, and noting
that the remaining contour and the pole at $s=-\tfrac12-\epsilon$ contribute
only at $\mathcal{O}(\epsilon^{0})$ inside the bracket, we find
\begin{equation}
\label{eq:J1111expanded}
\begin{split}
    J_{1111} = -\frac{i}{8\pi^{2}\,(-q^{2})\,(p_f\cdot u)^{1+2\epsilon}}
    \bigg[&\frac{1}{\epsilon^{2}}
    + \frac{\log\pi-\gamma_{E}+i\pi-L}{\epsilon} + \mathcal{O}(\epsilon^{0})\bigg] \ ,
\end{split}
\end{equation}
where $L \equiv \log\!\big[\sqrt{-q^{2}}/(2p_f\cdot u)\big] = \tfrac12\log\zeta$.
With the opposite $i0^{+}$ prescription on the worldline propagator, the $i\pi$ in the $1/\epsilon$ piece changes sign, while the overall sign is unaffected. Thus, using the time-symmetric propagator, we find the answer
\begin{equation}
\label{eq:J1111final}
\begin{split}
    J_{1111} = -\frac{i}{8\pi^{2}\,(-q^{2})\,(p_f\cdot u)^{1+2\epsilon}}
    \bigg[&\frac{1}{\epsilon^{2}}
    + \frac{\log\pi-\gamma_{E}-L}{\epsilon} + \mathcal{O}(\epsilon^{0})\bigg] \ .
\end{split}
\end{equation}

\section{$\phi \Phi \to \phi \Phi$ scattering}

\subsection{Fixing the lower-order counterterms}\label{app:counterterms}

\begin{figure}[h]
    \centering
    \begin{subfigure}{0.49\textwidth}
        \centering
        \begin{minipage}{0.24\textwidth}\centering
            \feynmandiagram [small, horizontal=a to b, inline=(a.base)] {
                a -- [graviton] b -- [half left, plain, thick] bm -- [scalar, plain, thick, half left] b,
            };
        \end{minipage}%
        \caption{Graviton one-point function.}
        \label{fig:GravitonTadpole}
    \end{subfigure}%
    \begin{subfigure}{0.49\textwidth}
        \centering
        \begin{minipage}{\textwidth}\centering
            \feynmandiagram [layered layout, horizontal=b to c] {
                a -- [plain, thick,] b -- [graviton, half left] c,
                b -- [plain, thick,] c,
                c -- [plain, thick,] d,
            };
        \end{minipage}%
        \caption{Massive scalar two-point amplitude.}
        \label{fig:TwoPtMassSc}
    \end{subfigure}\\[8pt]
\begin{subfigure}{\textwidth}
    \centering
    \begin{minipage}{0.19\textwidth}\centering
        \resizebox{!}{1.9cm}{%
        \feynmandiagram [small, horizontal=c to d] {
            b -- [plain, thick, half left] c
              -- [plain, thick, half left] b,
            c -- [graviton] d,
            e -- [graviton] b,
            f -- [graviton] b,
        };}
    \end{minipage}%
    \begin{minipage}{0.19\textwidth}\centering
        \resizebox{!}{1.9cm}{%
        \feynmandiagram [small, vertical=t2 to t3] {
            a -- [graviton] t1 -- [plain, thick] t2 -- [plain, thick] t3 -- [plain, thick] t1,
            t2 -- [graviton] p1,
            t3 -- [graviton] p2,
        };}
    \end{minipage}%
    \begin{minipage}{0.19\textwidth}\centering
        \resizebox{!}{1.9cm}{%
        \feynmandiagram [small, horizontal=b to c] {
            a -- [graviton] b,
            b -- [plain, thick, out=105, in=25, loop, min distance=1.5cm] b,
            b -- [graviton] c,
            b -- [graviton] d,
        };}
    \end{minipage}
    \caption{Graviton three-point amplitude. The first diagram has two further descendants
    obtained by exchanging the external legs.}
    \label{fig:GravitonThreePt}
\end{subfigure}\\[8pt]
    \begin{subfigure}{\textwidth}
    \centering
    \begin{minipage}{0.19\textwidth}\centering
        \resizebox{!}{1.9cm}{%
        \feynmandiagram [small, horizontal=c to d] {
            b -- [graviton, half left] c -- [graviton, half left] b,
            c -- [graviton] d,
            e -- [plain, thick] b,
            f -- [plain, thick] b,
        };}
    \end{minipage}%
    \begin{minipage}{0.19\textwidth}\centering
        \resizebox{!}{1.9cm}{%
        \feynmandiagram [small, vertical=t2 to t3] {
            a -- [graviton] t1 -- [graviton] t2 -- [plain, thick] t3 -- [graviton] t1,
            t2 -- [plain, thick] p1,
            t3 -- [plain, thick] p2,
        };}
    \end{minipage}%
    \begin{minipage}{0.19\textwidth}\centering
        \resizebox{!}{1.9cm}{%
        \feynmandiagram [small, vertical=c to b] {
            a -- [plain, thick] c -- [plain, thick] e -- [plain, thick] b -- [plain, thick] d,
            b -- [graviton] c,
            e -- [graviton] f,
        };}
    \end{minipage}%
    \begin{minipage}{0.19\textwidth}\centering
        \resizebox{!}{1.9cm}{%
        \begin{tikzpicture}\begin{feynman}
            \gravskeletonp
            \vertex (x) at (180-\scalarhalfangle:0.9);
            \diagram* {
                (a) -- [graviton] (d),
                (e) -- [plain, thick] (x) -- [plain, thick] (a),
                (x) -- [graviton, half left] (a),
                (f) -- [plain, thick] (a),
            };
        \end{feynman}\end{tikzpicture}}
    \end{minipage}%
    \begin{minipage}{0.19\textwidth}\centering
        \resizebox{!}{1.9cm}{%
        \begin{tikzpicture}\begin{feynman}
            \gravskeletonp
            \vertex (y) at (180+\scalarhalfangle:0.9);
            \diagram* {
                (a) -- [graviton] (d),
                (e) -- [plain, thick] (a),
                (f) -- [plain, thick] (y) -- [plain, thick] (a),
                (y) -- [graviton, half right] (a),
            };
        \end{feynman}\end{tikzpicture}}
    \end{minipage}
    \caption{Scalar--scalar--graviton amplitude.}
    \label{fig:ScaScaGravVertex}
\end{subfigure}\\[8pt]
    \caption{Diagrams fixing the counterterms. The solid line denotes the massive scalar.}
    \label{fig:Counterterms}
\end{figure}
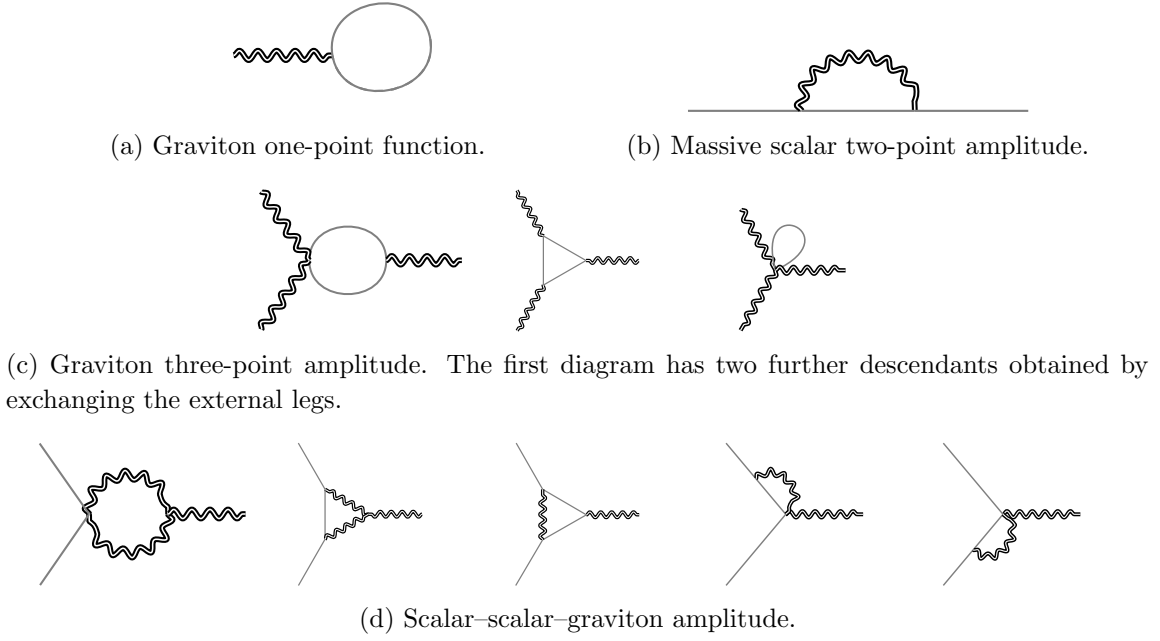

The diagrams in Fig.~\ref{fig:Counterterms} determine the lower-order counterterms. In the on-shell calculation, the graviton and ghost bubbles and triangles of the three-point amplitude, together with the massless-scalar vertex correction, are scaleless and vanish in dimensional regularization. Thus $\delta Z_\phi=0$.

The cosmological constant $\delta\Lambda_{\rm cc}$ is fixed by demanding that $h_{\mu\nu}$ acquire no one-point function (Fig.~\ref{fig:GravitonTadpole}), i.e., that the counterterm cancel the massive-scalar tadpole. Equivalently it may be read off from the on-shell graviton two-point amplitude to which the Einstein--Hilbert counterterm does not contribute, coming as it does with a factor of $k^{2}$. The contributing diagrams are obtained from Fig.~\ref{fig:GravitonBubbles} by removing the external legs. The one- and two-point determinations agree. With $\delta\Lambda_{\rm cc}$ fixed, the graviton three-point amplitude in Fig.~\ref{fig:GravitonThreePt} determines $\delta c_{\rm EH}$, and the massive scalar--scalar--graviton vertex in Fig.~\ref{fig:ScaScaGravVertex} determines $\delta Z_\Phi$. Canceling the ultraviolet divergences gives
\begin{equation}
\label{eq:Counterterms}
    \delta\Lambda_{\rm cc} = -\frac{\kappa^{2}M^{4}}{(16\pi)^{2}\epsilon}
      \ ,
    \quad
    \delta c_{\rm EH} = \frac{\kappa^{2}M^{2}}{384\pi^{2}\epsilon}
      \ ,
    \quad
    \delta Z_{\Phi} = -\frac{\kappa^{2}M^{2}}{16\pi^{2}\epsilon}
      \ .
\end{equation}
Finally, the one-loop correction to the massive scalar propagator (Fig.~\ref{fig:TwoPtMassSc}) shows that the mass does not run, i.e.,
\begin{equation}
    \delta_{M}=0\ .
\end{equation}

\subsection*{A note on IR divergences}

The one-loop $\Phi\Phi h$ amplitude carries an infrared divergence,
\begin{equation}
    \left.\mc{A}^{(1)}_{\Phi\Phi h}\right|_{\rm IR}
    = \frac{i\kappa^{3}M^{2}}{64\pi^{2}\epsilon}\,(p\cdot \varepsilon)^{2} \ ,
\end{equation}
which originates in the on-shell residue of the massive propagator. To see this, write the one-loop
1PI two-point function as $\Gamma(p^{2})$. In the gauge $\xi=2$ (we have checked the general case), \texttt{Package-X} \cite{Patel:2015tea, Patel:2016fam} gives the purely UV-divergent and finite parts to be
\begin{equation}
\label{eq:selfenergy}
    \Gamma(p^{2}) = \frac{i\kappa^{2}M^{2}\Delta}{32\pi^{2}}
    \left[-\frac{2}{\epsilon_{\rm UV}}
      + 2\log\frac{\pi\mu^{2}\Delta}{M^{4}}\,e^{\gamma_{\rm E}-2}
      + \frac{M^{2}}{p^{2}}\log\frac{M^{2}}{\Delta}\right] \ ,
    \qquad \Delta \equiv M^{2}-p^{2} \ .
\end{equation}
The finite part is non-analytic as $p^{2}\to M^{2}$, so the residue cannot be extracted by
differentiating the Laurent-expanded pole and then going on shell. Differentiating the unexpanded
integrand instead, or equivalently, rescaling $p^{\mu}\to\rho p^{\mu}$ and using
$\partial_{p^{2}}=(2\rho M^{2})^{-1}\partial_{\rho}$ at $\rho=1$, gives
\begin{equation}
\label{eq:residue}
    \Gamma'(M^{2}) = \frac{i\kappa^{2}M^{2}}{64\pi^{2}}
    \left[\frac{4}{\epsilon_{\rm UV}} - \frac{1}{\epsilon_{\rm IR}}
      + 3\log\frac{\mu^{2}}{\pi M^{2}} - 3\gamma_{\rm E} + 7\right] \ .
\end{equation}
The unit-residue condition $\Gamma'(M^{2})+\Gamma'_{\rm ct}(M^{2})=0$, with
$\Gamma_{\rm ct}=i\,\delta Z_{\Phi}(p^{2}-M^{2})+\cdots$, then fixes $\delta Z_{\Phi}$. Following the
standard convention, only the ultraviolet part is absorbed into the counterterm; the infrared part is
retained in the LSZ reduction, so that
\begin{equation}
\label{eq:Zphi}
    Z_{\Phi} = 1 - \frac{\kappa^{2}M^{2}}{16\pi^{2}\epsilon_{\rm UV}}
      + \frac{\kappa^{2}M^{2}}{64\pi^{2}\epsilon_{\rm IR}} + \mathcal{O}(\kappa^{4}) \ .
\end{equation}
The ultraviolet part agrees with the counterterm $\delta Z_\Phi$, and the on-shell residue also contains an infrared pole. An amplitude with $n$ external $\Phi$ legs is therefore corrected by
$\tfrac{n}{2}$ times the infrared part of Eq.~\eqref{eq:Zphi},
\begin{equation}
\label{eq:LSZ}
    \mc{M}^{(1)}_{n} = \mc{M}^{(1)}_{n,\,\text{amp}+\text{UV ct}}
      + \frac{n\,\kappa^{2}M^{2}}{128\pi^{2}\epsilon_{\rm IR}}\;\mc{M}^{(0)}_{n} \ .
\end{equation}
We will henceforth drop the subscripts on the regulator $\epsilon$, as in the main text.
For the three-point amplitude, $n=2$ and
$\mc{A}^{(0)}_{\Phi\Phi h}=-i\kappa\,(p\cdot \varepsilon)^{2}$, so the shift exactly cancels the loop pole,
\begin{equation}
    \left.\mc{A}^{(1)}_{\Phi\Phi h}\right|_{\rm IR}
    + \frac{\kappa^{2}M^{2}}{64\pi^{2}\epsilon}\,\mc{A}^{(0)}_{\Phi\Phi h} = 0 \ .
\end{equation}
The process of interest, $\phi\Phi\to\phi\Phi$, likewise has $n=2$ and no massive propagator at tree level, so Eq.~\eqref{eq:LSZ} applies there unchanged. Including the external-leg contribution gives the complete virtual infrared pole, which is subtracted to obtain the ultraviolet part.

\subsection{Isolation of the IR poles}\label{app:IR}

On-shell dimensional regularization can mix ultraviolet and infrared poles through scaleless integrals. We compare the virtual infrared contribution from the loop integrals and massive-scalar residues with the independently computed soft-graviton emission.

We label the incoming and outgoing massless scalars by $k,k'$ and the massive ones by $p,p'$, with
$p^2=p'^2=M^2$. The independent invariants are the half-Mandelstams
\begin{equation}
    \hat s=k\cdot p=k'\cdot p'=\tfrac12(s-M^2),
    \qquad
    \hat u=k\cdot p'=k'\cdot p=\tfrac12(M^2-u),
    \qquad
    \hat t=k\cdot k'=-\tfrac12 t,
    \label{eq:invariants}
\end{equation}
all positive in the physical region, subject to the constraint
\begin{equation}
    \hat t=\hat s-\hat u \ .
    \label{eq:constraint}
\end{equation}
The remaining invariant $p\cdot p'=M^2+\hat t$ is traded for
the relative rapidity $\zeta$ of the massive legs,
\begin{equation}
    p\cdot p'=M^2\cosh\zeta
    \quad\Longleftrightarrow\quad
    \hat t=2M^2\sinh^2\tfrac{\zeta}{2} \ ,
    \label{eq:rapidity}
\end{equation}
so that $2(p\cdot p')^2-M^4=M^4\cosh2\zeta$. The tree amplitude is
$\mathcal{M}^{(0)}=i\kappa^2\hat s\hat u/2\hat t$.

All IR poles below are proportional to $\mathcal{M}^{(0)}$, and we extract them through a single
soft function $\mathcal{W}$,\footnote{$\mathcal{W}$ has mass dimension two; it is the invariant
multiplying $\kappa^{2}$.} defined by
\begin{equation}
    \mathcal{M}^{(1)}\Big|_{\rm IR}
    = -\frac{\kappa^2}{64\pi^2\epsilon}\,\mathcal{W}\,\mathcal{M}^{(0)} \ .
    \label{eq:W-def}
\end{equation}
We absorb the common factor of $\hat t$ into the definition of $\mathcal W$.

\subsubsection*{Virtual poles}

For the truncated one-loop graphs, \texttt{Package-X} gives Eq.~\eqref{eq:W-def} with
\begin{equation}
    \mathcal{W}_{\rm 1PI}
    = M^2\cosh2\zeta\,\frac{\zeta}{\sinh\zeta}
      +4\hat s\log\frac{M^2}{-\hat s}
      -2\hat t\log\frac{2M^2}{\hat t}
      -4\hat u\log\frac{M^2}{\hat u} \ .
    \label{eq:W-1PI}
\end{equation}
The loop measure supplies the factor $\mu^{2\epsilon}$. A $\log\mu$ term in the single pole would therefore be proportional to the double-pole coefficient, which cancels in the sum. The soft-emission calculation below checks this cancellation~\cite{Weinberg:1965nx, Akhoury:2011kq}.

Each of the two external massive legs contributes an on-shell residue factor, giving
$\mathcal{W}_Z=-M^2$; the massless legs carry none, their self-energies being scaleless. The complete
virtual pole is therefore Eq.~\eqref{eq:W-def} with
\begin{equation}
    \mathcal{W}_V = \mathcal{W}_{\rm 1PI}+\mathcal{W}_Z
    = M^2\left[\cosh2\zeta\,\frac{\zeta}{\sinh\zeta}-1\right]
      +4\hat s\log\frac{M^2}{-\hat s}
      -2\hat t\log\frac{2M^2}{\hat t}
      -4\hat u\log\frac{M^2}{\hat u} \ .
    \label{eq:W-virtual}
\end{equation}
The infrared divergence in the inclusive rate must cancel against unresolved soft emission by the Bloch--Nordsieck theorem~\cite{Bloch:1937pw,Kinoshita:1962ur,Lee:1964is}. We compute the emission contribution independently below.

\subsubsection*{Soft-graviton emission}

Writing $p_i\in\{k,k',p,p'\}$ with signature $\sigma_i=-1$ ($+1$) for incoming (outgoing) legs, the
leading soft factor multiplying $|\mathcal{M}^{(0)}|^2$ is \cite{Weinberg:1964ew, Weinberg:1965nx}
\begin{equation}
    \mathcal{S}(q)
    = \frac{\kappa^2}{4}\sum_{i,j}\sigma_i\sigma_j\,
      \frac{(p_i\cdot p_j)^2-\frac{1}{D-2}p_i^2p_j^2}{(p_i\cdot q)(p_j\cdot q)} \ ,
    \label{eq:soft-factor}
\end{equation}
the gauge-dependent terms having dropped out by $\sum_i\sigma_ip_i^\mu=0$. The integrated unresolved
factor is
\begin{equation}
    \mathcal{R}_{\rm soft}=\int_{u\cdot q<\omega_c}\!\!\diff\Phi_1(q)\,\mathcal{S}(q) \ ,
    \qquad
    \diff\Phi_1(q)\equiv\mu^{2\epsilon}\frac{\diff^{D-1}\vec q}{(2\pi)^{D-1}2q^0} \ ,
\end{equation}
with $u^\mu$ the laboratory-frame velocity and $\omega_c$ the detector resolution.

Every term reduces to the dipole family
$\mathcal{J}_{ij}=\int\diff\Phi_1(q)\,[(p_i\cdot q)(p_j\cdot q)]^{-1}$. Writing $q^\mu=\omega n^\mu$
with $\omega=u\cdot q$, the radial integral $\int_0^{\omega_c}\!\diff\omega\,\omega^{-1-2\epsilon}$ is
common to all of them and only the angular integral distinguishes the three classes. With
$E_i\equiv p_i\cdot u$ and $\mathcal{L}\equiv2\log(\omega_c/\mu)+\gamma_E-\log\pi$,
\begin{align}
    \mathcal{J}_{kk'} &= \frac{1}{8\pi^2\hat t}
      \left[\frac{1}{\epsilon^2}
        -\frac{1}{\epsilon}\Big(\mathcal{L}+\log 2+\log\frac{\hat t}{2E_kE_{k'}}\Big)\right] ,
    \nonumber\\
    \mathcal{J}_{kp} &= \frac{1}{16\pi^2(k\cdot p)}
      \left[\frac{1}{\epsilon^2}
        -\frac{1}{\epsilon}\Big(\mathcal{L}+2\log\frac{k\cdot p}{M E_k}\Big)\right] ,
    \label{eq:dipoles}\\
    \mathcal{J}_{ij} &= -\frac{1}{8\pi^2M^2\epsilon}\,\frac{\zeta_{ij}}{\sinh\zeta_{ij}} ,
      \qquad p_i^2=p_j^2=M^2 ,
    \nonumber
\end{align}
each up to $O(\epsilon^0)$, with $\cosh\zeta_{ij}=(p_i\cdot p_j)/M^2$. The mixed case covers all four
pairs, with $k\cdot p\to\hat s$ for $\mathcal{J}_{kp},\mathcal{J}_{k'p'}$ and $\to\hat u$ for
$\mathcal{J}_{kp'},\mathcal{J}_{k'p}$. Only the massless energy appears, since $E$ for the massive leg is fixed by its own rest frame. The purely massive class has no collinear divergence, so the
laboratory frame enters only at $O(\epsilon^0)$, and the self-eikonal term is the smooth limit
$\zeta_{ij}\to0$, $\mathcal{Z}_p\equiv\mathcal{J}_{pp}=-1/(8\pi^2M^2\epsilon)$. The diagonal massless
terms vanish, their gravitational numerator being zero.

The trace term in Eq.~\eqref{eq:soft-factor} vanishes whenever a leg is massless and multiplies a
single pole otherwise, so $1/(D-2)\to1/2$ throughout. Assembling,
\begin{equation}
    \mathcal{R}_{\rm soft}
    = \frac{\kappa^2}{4}\Big[
      -2\hat t^{\,2}\mathcal{J}_{kk'}
      +2\hat s^{\,2}\big(\mathcal{J}_{kp}+\mathcal{J}_{k'p'}\big)
      -2\hat u^{\,2}\big(\mathcal{J}_{kp'}+\mathcal{J}_{k'p}\big)
      +M^4\mathcal{Z}_p
      -M^4\cosh2\zeta\,\mathcal{J}_{pp'}\Big] .
    \label{eq:soft-master}
\end{equation}
The double pole carries the coefficient $-\hat t+\hat s-\hat u$ and so vanishes by
Eq.~\eqref{eq:constraint}. This is the statement that in gravity, soft
emission is unaccompanied by collinear divergences. The single pole takes the form
\begin{equation}
    \mathcal{R}_{\rm soft}\Big|_{\rm IR}
    = \frac{\kappa^2}{32\pi^2\epsilon}\,\mathcal{W}_R \ ,
    \quad
    \mathcal{W}_R
    = M^2\left[\cosh2\zeta\,\frac{\zeta}{\sinh\zeta}-1\right]
      +4\hat s\log\frac{M^2}{\hat s}
      -2\hat t\log\frac{2M^2}{\hat t}
      -4\hat u\log\frac{M^2}{\hat u} \ .
    \label{eq:W-real}
\end{equation}
Note that the surviving pole is Lorentz invariant and independent of both $\omega_c$ and
$\mu$.

\subsubsection*{Cancellation}

Cancellation at the level of the cross section requires $2\mathrm{Re}\big[\mathcal{M}^{(0)*}\mathcal{M}^{(1)}\big]_{\rm IR} =-|\mathcal{M}^{(0)}|^2\mathcal{R}_{\rm soft}|_{\rm IR}$. Inserting Eqs.~\eqref{eq:W-def} and \eqref{eq:W-real}, the common factor drops out and the condition reduces to $\mathrm{Re}\,\mathcal{W}_V=\mathcal{W}_R$. The independently computed virtual and real-emission contributions in Eqs.~\eqref{eq:W-virtual} and \eqref{eq:W-real} satisfy this relation. They differ only in their
imaginary parts,
\begin{equation}
    \mathcal{W}_V-\mathcal{W}_R
    = 4\hat s\left[\log\frac{M^2}{-\hat s}-\log\frac{M^2}{\hat s}\right]
    = 4\pi i\,\hat s
    \quad\Longrightarrow\quad
    \mathcal{M}^{(1)}\Big|_{\rm IR}-\mathcal{M}^{(1)}_{\rm KLN}\Big|_{\rm IR}
    = -\frac{i\kappa^2\hat s}{16\pi\epsilon}\,\mathcal{M}^{(0)} \ .
\end{equation}
This residual divergence is a pure phase, the gravitational Coulomb phase~\cite{Weinberg:1965nx,Kabat:1992tb}. The Bloch--Nordsieck theorem constrains only $\mathrm{Re}\big[\mathcal{M}^{(0)*}\mathcal{M}^{(1)}\big]$, and a phase drops out of $|\mathcal{M}|^2$.

\bibliography{ref}

@article{Mogull:2020sak,
    author = "Mogull, Gustav and Plefka, Jan and Steinhoff, Jan",
    title = "{Classical black hole scattering from a worldline quantum field theory}",
    eprint = "2010.02865",
    archivePrefix = "arXiv",
    primaryClass = "hep-th",
    reportNumber = "UUITP-37/20, HU-EP-20/22-RTG",
    doi = "10.1007/JHEP02(2021)048",
    journal = "JHEP",
    volume = "02",
    pages = "048",
    year = "2021"
}

@article{Patel:2015tea,
    author = "Patel, Hiren H.",
    title = "{Package-X: A Mathematica package for the analytic calculation of one-loop integrals}",
    eprint = "1503.01469",
    archivePrefix = "arXiv",
    primaryClass = "hep-ph",
    doi = "10.1016/j.cpc.2015.08.017",
    journal = "Comput. Phys. Commun.",
    volume = "197",
    pages = "276--290",
    year = "2015"
}

@article{Patel:2016fam,
    author = "Patel, Hiren H.",
    title = "{Package-X 2.0: A Mathematica package for the analytic calculation of one-loop integrals}",
    eprint = "1612.00009",
    archivePrefix = "arXiv",
    primaryClass = "hep-ph",
    doi = "10.1016/j.cpc.2017.04.015",
    journal = "Comput. Phys. Commun.",
    volume = "218",
    pages = "66--70",
    year = "2017"
}

@article{Weinberg:1965nx,
    author = "Weinberg, Steven",
    title = "{Infrared photons and gravitons}",
    doi = "10.1103/PhysRev.140.B516",
    journal = "Phys. Rev.",
    volume = "140",
    pages = "B516--B524",
    year = "1965"
}

@article{Akhoury:2011kq,
    author = "Akhoury, Ratindranath and Saotome, Ryo and Sterman, George",
    title = "{Collinear and Soft Divergences in Perturbative Quantum Gravity}",
    eprint = "1109.0270",
    archivePrefix = "arXiv",
    primaryClass = "hep-th",
    reportNumber = "YITP-SB-11-31",
    doi = "10.1103/PhysRevD.84.104040",
    journal = "Phys. Rev. D",
    volume = "84",
    pages = "104040",
    year = "2011"
}

@article{Weinberg:1964ew,
    author = "Weinberg, Steven",
    title = "{Photons and Gravitons in  $S$-Matrix Theory: Derivation of Charge Conservation and Equality of Gravitational and Inertial Mass}",
    doi = "10.1103/PhysRev.135.B1049",
    journal = "Phys. Rev.",
    volume = "135",
    pages = "B1049--B1056",
    year = "1964"
}

@article{Latosh:2025vax,
    author = "Latosh, Boris",
    title = "{FeynGrav 4.0}",
    eprint = "2510.17320",
    archivePrefix = "arXiv",
    primaryClass = "hep-th",
    doi = "10.1016/j.cpc.2026.110199",
    journal = "Comput. Phys. Commun.",
    volume = "326",
    pages = "110199",
    year = "2026"
}

@article{Porto:2016zng,
    author = "Porto, Rafael A.",
    title = "{The Tune of Love and the Nature(ness) of Spacetime}",
    eprint = "1606.08895",
    archivePrefix = "arXiv",
    primaryClass = "gr-qc",
    doi = "10.1002/prop.201600064",
    journal = "Fortsch. Phys.",
    volume = "64",
    number = "10",
    pages = "723--729",
    year = "2016"
}

@article{Hui:2020xxx,
    author = "Hui, Lam and Joyce, Austin and Penco, Riccardo and Santoni, Luca and Solomon, Adam R.",
    title = "{Static response and Love numbers of Schwarzschild black holes}",
    eprint = "2010.00593",
    archivePrefix = "arXiv",
    primaryClass = "hep-th",
    doi = "10.1088/1475-7516/2021/04/052",
    journal = "JCAP",
    volume = "04",
    pages = "052",
    year = "2021"
}

@article{Charalambous:2021mea,
    author = "Charalambous, Panagiotis and Dubovsky, Sergei and Ivanov, Mikhail M.",
    title = "{On the Vanishing of Love Numbers for Kerr Black Holes}",
    eprint = "2102.08917",
    archivePrefix = "arXiv",
    primaryClass = "hep-th",
    reportNumber = "INR-TH-2021-001",
    doi = "10.1007/JHEP05(2021)038",
    journal = "JHEP",
    volume = "05",
    pages = "038",
    year = "2021"
}

@article{Charalambous:2021kcz,
    author = "Charalambous, Panagiotis and Dubovsky, Sergei and Ivanov, Mikhail M.",
    title = "{Hidden Symmetry of Vanishing Love Numbers}",
    eprint = "2103.01234",
    archivePrefix = "arXiv",
    primaryClass = "hep-th",
    reportNumber = "INR-TH-2021-003",
    doi = "10.1103/PhysRevLett.127.101101",
    journal = "Phys. Rev. Lett.",
    volume = "127",
    number = "10",
    pages = "101101",
    year = "2021"
}

@article{Charalambous:2022rre,
    author = "Charalambous, Panagiotis and Dubovsky, Sergei and Ivanov, Mikhail M.",
    title = "{Love symmetry}",
    eprint = "2209.02091",
    archivePrefix = "arXiv",
    primaryClass = "hep-th",
    doi = "10.1007/JHEP10(2022)175",
    journal = "JHEP",
    volume = "10",
    pages = "175",
    year = "2022"
}

@article{Parra-Martinez:2025bcu,
    author = "Parra-Martinez, Julio and Podo, Alessandro",
    title = "{Naturalness of vanishing black-hole tides}",
    eprint = "2510.20694",
    archivePrefix = "arXiv",
    primaryClass = "hep-th",
    month = "10",
    year = "2025"
}

@article{Saketh:2024juq,
    author = "Saketh, M. V. S. and Zhou, Zihan and Ghosh, Suprovo and Steinhoff, Jan and Chatterjee, Debarati",
    title = "{Investigating tidal heating in neutron stars via gravitational Raman scattering}",
    eprint = "2407.08327",
    archivePrefix = "arXiv",
    primaryClass = "gr-qc",
    doi = "10.1103/PhysRevD.110.103001",
    journal = "Phys. Rev. D",
    volume = "110",
    pages = "103001",
    year = "2024"
}

@misc{Chakrabarti:2013lua,
    author = "Chakrabarti, Sayan and Delsate, T{\'e}rence and Steinhoff, Jan",
    title = "{New perspectives on neutron star and black hole spectroscopy and dynamic tides}",
    eprint = "1304.2228",
    archivePrefix = "arXiv",
    primaryClass = "gr-qc",
    month = "4",
    year = "2013"
}

@article{Saketh:2023bul,
    author = "Saketh, M. V. S. and Zhou, Zihan and Ivanov, Mikhail M.",
    title = "{Dynamical tidal response of Kerr black holes from scattering amplitudes}",
    eprint = "2307.10391",
    archivePrefix = "arXiv",
    primaryClass = "hep-th",
    doi = "10.1103/PhysRevD.109.064058",
    journal = "Phys. Rev. D",
    volume = "109",
    number = "6",
    pages = "064058",
    year = "2024"
}

@article{Mandal:2023hqa,
    author = "Mandal, Manoj K. and Mastrolia, Pierpaolo and Silva, Hector O. and Patil, Raj and Steinhoff, Jan",
    title = "{Renormalizing Love: tidal effects at the third post-Newtonian order}",
    eprint = "2308.01865",
    archivePrefix = "arXiv",
    primaryClass = "hep-th",
    reportNumber = "HU-EP-23/43-RTG",
    doi = "10.1007/JHEP02(2024)188",
    journal = "JHEP",
    volume = "02",
    pages = "188",
    year = "2024"
}

@article{Porto:2016pyg,
    author = "Porto, Rafael A.",
    title = "{The effective field theorist{\textquoteright}s approach to gravitational dynamics}",
    eprint = "1601.04914",
    archivePrefix = "arXiv",
    primaryClass = "hep-th",
    doi = "10.1016/j.physrep.2016.04.003",
    journal = "Phys. Rept.",
    volume = "633",
    pages = "1--104",
    year = "2016"
}

@article{Levi:2018nxp,
    author = "Levi, Mich{\`e}le",
    title = "{Effective Field Theories of Post-Newtonian Gravity: A comprehensive review}",
    eprint = "1807.01699",
    archivePrefix = "arXiv",
    primaryClass = "hep-th",
    doi = "10.1088/1361-6633/ab12bc",
    journal = "Rept. Prog. Phys.",
    volume = "83",
    number = "7",
    pages = "075901",
    year = "2020"
}

@article{Goldberger:2004jt,
    author = "Goldberger, Walter D. and Rothstein, Ira Z.",
    title = "{An Effective field theory of gravity for extended objects}",
    eprint = "hep-th/0409156",
    archivePrefix = "arXiv",
    reportNumber = "UCSD-PTH-04-17, CMU-HEP-04-06",
    doi = "10.1103/PhysRevD.73.104029",
    journal = "Phys. Rev. D",
    volume = "73",
    pages = "104029",
    year = "2006"
}

@article{Saketh:2022xjb,
    author = "Saketh, M. V. S. and Steinhoff, Jan and Vines, Justin and Buonanno, Alessandra",
    title = "{Modeling horizon absorption in spinning binary black holes using effective worldline theory}",
    eprint = "2212.13095",
    archivePrefix = "arXiv",
    primaryClass = "gr-qc",
    doi = "10.1103/PhysRevD.107.084006",
    journal = "Phys. Rev. D",
    volume = "107",
    number = "8",
    pages = "084006",
    year = "2023"
}

@article{Aoude:2020onz,
    author = "Aoude, Rafael and Haddad, Kays and Helset, Andreas",
    title = "{On-shell heavy particle effective theories}",
    eprint = "2001.09164",
    archivePrefix = "arXiv",
    primaryClass = "hep-th",
    reportNumber = "SAGEX-20-02-E, MITP/20-002",
    doi = "10.1007/JHEP05(2020)051",
    journal = "JHEP",
    volume = "05",
    pages = "051",
    year = "2020"
}

@article{Ivanov:2024sds,
    author = "Ivanov, Mikhail M. and Li, Yue-Zhou and Parra-Martinez, Julio and Zhou, Zihan",
    title = "{Gravitational Raman Scattering in Effective Field Theory: A Scalar Tidal Matching at O(G3)}",
    eprint = "2401.08752",
    archivePrefix = "arXiv",
    primaryClass = "hep-th",
    reportNumber = "MIT-CTP/5664",
    doi = "10.1103/PhysRevLett.132.131401",
    journal = "Phys. Rev. Lett.",
    volume = "132",
    number = "13",
    pages = "131401",
    year = "2024",
    note = "[Erratum: Phys.Rev.Lett. 134, 159901 (2025)]"
}

@article{Saketh:2022wap,
    author = "Saketh, M. V. S. and Vines, Justin",
    title = "{Scattering of gravitational waves off spinning compact objects with an effective worldline theory}",
    eprint = "2208.03170",
    archivePrefix = "arXiv",
    primaryClass = "gr-qc",
    doi = "10.1103/PhysRevD.106.124026",
    journal = "Phys. Rev. D",
    volume = "106",
    number = "12",
    pages = "124026",
    year = "2022"
}

@misc{Ivanov:2026icp,
    author = "Ivanov, Mikhail M. and Li, Yue-Zhou and Parra-Martinez, Julio and Zhou, Zihan",
    title = "{Gravitational Raman Scattering: a Systematic Toolkit for Tidal Effects in General Relativity}",
    eprint = "2602.06951",
    archivePrefix = "arXiv",
    primaryClass = "hep-th",
    reportNumber = "MIT-CTP/6001",
    month = "2",
    year = "2026"
}

@article{Ivanov:2022qqt,
    author = "Ivanov, Mikhail M. and Zhou, Zihan",
    title = "{Vanishing of Black Hole Tidal Love Numbers from Scattering Amplitudes}",
    eprint = "2209.14324",
    archivePrefix = "arXiv",
    primaryClass = "hep-th",
    doi = "10.1103/PhysRevLett.130.091403",
    journal = "Phys. Rev. Lett.",
    volume = "130",
    number = "9",
    pages = "091403",
    year = "2023"
}

@article{Steinhoff:2016rfi,
    author = "Steinhoff, Jan and Hinderer, Tanja and Buonanno, Alessandra and Taracchini, Andrea",
    title = "{Dynamical Tides in General Relativity: Effective Action and Effective-One-Body Hamiltonian}",
    eprint = "1608.01907",
    archivePrefix = "arXiv",
    primaryClass = "gr-qc",
    doi = "10.1103/PhysRevD.94.104028",
    journal = "Phys. Rev. D",
    volume = "94",
    number = "10",
    pages = "104028",
    year = "2016"
}

@article{Goldberger:2005cd,
    author = "Goldberger, Walter D. and Rothstein, Ira Z.",
    title = "{Dissipative effects in the worldline approach to black hole dynamics}",
    eprint = "hep-th/0511133",
    archivePrefix = "arXiv",
    doi = "10.1103/PhysRevD.73.104030",
    journal = "Phys. Rev. D",
    volume = "73",
    pages = "104030",
    year = "2006"
}

@article{Goldberger:2020fot,
    author = "Goldberger, Walter D. and Li, Jingping and Rothstein, Ira Z.",
    title = "{Non-conservative effects on spinning black holes from world-line effective field theory}",
    eprint = "2012.14869",
    archivePrefix = "arXiv",
    primaryClass = "hep-th",
    doi = "10.1007/JHEP06(2021)053",
    journal = "JHEP",
    volume = "06",
    pages = "053",
    year = "2021"
}

@article{Caron-Huot:2025tlq,
    author = "Caron-Huot, Simon and Correia, Miguel and Isabella, Giulia and Solon, Mikhail",
    title = "{Gravitational Wave Scattering via the Born Series: Scalar Tidal Matching to O(G7) and Beyond}",
    eprint = "2503.13593",
    archivePrefix = "arXiv",
    primaryClass = "hep-th",
    doi = "10.1103/qd3c-nfz6",
    journal = "Phys. Rev. Lett.",
    volume = "135",
    number = "19",
    pages = "191601",
    year = "2025"
}

@article{Damour:2009vw,
  author = {Damour, Thibault and Nagar, Alessandro},
  title = {Relativistic tidal properties of neutron stars},
  journal = {Phys. Rev. D},
  volume = {80},
  pages = {084035},
  year = {2009},
  eprint = {0906.0096}
}

@article{Jakobsen:2023pvx,
    author = "Jakobsen, Gustav Uhre and Mogull, Gustav and Plefka, Jan and Sauer, Benjamin",
    title = "{Tidal effects and renormalization at fourth post-Minkowskian order}",
    eprint = "2312.00719",
    archivePrefix = "arXiv",
    primaryClass = "hep-th",
    reportNumber = "HU-EP-23/63-RTG",
    doi = "10.1103/PhysRevD.109.L041504",
    journal = "Phys. Rev. D",
    volume = "109",
    number = "4",
    pages = "L041504",
    year = "2024"
}

@article{Kosmopoulos:2025rfj,
    author = "Kosmopoulos, Dimitrios and Perrone, Davide and Solon, Mikhail",
    title = "{Dynamical Love Numbers for Black Holes and Beyond from Shell Effective Field Theory}",
    eprint = "2512.04002",
    archivePrefix = "arXiv",
    primaryClass = "hep-th",
    reportNumber = "CERN-TH-2025-247, CERN-TH-2025-247",
    doi = "10.1103/mtqm-xz2k",
    journal = "Phys. Rev. Lett.",
    volume = "136",
    number = "21",
    pages = "211401",
    year = "2026"
}

@misc{Solon:2026ubm,
    author = "Solon, Mikhail P.",
    title = "{Universal Closed Form for Dynamical Love Numbers of Black Holes}",
    eprint = "2606.19281",
    archivePrefix = "arXiv",
    primaryClass = "hep-th",
    month = "6",
    year = "2026"
}

@article{Bjerrum-Bohr:2025bqg,
    author = "Bjerrum-Bohr, N. Emil J. and Chen, Gang and Eriksen, Carl Jordan and Shah, Nabha",
    title = "{The gravitational Compton amplitude from flat and curved spacetimes at second post-Minkowskian order}",
    eprint = "2506.19705",
    archivePrefix = "arXiv",
    primaryClass = "hep-th",
    doi = "10.1007/JHEP10(2025)235",
    journal = "JHEP",
    volume = "10",
    pages = "235",
    year = "2025"
}

@article{Bautista:2026qse,
    author = "Bautista, Yilber Fabian and Driesse, Mathias and Haddad, Kays and Jakobsen, Gustav Uhre",
    title = "{Gravitational wave scattering in spinless WQFT}",
    eprint = "2602.06125",
    archivePrefix = "arXiv",
    primaryClass = "hep-th",
    reportNumber = "HU-EP-26/05",
    doi = "10.1007/JHEP05(2026)252",
    journal = "JHEP",
    volume = "05",
    pages = "252",
    year = "2026"
}

@misc{Bjerrum-Bohr:2026fhs,
    author = "Bjerrum-Bohr, N. Emil J. and Chen, Gang and Jordan Eriksen, Carl and Shah, Nabha",
    title = "{The gravitational Compton amplitude at third post-Minkowskian order}",
    eprint = "2602.06947",
    archivePrefix = "arXiv",
    primaryClass = "hep-th",
    reportNumber = "HU-EP-26/07-RTG",
    month = "2",
    year = "2026"
}

@article{Binnington:2009bb,
    author = "Binnington, Taylor and Poisson, Eric",
    title = "{Relativistic theory of tidal Love numbers}",
    eprint = "0906.1366",
    archivePrefix = "arXiv",
    primaryClass = "gr-qc",
    doi = "10.1103/PhysRevD.80.084018",
    journal = "Phys. Rev. D",
    volume = "80",
    pages = "084018",
    year = "2009"
}

@article{Riva:2023Love,
    author = "Riva, Massimiliano Maria and Santoni, Luca and Savi{\'c}, Nikola and Vernizzi, Filippo",
    title = "{Vanishing of Nonlinear Tidal Love Numbers of Schwarzschild Black Holes}",
    eprint = "2312.05065",
    archivePrefix = "arXiv",
    primaryClass = "gr-qc",
    doi = "10.1016/j.physletb.2024.138710",
    journal = "Phys. Lett. B",
    volume = "854",
    pages = "138710",
    year = "2024"
}

@article{Iteanu:2024Love,
    author = "Iteanu, Simon and Riva, Massimiliano Maria and Santoni, Luca and Savi{\'c}, Nikola and Vernizzi, Filippo",
    title = "{Vanishing of Quadratic Love Numbers of Schwarzschild Black Holes}",
    eprint = "2410.03542",
    archivePrefix = "arXiv",
    primaryClass = "gr-qc",
    doi = "10.1007/JHEP02(2025)174",
    journal = "JHEP",
    volume = "02",
    pages = "174",
    year = "2025"
}

@article{Brunello:2026lzf,
    author = "Brunello, Giacomo and Meo, Mario and Smith, Sid",
    title = "{Gravitational Compton scattering at the fifth post-Minkowskian order}",
    eprint = "2608.17946",
    archivePrefix = "arXiv",
    primaryClass = "hep-th",
    month = "8",
    year = "2026"
}

@article{Kol:2011vg,
    author = "Kol, Barak and Smolkin, Michael",
    title = "{Black hole stereotyping: Induced gravito-static polarization}",
    eprint = "1110.3764",
    archivePrefix = "arXiv",
    primaryClass = "hep-th",
    doi = "10.1007/JHEP02(2012)010",
    journal = "JHEP",
    volume = "02",
    pages = "010",
    year = "2012"
}

@article{Ivanov:2022hlo,
    author = "Ivanov, Mikhail M. and Zhou, Zihan",
    title = "{Revisiting the matching of black hole tidal responses: A systematic study of relativistic and logarithmic corrections}",
    eprint = "2208.08459",
    archivePrefix = "arXiv",
    primaryClass = "hep-th",
    doi = "10.1103/PhysRevD.107.084030",
    journal = "Phys. Rev. D",
    volume = "107",
    number = "8",
    pages = "084030",
    year = "2023"
}

@article{Fang:2005qq,
    author = "Fang, Hua and Lovelace, Geoffrey",
    title = "{Tidal coupling of a Schwarzschild black hole and circularly orbiting moon}",
    eprint = "gr-qc/0505156",
    archivePrefix = "arXiv",
    doi = "10.1103/PhysRevD.72.124016",
    journal = "Phys. Rev. D",
    volume = "72",
    pages = "124016",
    year = "2005"
}

@article{Damour:2009va,
    author = "Damour, Thibault and Lecian, Orchidea Maria",
    title = "{On the gravitational polarizability of black holes}",
    eprint = "0906.3003",
    archivePrefix = "arXiv",
    primaryClass = "gr-qc",
    reportNumber = "IHES-P-09-28",
    doi = "10.1103/PhysRevD.80.044017",
    journal = "Phys. Rev. D",
    volume = "80",
    pages = "044017",
    year = "2009"
}

@article{Gurlebeck:2015xpa,
    author = {G{\"u}rlebeck, Norman},
    title = "{No-hair theorem for Black Holes in Astrophysical Environments}",
    eprint = "1503.03240",
    archivePrefix = "arXiv",
    primaryClass = "gr-qc",
    doi = "10.1103/PhysRevLett.114.151102",
    journal = "Phys. Rev. Lett.",
    volume = "114",
    number = "15",
    pages = "151102",
    year = "2015"
}

@article{Poisson:2021yau,
    author = "Poisson, Eric",
    title = "{Tidally induced multipole moments of a nonrotating black hole vanish to all post-Newtonian orders}",
    eprint = "2108.07328",
    archivePrefix = "arXiv",
    primaryClass = "gr-qc",
    doi = "10.1103/PhysRevD.104.104062",
    journal = "Phys. Rev. D",
    volume = "104",
    number = "10",
    pages = "104062",
    year = "2021"
}

@article{LeTiec:2020spy,
    author = "Le Tiec, Alexandre and Casals, Marc",
    title = "{Spinning Black Holes Fall in Love}",
    eprint = "2007.00214",
    archivePrefix = "arXiv",
    primaryClass = "gr-qc",
    doi = "10.1103/PhysRevLett.126.131102",
    journal = "Phys. Rev. Lett.",
    volume = "126",
    number = "13",
    pages = "131102",
    year = "2021"
}

@article{LeTiec:2020bos,
    author = "Le Tiec, Alexandre and Casals, Marc and Franzin, Edgardo",
    title = "{Tidal Love Numbers of Kerr Black Holes}",
    eprint = "2010.15795",
    archivePrefix = "arXiv",
    primaryClass = "gr-qc",
    doi = "10.1103/PhysRevD.103.084021",
    journal = "Phys. Rev. D",
    volume = "103",
    number = "8",
    pages = "084021",
    year = "2021"
}

@article{Chia:2020yla,
    author = "Chia, Horng Sheng",
    title = "{Tidal deformation and dissipation of rotating black holes}",
    eprint = "2010.07300",
    archivePrefix = "arXiv",
    primaryClass = "gr-qc",
    doi = "10.1103/PhysRevD.104.024013",
    journal = "Phys. Rev. D",
    volume = "104",
    number = "2",
    pages = "024013",
    year = "2021"
}

@article{Hui:2021vcv,
    author = "Hui, Lam and Joyce, Austin and Penco, Riccardo and Santoni, Luca and Solomon, Adam R.",
    title = "{Ladder symmetries of black holes. Implications for love numbers and no-hair theorems}",
    eprint = "2105.01069",
    archivePrefix = "arXiv",
    primaryClass = "hep-th",
    doi = "10.1088/1475-7516/2022/01/032",
    journal = "JCAP",
    volume = "01",
    number = "01",
    pages = "032",
    year = "2022"
}

@article{Katagiri:2022vyz,
    author = "Katagiri, Takuya and Kimura, Masashi and Nakano, Hiroyuki and Omukai, Kazuyuki",
    title = "{Vanishing Love numbers of black holes in general relativity: From spacetime conformal symmetry of a two-dimensional reduced geometry}",
    eprint = "2209.10469",
    archivePrefix = "arXiv",
    primaryClass = "gr-qc",
    doi = "10.1103/PhysRevD.107.124030",
    journal = "Phys. Rev. D",
    volume = "107",
    number = "12",
    pages = "124030",
    year = "2023"
}

@article{Berens:2022ebl,
    author = "Berens, Roman and Hui, Lam and Sun, Zimo",
    title = "{Ladder symmetries of black holes and de Sitter space: love numbers and quasinormal modes}",
    eprint = "2212.09367",
    archivePrefix = "arXiv",
    primaryClass = "hep-th",
    doi = "10.1088/1475-7516/2023/06/056",
    journal = "JCAP",
    volume = "06",
    pages = "056",
    year = "2023"
}

@article{Rai:2024lho,
    author = "Rai, Mudit and Santoni, Luca",
    title = {{Ladder symmetries and Love numbers of Reissner-Nordstr{\"o}m black holes}},
    eprint = "2404.06544",
    archivePrefix = "arXiv",
    primaryClass = "gr-qc",
    doi = "10.1007/JHEP07(2024)098",
    journal = "JHEP",
    volume = "07",
    pages = "098",
    year = "2024"
}

@article{Charalambous:2023jgq,
    author = "Charalambous, Panagiotis and Ivanov, Mikhail M.",
    title = "{Scalar Love numbers and Love symmetries of 5-dimensional Myers-Perry black holes}",
    eprint = "2303.16036",
    archivePrefix = "arXiv",
    primaryClass = "hep-th",
    doi = "10.1007/JHEP07(2023)222",
    journal = "JHEP",
    volume = "07",
    pages = "222",
    year = "2023"
}

@article{Gounis:2024hcm,
    author = "Gounis, L. -R. and Kehagias, A. and Riotto, A.",
    title = "{The vanishing of the non-linear static love number of Kerr black holes and the role of symmetries}",
    eprint = "2412.08249",
    archivePrefix = "arXiv",
    primaryClass = "gr-qc",
    doi = "10.1088/1475-7516/2025/03/002",
    journal = "JCAP",
    volume = "03",
    pages = "002",
    year = "2025"
}

@article{Creci:2021rkz,
    author = "Creci, Gast{\'o}n and Hinderer, Tanja and Steinhoff, Jan",
    title = "{Tidal response from scattering and the role of analytic continuation}",
    eprint = "2108.03385",
    archivePrefix = "arXiv",
    primaryClass = "gr-qc",
    doi = "10.1103/PhysRevD.104.124061",
    journal = "Phys. Rev. D",
    volume = "104",
    number = "12",
    pages = "124061",
    year = "2021",
    note = "[Erratum: Phys.Rev.D 105, 109902 (2022)]"
}

@article{Bautista:2021wfy,
    author = "Bautista, Yilber Fabian and Guevara, Alfredo and Kavanagh, Chris and Vines, Justin",
    title = "{Scattering in black hole backgrounds and higher-spin amplitudes. Part I}",
    eprint = "2107.10179",
    archivePrefix = "arXiv",
    primaryClass = "hep-th",
    doi = "10.1007/JHEP03(2023)136",
    journal = "JHEP",
    volume = "03",
    pages = "136",
    year = "2023"
}

@article{Bautista:2022wjf,
    author = "Bautista, Yilber Fabian and Guevara, Alfredo and Kavanagh, Chris and Vines, Justin",
    title = "{Scattering in black hole backgrounds and higher-spin amplitudes. Part II}",
    eprint = "2212.07965",
    archivePrefix = "arXiv",
    primaryClass = "hep-th",
    doi = "10.1007/JHEP05(2023)211",
    journal = "JHEP",
    volume = "05",
    pages = "211",
    year = "2023"
}

@article{Bautista:2023sdf,
    author = "Bautista, Yilber Fabian and Bonelli, Giulio and Iossa, Cristoforo and Tanzini, Alessandro and Zhou, Zihan",
    title = "{Black hole perturbation theory meets CFT2: Kerr-Compton amplitudes from Nekrasov-Shatashvili functions}",
    eprint = "2312.05965",
    archivePrefix = "arXiv",
    primaryClass = "hep-th",
    doi = "10.1103/PhysRevD.109.084071",
    journal = "Phys. Rev. D",
    volume = "109",
    number = "8",
    pages = "084071",
    year = "2024"
}

@article{Perry:2023wmm,
    author = "Perry, Malcolm and Rodriguez, Maria J.",
    title = "{Dynamical Love Numbers for Kerr Black Holes}",
    eprint = "2310.03660",
    archivePrefix = "arXiv",
    primaryClass = "gr-qc",
    month = "10",
    year = "2023"
}

@article{Chakraborty:2023zed,
    author = "Chakraborty, Sumanta and Maggio, Elisa and Silvestrini, Michela and Pani, Paolo",
    title = "{Dynamical tidal Love numbers of Kerr-like compact objects}",
    eprint = "2310.06023",
    archivePrefix = "arXiv",
    primaryClass = "gr-qc",
    doi = "10.1103/PhysRevD.110.084042",
    journal = "Phys. Rev. D",
    volume = "110",
    number = "8",
    pages = "084042",
    year = "2024"
}

@article{Chakraborty:2026dox,
    author = "Chakraborty, Sumanta and Saketh, M. V. S. and Hinderer, Tanja and Steinhoff, Jan",
    title = "{Dynamical tidal Love numbers of black holes under generic perturbations: Connecting black hole perturbation theory with effective field theory}",
    eprint = "2605.00693",
    archivePrefix = "arXiv",
    primaryClass = "gr-qc",
    doi = "10.1103/lfq4-49k1",
    journal = "Phys. Rev. D",
    volume = "114",
    number = "6",
    pages = "064011",
    year = "2026"
}

@article{Wang:2026qst,
    author = "Wang, Luohan and Lehner, Luis and Micol, Mait{\'a} and Sturani, Riccardo",
    title = "{Matching Tidal Deformability (Wilson) Coefficients to Black Hole Love Numbers in Higher-Curvature Gravity}",
    eprint = "2604.04259",
    archivePrefix = "arXiv",
    primaryClass = "gr-qc",
    month = "4",
    year = "2026"
}

@article{Rodriguez:2026iot,
    author = "Rodr{\'\i}guez, Mar{\'\i}a J. and Santoni, Luca and Solomon, Adam R.",
    title = "{Love numbers of black holes and compact objects}",
    eprint = "2604.08653",
    archivePrefix = "arXiv",
    primaryClass = "gr-qc",
    month = "4",
    year = "2026"
}

@article{Bern:2020uwk,
    author = "Bern, Zvi and Parra-Martinez, Julio and Roiban, Radu and Sawyer, Eric and Shen, Chia-Hsien",
    title = "{Leading Nonlinear Tidal Effects and Scattering Amplitudes}",
    eprint = "2010.08559",
    archivePrefix = "arXiv",
    primaryClass = "hep-th",
    reportNumber = "CALT-TH/2020-041",
    doi = "10.1007/JHEP05(2021)188",
    journal = "JHEP",
    volume = "05",
    pages = "188",
    year = "2021"
}

@article{Cheung:2020sdj,
    author = "Cheung, Clifford and Solon, Mikhail P.",
    title = "{Tidal Effects in the Post-Minkowskian Expansion}",
    eprint = "2006.06665",
    archivePrefix = "arXiv",
    primaryClass = "hep-th",
    reportNumber = "CALT-TH 2020-025",
    doi = "10.1103/PhysRevLett.125.191601",
    journal = "Phys. Rev. Lett.",
    volume = "125",
    number = "19",
    pages = "191601",
    year = "2020"
}

@article{Haddad:2020que,
    author = "Haddad, Kays and Helset, Andreas",
    title = "{Tidal effects in quantum field theory}",
    eprint = "2008.04920",
    archivePrefix = "arXiv",
    primaryClass = "hep-th",
    doi = "10.1007/JHEP12(2020)024",
    journal = "JHEP",
    volume = "12",
    pages = "024",
    year = "2020"
}

@article{Kalin:2020lmz,
    author = {K{\"a}lin, Gregor and Liu, Zhengwen and Porto, Rafael A.},
    title = "{Conservative Tidal Effects in Compact Binary Systems to Next-to-Leading Post-Minkowskian Order}",
    eprint = "2008.06047",
    archivePrefix = "arXiv",
    primaryClass = "hep-th",
    reportNumber = "DESY-20-131, DESY 20-131, SLAC-PUB-17555",
    doi = "10.1103/PhysRevD.102.124025",
    journal = "Phys. Rev. D",
    volume = "102",
    pages = "124025",
    year = "2020"
}

@article{Bini:2020flp,
    author = "Bini, Donato and Damour, Thibault and Geralico, Andrea",
    title = "{Scattering of tidally interacting bodies in post-Minkowskian gravity}",
    eprint = "2001.00352",
    archivePrefix = "arXiv",
    primaryClass = "gr-qc",
    doi = "10.1103/PhysRevD.101.044039",
    journal = "Phys. Rev. D",
    volume = "101",
    number = "4",
    pages = "044039",
    year = "2020"
}

@article{Pereniguez:2021xcj,
    author = "Pere{\~n}iguez, David and Cardoso, Vitor",
    title = "{Love numbers and magnetic susceptibility of charged black holes}",
    eprint = "2112.08400",
    archivePrefix = "arXiv",
    primaryClass = "gr-qc",
    doi = "10.1103/PhysRevD.105.044026",
    journal = "Phys. Rev. D",
    volume = "105",
    number = "4",
    pages = "044026",
    year = "2022"
}

@article{Damgaard:2019lfh,
    author = "Damgaard, Poul H. and Haddad, Kays and Helset, Andreas",
    title = "{Heavy Black Hole Effective Theory}",
    eprint = "1908.10308",
    archivePrefix = "arXiv",
    primaryClass = "hep-ph",
    doi = "10.1007/JHEP11(2019)070",
    journal = "JHEP",
    volume = "11",
    pages = "070",
    year = "2019"
}

@article{Brandhuber:2021eyq,
    author = "Brandhuber, Andreas and Chen, Gang and Travaglini, Gabriele and Wen, Congkao",
    title = "{Classical gravitational scattering from a gauge-invariant double copy}",
    eprint = "2108.04216",
    archivePrefix = "arXiv",
    primaryClass = "hep-th",
    reportNumber = "QMUL-PH-21-18, SAGEX-21-07",
    doi = "10.1007/JHEP10(2021)118",
    journal = "JHEP",
    volume = "10",
    pages = "118",
    year = "2021"
}

@article{Donoghue:1994dn,
    author = "Donoghue, John F.",
    title = "{General relativity as an effective field theory: The leading quantum corrections}",
    eprint = "gr-qc/9405057",
    archivePrefix = "arXiv",
    reportNumber = "UMHEP-408",
    doi = "10.1103/PhysRevD.50.3874",
    journal = "Phys. Rev. D",
    volume = "50",
    pages = "3874--3888",
    year = "1994"
}

@article{Bjerrum-Bohr:2002gqz,
    author = "Bjerrum-Bohr, N. E. J and Donoghue, John F. and Holstein, Barry R.",
    title = "{Quantum gravitational corrections to the nonrelativistic scattering potential of two masses}",
    eprint = "hep-th/0211072",
    archivePrefix = "arXiv",
    doi = "10.1103/PhysRevD.71.069903",
    journal = "Phys. Rev. D",
    volume = "67",
    pages = "084033",
    year = "2003",
    note = "[Erratum: Phys.Rev.D 71, 069903 (2005)]"
}

@article{Bjerrum-Bohr:2014zsa,
    author = "Bjerrum-Bohr, N. E. J. and Donoghue, John F. and Holstein, Barry R. and Plant{\'e}, Ludovic and Vanhove, Pierre",
    title = "{Bending of Light in Quantum Gravity}",
    eprint = "1410.7590",
    archivePrefix = "arXiv",
    primaryClass = "hep-th",
    reportNumber = "IPHT-T14-108, IHES-P-14-33, ACFI-T14-21",
    doi = "10.1103/PhysRevLett.114.061301",
    journal = "Phys. Rev. Lett.",
    volume = "114",
    number = "6",
    pages = "061301",
    year = "2015"
}

@article{Bjerrum-Bohr:2016hpa,
    author = "Bjerrum-Bohr, N. E. J. and Donoghue, John F. and Holstein, Barry R. and Plante, Ludovic and Vanhove, Pierre",
    title = "{Light-like Scattering in Quantum Gravity}",
    eprint = "1609.07477",
    archivePrefix = "arXiv",
    primaryClass = "hep-th",
    reportNumber = "IPHT-T16-082, ACFI-T16-23",
    doi = "10.1007/JHEP11(2016)117",
    journal = "JHEP",
    volume = "11",
    pages = "117",
    year = "2016"
}

@article{Beneke:1997zp,
    author = "Beneke, M. and Smirnov, Vladimir A.",
    title = "{Asymptotic expansion of Feynman integrals near threshold}",
    eprint = "hep-ph/9711391",
    archivePrefix = "arXiv",
    reportNumber = "CERN-TH-97-315",
    doi = "10.1016/S0550-3213(98)00138-2",
    journal = "Nucl. Phys. B",
    volume = "522",
    pages = "321--344",
    year = "1998"
}

@article{tHooft:1974toh,
    author = "'t Hooft, Gerard and Veltman, M. J. G.",
    title = "{One-loop divergencies in the theory of gravitation}",
    doi = "10.1142/9789814539395_0001",
    journal = "Ann. Inst. H. Poincare Phys. Theor. A",
    volume = "20",
    number = "1",
    pages = "69--94",
    year = "1974"
}

@inproceedings{Goldberger:2022ebt,
    author = "Goldberger, Walter D.",
    title = "{Effective field theories of gravity and compact binary dynamics: A Snowmass 2021 whitepaper}",
    booktitle = "{Snowmass 2021}",
    eprint = "2206.14249",
    archivePrefix = "arXiv",
    primaryClass = "hep-th",
    month = "6",
    year = "2022"
}

@article{Chetyrkin:1981qh,
    author = "Chetyrkin, K. G. and Tkachov, F. V.",
    title = "{Integration by parts: The algorithm to calculate $\beta$-functions in 4 loops}",
    doi = "10.1016/0550-3213(81)90199-1",
    journal = "Nucl. Phys. B",
    volume = "192",
    pages = "159--204",
    year = "1981"
}

@article{Passarino:1978jh,
    author = "Passarino, G. and Veltman, M. J. G.",
    title = "{One Loop Corrections for e+ e- Annihilation Into mu+ mu- in the Weinberg Model}",
    reportNumber = "Print-79-0284 (UTRECHT)",
    doi = "10.1016/0550-3213(79)90234-7",
    journal = "Nucl. Phys. B",
    volume = "160",
    pages = "151--207",
    year = "1979"
}

@book{Smirnov:2012gma,
    author = "Smirnov, Vladimir A.",
    title = "{Analytic tools for Feynman integrals}",
    doi = "10.1007/978-3-642-34886-0",
    volume = "250",
    year = "2012"
}

@article{Bern:1994zx,
    author = "Bern, Zvi and Dixon, Lance J. and Dunbar, David C. and Kosower, David A.",
    title = "{One loop n point gauge theory amplitudes, unitarity and collinear limits}",
    eprint = "hep-ph/9403226",
    archivePrefix = "arXiv",
    reportNumber = "SLAC-PUB-6415, SACLAY-SPH-T-94-20, UCLA-TEP-94-4, SWAT-94-17",
    doi = "10.1016/0550-3213(94)90179-1",
    journal = "Nucl. Phys. B",
    volume = "425",
    pages = "217--260",
    year = "1994"
}

@article{Bloch:1937pw,
    author = "Bloch, F. and Nordsieck, A.",
    title = "{Note on the Radiation Field of the electron}",
    reportNumber = "RX-1199",
    doi = "10.1103/PhysRev.52.54",
    journal = "Phys. Rev.",
    volume = "52",
    pages = "54--59",
    year = "1937"
}

@article{Kinoshita:1962ur,
    author = "Kinoshita, T.",
    title = "{Mass singularities of Feynman amplitudes}",
    doi = "10.1063/1.1724268",
    journal = "J. Math. Phys.",
    volume = "3",
    pages = "650--677",
    year = "1962"
}

@article{Lee:1964is,
    author = "Lee, T. D. and Nauenberg, M.",
    editor = "Feinberg, G.",
    title = "{Degenerate Systems and Mass Singularities}",
    doi = "10.1103/PhysRev.133.B1549",
    journal = "Phys. Rev.",
    volume = "133",
    pages = "B1549--B1562",
    year = "1964"
}

@article{Kabat:1992tb,
    author = "Kabat, Daniel N. and Ortiz, Miguel",
    title = "{Eikonal quantum gravity and Planckian scattering}",
    eprint = "hep-th/9203082",
    archivePrefix = "arXiv",
    reportNumber = "MIT-CTP-2069",
    doi = "10.1016/0550-3213(92)90627-N",
    journal = "Nucl. Phys. B",
    volume = "388",
    pages = "570--592",
    year = "1992"
}

@article{Penna:2018gfx,
    author = "Penna, Robert F.",
    title = "{Near-horizon Carroll symmetry and black hole Love numbers}",
    eprint = "1812.05643",
    archivePrefix = "arXiv",
    primaryClass = "hep-th",
    month = "12",
    year = "2018"
}

@article{BenAchour:2022uqo,
    author = "Ben Achour, Jibril and Livine, Etera R. and Mukohyama, Shinji and Uzan, Jean-Philippe",
    title = "{Hidden symmetry of the static response of black holes: applications to Love numbers}",
    eprint = "2202.12828",
    archivePrefix = "arXiv",
    primaryClass = "gr-qc",
    reportNumber = "YITP-22-17, IPMU22-0003",
    doi = "10.1007/JHEP07(2022)112",
    journal = "JHEP",
    volume = "07",
    pages = "112",
    year = "2022"
}

@article{Sharma:2024hlz,
    author = "Sharma, Chanchal and Ghosh, Rajes and Sarkar, Sudipta",
    title = "{Exploring ladder symmetry and Love numbers for static and rotating black holes}",
    eprint = "2401.00703",
    archivePrefix = "arXiv",
    primaryClass = "gr-qc",
    doi = "10.1103/PhysRevD.109.L041505",
    journal = "Phys. Rev. D",
    volume = "109",
    number = "4",
    pages = "L041505",
    year = "2024"
}

@article{Charalambous:2024tdj,
    author = "Charalambous, Panagiotis",
    title = "{Love numbers and Love symmetries for p-form and gravitational perturbations of higher-dimensional spherically symmetric black holes}",
    eprint = "2402.07574",
    archivePrefix = "arXiv",
    primaryClass = "hep-th",
    doi = "10.1007/JHEP04(2024)122",
    journal = "JHEP",
    volume = "04",
    pages = "122",
    year = "2024"
}

@mastersthesis{Charalambous:2024gpf,
    author = "Charalambous, Panagiotis",
    title = "{Magic zeroes in the black hole response problem and a Love symmetry resolution}",
    eprint = "2404.17030",
    archivePrefix = "arXiv",
    primaryClass = "hep-th",
    type = "Other thesis",
    month = "4",
    year = "2024"
}

@article{Gray:2024qys,
    author = "Gray, Finnian and Keeler, Cynthia and Kubiznak, David and Martin, Victoria",
    title = "{Love symmetry in higher-dimensional rotating black hole spacetimes}",
    eprint = "2409.05964",
    archivePrefix = "arXiv",
    primaryClass = "gr-qc",
    doi = "10.1007/JHEP03(2025)036",
    journal = "JHEP",
    volume = "03",
    pages = "036",
    year = "2025"
}

@article{Combaluzier-Szteinsznaider:2024sgb,
    author = "Combaluzier-Szteinsznaider, Oscar and Hui, Lam and Santoni, Luca and Solomon, Adam R. and Wong, Sam S. C.",
    title = "{Symmetries of vanishing nonlinear Love numbers of Schwarzschild black holes}",
    eprint = "2410.10952",
    archivePrefix = "arXiv",
    primaryClass = "gr-qc",
    doi = "10.1007/JHEP03(2025)124",
    journal = "JHEP",
    volume = "03",
    pages = "124",
    year = "2025"
}

@article{Charalambous:2025ekl,
    author = "Charalambous, Panagiotis and Dubovsky, Sergei and Ivanov, Mikhail M.",
    title = "{Love numbers of black p-branes: fine tuning, Love symmetries, and their geometrization}",
    eprint = "2502.02694",
    archivePrefix = "arXiv",
    primaryClass = "hep-th",
    reportNumber = "MIT-CTP/5836",
    doi = "10.1007/JHEP06(2025)180",
    journal = "JHEP",
    volume = "06",
    pages = "180",
    year = "2025"
}

@article{Berens:2025okm,
    author = "Berens, Roman and Hui, Lam and McLoughlin, Daniel and Penco, Riccardo and Staunton, John",
    title = "{Geometric symmetries for the vanishing of the black hole tidal Love numbers}",
    eprint = "2510.18952",
    archivePrefix = "arXiv",
    primaryClass = "hep-th",
    doi = "10.1088/1475-7516/2026/05/096",
    journal = "JCAP",
    volume = "05",
    pages = "096",
    year = "2026"
}

@article{Sharma:2025xii,
    author = "Sharma, Chanchal and Roy, Shuvayu and Sarkar, Sudipta",
    title = "{Ladder symmetry: The necessary and sufficient condition for vanishing Love numbers}",
    eprint = "2511.09670",
    archivePrefix = "arXiv",
    primaryClass = "gr-qc",
    doi = "10.1103/44dg-smt2",
    journal = "Phys. Rev. D",
    volume = "113",
    number = "2",
    pages = "024066",
    year = "2026"
}

@article{Guevara:2025psg,
    author = "Guevara, Alfredo and Kol, Uri",
    title = "{New Near Extremal Black Holes and Love Symmetry}",
    eprint = "2511.18637",
    archivePrefix = "arXiv",
    primaryClass = "hep-th",
    month = "11",
    year = "2025"
}

@article{DeLuca:2025zqr,
    author = "De Luca, Valerio and Khek, Brandon and Khoury, Justin and Trodden, Mark",
    title = "{Hidden symmetries for tidal Love numbers: Generalities and applications to analog black holes}",
    eprint = "2512.06082",
    archivePrefix = "arXiv",
    primaryClass = "gr-qc",
    doi = "10.1103/5bdk-xclx",
    journal = "Phys. Rev. D",
    volume = "113",
    number = "4",
    pages = "044006",
    year = "2026"
}

@article{Cvetic:2026wht,
    author = "Cveti{\v{c}}, M. and Liao, M. A. and Stetsko, M. M.",
    title = "{Tidal perturbations and Love symmetry for five-dimensional charged rotating black holes}",
    eprint = "2601.20514",
    archivePrefix = "arXiv",
    primaryClass = "hep-th",
    doi = "10.1103/xtcv-bsdn",
    journal = "Phys. Rev. D",
    volume = "113",
    number = "8",
    pages = "085008",
    year = "2026"
}

@article{Correia:2026utp,
    author = "Correia, Miguel and Isabella, Giulia and Wolz, Anna M.",
    title = "{Gravitational Compton Amplitude to All Orders in Perturbation Theory}",
    eprint = "2608.26284",
    archivePrefix = "arXiv",
    primaryClass = "hep-th",
    month = "8",
    year = "2026"
}

@article{Barack:2023oqp,
    author = "Barack, Leor and others",
    title = "{Comparison of post-Minkowskian and self-force expansions: Scattering in a scalar charge toy model}",
    eprint = "2304.09200",
    archivePrefix = "arXiv",
    primaryClass = "hep-th",
    doi = "10.1103/PhysRevD.108.024025",
    journal = "Phys. Rev. D",
    volume = "108",
    number = "2",
    pages = "024025",
    year = "2023"
}

@article{Chakraborty:2026qru,
    author = "Chakraborty, Sumanta and Pani, Paolo",
    title = "{Tidal Response of Compact Objects}",
    eprint = "2604.08679",
    archivePrefix = "arXiv",
    primaryClass = "gr-qc",
    month = "4",
    year = "2026"
}

@article{Barbosa:2025uau,
    author = "Barbosa, Sergio and Brax, Philippe and Fichet, Sylvain and de Souza, Lucas",
    title = "{Running Love numbers and the Effective Field Theory of gravity}",
    eprint = "2501.18684",
    archivePrefix = "arXiv",
    primaryClass = "hep-th",
    doi = "10.1088/1475-7516/2025/07/071",
    journal = "JCAP",
    volume = "07",
    pages = "071",
    year = "2025"
}

@article{Barbosa:2026qcv,
    author = "Barbosa, Sergio and Fichet, Sylvain and de Souza, Lucas",
    title = "{Running Love numbers of charged black holes}",
    eprint = "2602.00349",
    archivePrefix = "arXiv",
    primaryClass = "hep-th",
    doi = "10.1088/1475-7516/2026/09/009",
    journal = "JCAP",
    volume = "09",
    pages = "009",
    year = "2026"
}

@article{Ghosh:2026vig,
    author = "Ghosh, Rajes and Bhatt, Rajendra Prasad and Chakraborty, Sumanta and Bose, Sukanta",
    title = "{Universal Ladder Structure Across Scales: From Quantum to Black Hole Physics}",
    eprint = "2604.06249",
    archivePrefix = "arXiv",
    primaryClass = "gr-qc",
    month = "4",
    year = "2026"
}

\end{document}